\documentclass[twocolumn]{aastex63}

\usepackage{lineno}
\received{January 20, 2022}
\revised{March 3, 2022}
\accepted{March 3, 2022}
\submitjournal{ApJ}

\shorttitle{Radio continuum emission from young nearby stars}
\shortauthors{Launhardt et al.}

\begin{document}

\title{Non-thermal radio continuum emission from young nearby stars}

\correspondingauthor{Ralf Launhardt}
\email{rl@mpia.de}

\author{Ralf Launhardt} 
\affiliation{Max-Planck-Institut f\"ur Astronomie, K\"onigstuhl 17, 69117 Heidelberg, Germany}

\author{Laurent Loinard}
\affiliation{Instituto de Radioastronom\'{\i}a y Astrof\'{\i}sica, Universidad Nacional Aut\'onoma de M\'exico Apartado Postal 3-72, 58090, Morelia, Michoac\'an, Mexico}
\affiliation{Instituto de Astronom\'{\i}a, Universidad Nacional Aut\'onoma de M\'exico, Apartado Postal 70-264, Ciudad de M\'exico 04510, Mexico}

\author{Sergio A. Dzib}
\affiliation{Max-Planck-Institut f\"ur Radioastronomie, Auf dem H\"ugel 69, 53121 Bonn, Germany}

\author{Jan Forbrich}
\affiliation{Centre for Astrophysics Research, University of Hertfordshire, College Lane, Hatfield AL10 9AB, UK}
\affiliation{Center for Astrophysics $\vert$\,Havard\,\&\,Smithsonian, 60\ Garden Street, Cambridge, MA 02138, USA}

\author{Geoffrey C. Bower}
\affiliation{Academia Sinica Institute of Astronomy and Astrophysics, 645 N. A'ohoku Place, Hilo, HI 96720, USA}

\author{Thomas K. Henning} 
\affiliation{Max-Planck-Institut f\"ur Astronomy, K\"onigstuhl 17, 69117 Heidelberg, Germany}

\author{Amy J. Mioduszewski}
\affiliation{National Radio Astronomy Observatory, Array Operations Center, 1003 Lopezville Road, Socorro, NM\,87801, USA}

\author{Sabine Reffert}
\affiliation{Landessternwarte, Zentrum f\"ur Astronomie der Universit\"at Heidelberg, K\"onigstuhl 12, 69117, Heidelberg, Germany}




\begin{abstract}
Young and magnetically active low-mass stars often exhibit non-thermal coronal radio emission due to the gyration of electrons in their magnetized chromospheres. This emission is easily detectable at centimeter wavelengths with the current sensitivity of large radio interferometers like the VLA. With the aim of identifying nearby stars adequate for future accurate radio astrometric monitoring using Very Long Baseline Interferometry (VLBI), we have used the VLA in its B\, configuration to search for radio emission at \mbox{$\nu \simeq 6$\,GHz} \mbox{($\lambda \simeq 5$\,cm)} toward a sample of 170 nearby ($<$130\,pc) mostly young (5\,--\,500\,Myr) stars of spectral types between F4 and M2. At our mean $3\sigma$ detection limit of $\simeq$\,50\,$\mu$Jy, we identify 31 young stars with coronal radio emission (an 18\% system detection rate) and more than 600 background (most likely extra-galactic) sources. Among the targeted stars, we find a significant decline of the detection rate with age from 56$\pm$20\% for stars with ages $\le10$\,Myr to 10$\pm$3\% for stars with ages 100\,--\,200\,Myr. No star older than 200\,Myr was detected. The detection rate also declines with $T_{\rm eff}$ from 36$\pm$10\% for stars with $T_{\rm eff}<4000$\,K to 13$\pm$3\% for earlier spectral types with $T_{\rm eff}>5000$\,K. The binarity fraction among the radio-bright stars is at least twice as high as among the radio-quiet stars. The radio-bright nearby young stars identified here provide an interesting sample for future astrometric studies using VLBI arrays aimed at searching for hitherto unknown tight binary components or even exoplanets.
\end{abstract}


\keywords{radiation mechanisms: non-thermal --- 
          stars: pre-main sequence --- 
          radio continuum: stars --- 
          magnetic fields --- 
          techniques: interferometric}



\section{Introduction} \label{sec:intro}

Young low-mass stars are often magnetically and chromospherically active due to their convective envelopes that enable dynamo processes \citep{bouvier2014}. One of the many possible manifestations of this activity is the existence of coronal radio emission detectable at centimeter wavelengths. Such radio emission is typically highly variable with flares and is largely accepted to result from the gyration of electrons in dynamo-driven stellar magnetic fields \citep{feigelson1985,Dulk1985,guedel2002}. Gyro-synchrotron radiation, where the population of electrons is mildly relativistic, seems to be the most common situation, but maser-amplified cyclotron emission (where the electrons are non-relativistic) and synchrotron radiation (associated with highly relativistic electrons) have been suggested in a few cases \citep[][]{dzib2010,deller2013}. Regardless of the exact emission mechanism, the emission is normally confined to regions extending at most a few stellar radii \citep{smith2003,massi2006,Torres2012}, making high resolution Very Long Baseline Interferometry (VLBI) observations of coronal radio emission a powerful tool to study the astrometry of low-mass young stars \citep[e.g.,][]{ortiz-leon2017a, dzib2021}. Furthermore, milli-arcsecond resolution VLBI observations would efficiently filter out any possible thermal free-free contribution to the radio fluxes, since even faint detections will have corresponding brightness temperatures of several $10^6$\,K \citep{forbrich2021}.

The {\it Gould's Belt Very Large Array survey} of nearby regions \citep{loinard2011,dzib2013,dzib2015,kounkel2014,ortiz-leon2015,pech2016} has shown that, with the current level of sensitivity reachable in large-scale surveys ($\sigma\approx20\,\mu$Jy), the fraction of young stellar objects (YSOs) with coronal emission that can be detected at radio wavelengths is of order 10-30\% in regions of on-going star-formation located within a few hundred parsecs around the Sun, such as Taurus, Ophiuchus, or Perseus. In such regions, the most frequently detected stars with coronal radio emission are young, but no longer embedded, Class\,III weak-line T\,Tau stars (TTS) of spectral types K and earlier, but also somewhat younger Class\,II classical TTS, often with IR excess originating from circumstellar disks.\footnote{Note that YSOs, especially when at earlier evolutionary stages (Class\,0 and I), can also emit at radio wavelengths because of thermal free-free emission associated with shock-ionized material in, e.g., jets, winds, or accretion flows \citep[e.g.,][]{andre1987,anglada1992,eisloeffel2000,anglada2018}. We may not always be able to disentangle these mechanisms and they can also occur simultaneously. However, these mechanisms are not considered here as they require the presence of dense circumstellar material, not expected to be present in stars with ages larger than a few Myr.} 

Although it has been shown that non-thermal (coronal) radio emission is predominantly detected towards low-mass (spectral types M, K, and G) Class\,II classical TTS and Class\,III weak-line TTS \citep[e.g.,][]{pech2016}, little is known empirically about the evolution with age of coronal radio emission. Theoretically, such emission is expected to decrease with age, since it depends on the stellar magnetic activity, which in turn is linked to stellar rotation period that  increases on a timescale of a few 100\,Myr due to magnetic braking \citep[e.g.,][]{guedel2002,guedel2004,bouvier2014}. Thus, it remains to be shown how frequently weak-lined T\,Tauri stars with ages between 10 and a few 100\,Myr exhibit strong coronal emission and if the emission indeed declines on that timescale.

In this article, we present the results of a survey for coronal radio emission from young low-mass stars in the immediate solar neighbourhood at distances between 8 and 130\,pc, i.e., inside the Local Bubble \citep{zucker2022} and Gould's Belt and thus complementary to the {\it Gould's Belt Very Large Array survey}. The ultimate goal of our survey was to identify suitable target stars for an anticipated astrometric planet search with the Very Long Baseline Array \citep[VLBA; see., e.g.,][]{Bower2009,curiel2020}. At the same time, our survey investigates systematically, for the first time, the coronal radio emission from the most nearby young solar-mass stars around the Sun.

The paper is structured as follows. 
In Sect.\,\ref{sec:sources}, we introduce the target selection and source list of our survey. 
Section\,\ref{sec:obs} describes the Karl G.\ Jansky Very Large Array (VLA) observations and data reduction.
The results of the survey are presented in Sect.\,\ref{sec:res}
and the relation between radio emission and stellar properties are discussed in Sect.\,\ref{sec:dis}.
Finally, the paper is summarised in Sect.\,\ref{sec:sum}.


\section{Target selection and source list} \label{sec:sources}


\begin{figure}[htb!]
\epsscale{.95}
\plotone{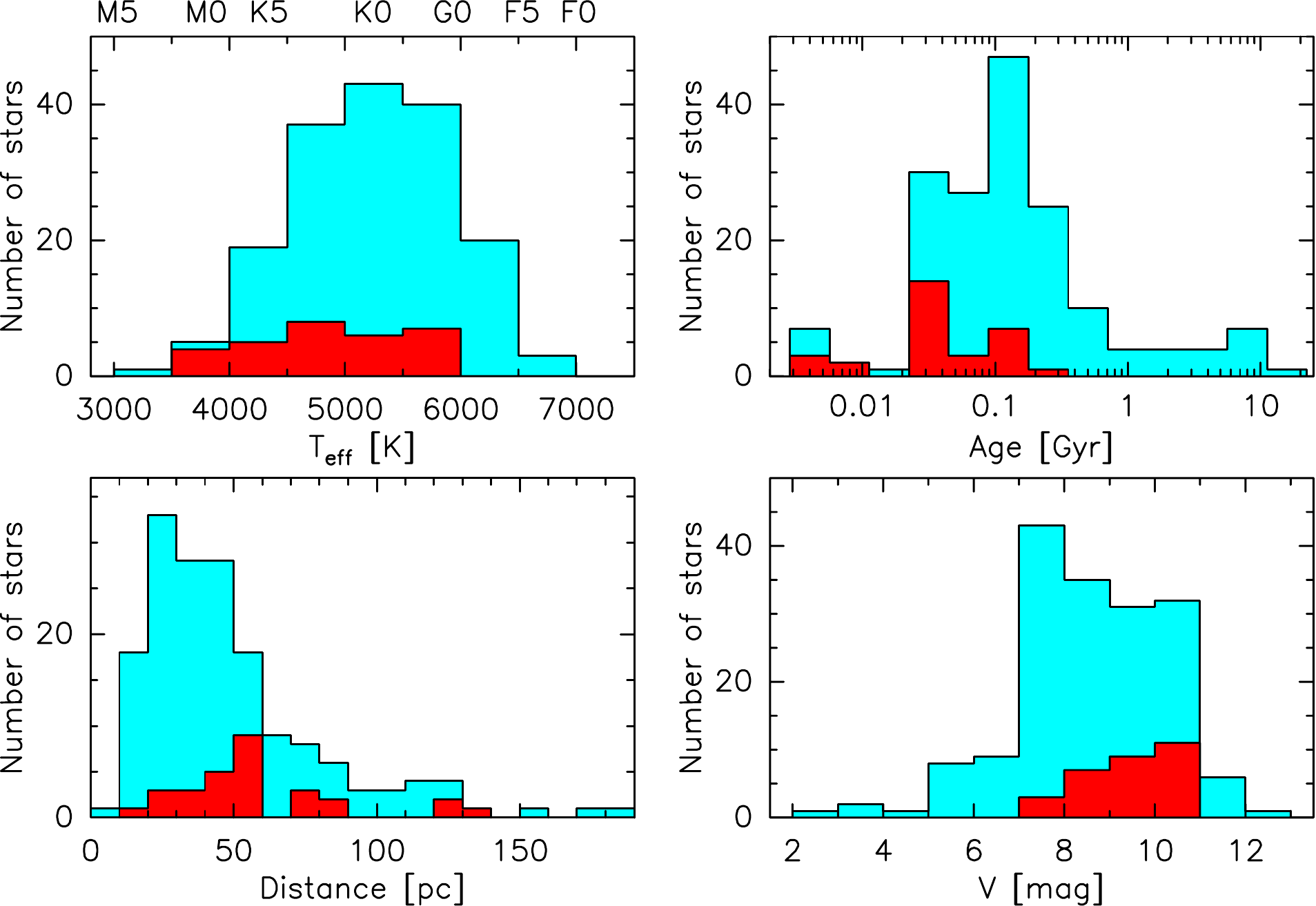}
\caption{\label{fig:sourcedist} Distribution of stellar effective temperatures, ages, distances, and $V$\,magnitudes for the observed (light blue histograms) and detected (at 6\,GHz) stars (red histograms).}
\end{figure}

\begin{figure}[htb!]
\epsscale{.9}
\plotone{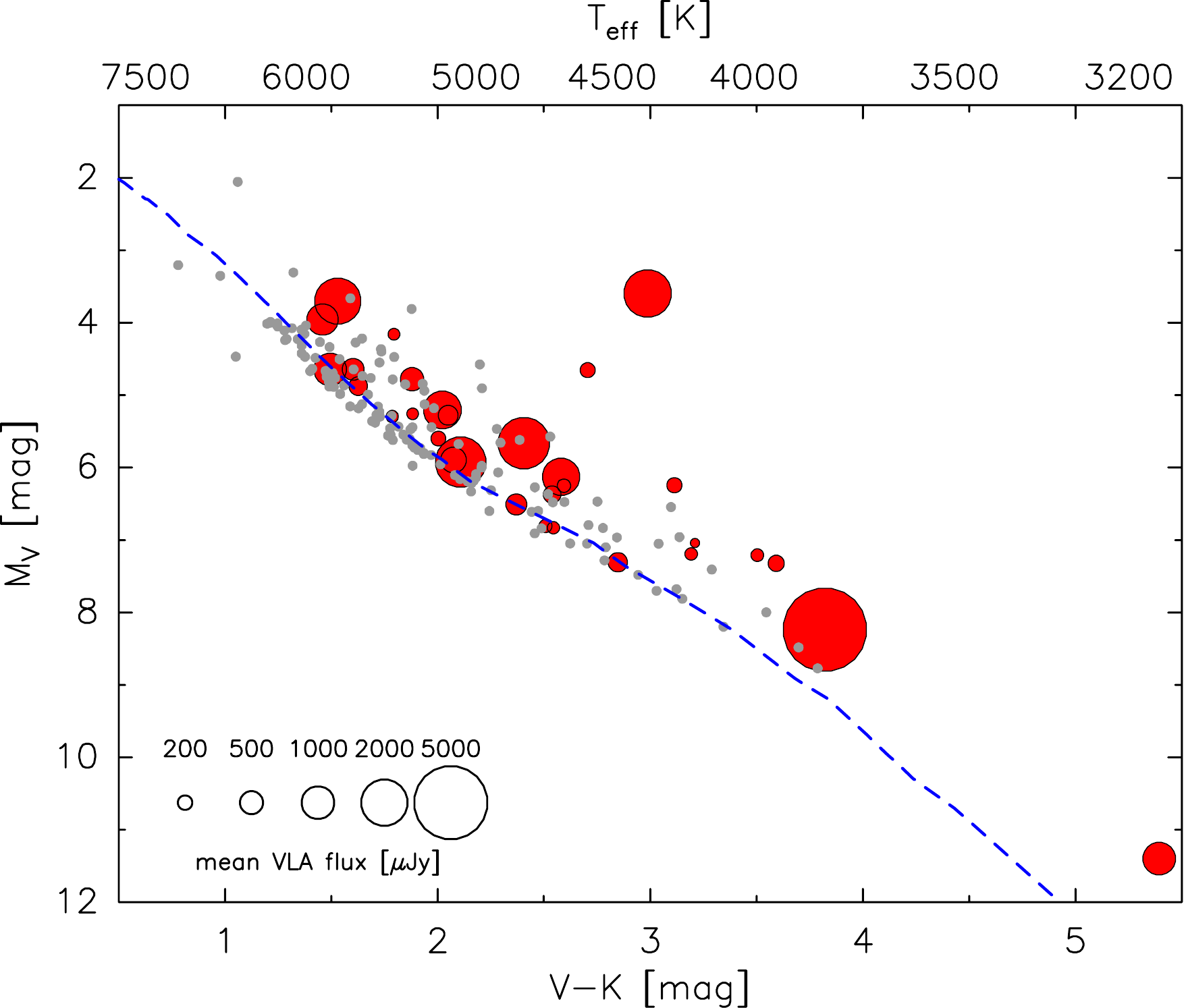}
\caption{\label{fig:cmd} 
Colour-magnitude diagram of observed target stars. Non-detected stars (at 6\,GHz) are marked by grey dots, detected stars are marked by filled red circles. The symbol size scales with the mean VLA flux following the scale displayed in the lower left corner. To guide the eye, the main sequence from \citet{pm2013} is marked by a dashed blue curve and the effective temperatures corresponding to the \mbox{$V-K$} colour of main-sequence stars are marked on top.}
\end{figure}

Owing to the findings about the coronal gyro-synchrotron radio emission mentioned above and our goal to identify targets for an anticipated astrometric planet search, we restricted our source list to stars within 130\,pc around the Sun (based on {\it Hipparcos} parallaxes) that have published age estimates below 300\,Myr, and to stars with roughly solar masses in the spectral range late K to mid F. 
Due to the latitude distribution of the VLBA telescopes (for the anticipated astrometric monitoring), we further restricted the sample to targets with Decl.\,$\ge-20\degr$. Finally, we rejected known spectroscopic binaries for which the later astrometric analysis would be complicated or even impossible. In some individual cases, where, e.g., the distance, or spectral type had a non-negligible uncertainty, the boundaries of these selection criteria were not applied very strictly.

With these selection criteria, we have compiled our target list from the body of literature devoted to the kinematics of young stars in the solar neighbourhood \citep{montes2001,wichmann2003,zuckerman2004,lopez2006,dasilva2009,montes2009,maldonado2010,zuckerman2011,nakajima2012}, and studies of the evolution of protoplanetary disks and debris disks \citep{meyer2006,carpenter2009}. The resulting list contains in total 220 stars, of which 169 stars were observed with the VLA. 
The remaining $\approx$50 stars in the R.A. range $17.5\pm2$\,hrs could not be scheduled for observations due to over-subscription of the Galactic center R.A. range. 

While the original target selection was based on the criteria and resources described above, we later re-evaluated the stellar effective temperatures, $T_{\rm eff}$, distances, $d$, and ages, $t_{\ast}$. 
$T_{\rm eff}$\ was derived by fitting simultaneous stellar \citep[\textsc{phoenix};][]{husser2013} and blackbody models (to account for disk excess emission where present) to the observed photometry and spectra as described in \citet{launhardt2020}. 
The distances are re-derived from the {\it Gaia}-DR2 (hereafter GDR2) parallaxes \citep{gaia_dr2} according to the formalism described by \citet{bailer2018}. For four targets, which have no valid GDR2 parallax solutions, distances are derived by other means as indicated in Tab.\,\ref{tbl-sourcelist}.

To re-evaluate the stellar ages, we first checked each target for membership of known young associations (moving groups) using the banyan $\Sigma$\ tool\footnote{http://www.exoplanetes.umontreal.ca/banyan/banyansigma.php} \citep{gagne2018}. If the membership probability was $\geq$80\%, then we assigned the stellar age $t_{\ast}$\ to be the mean age of the association\footnote{For five stars with membership probabilities between 55 and 75\%, we also assigned the mean age of the association because these ages are widely used in the literature. These stars are marked in Table\,\ref{tbl-sourcelist}.}. In addition, we adopted association ages for 31 more targets that were identified by other authors \citep[][and others]{montes2001,dasilva2009} as likely association members. In total we thus assigned association ages to 99 of our 169 observed target stars. For the remaining 70 field stars, ages were assigned by compiling various literature estimates \citep[particularly from large surveys with diverse age-determination methods, e.g.,][]{baffles2020}. If multiple valid age estimates were available, then we adopted an age (and conservative uncertainty range) that accounted for the spread in estimates. 

The complete list of observed stars with their basic properties is compiled in Table\,\ref{tbl-sourcelist}. Stars with VLA detections are marked in boldface. Figure\,\ref{fig:sourcedist} shows the distributions of $T_{\rm eff}$\ stellar effective temperatures and corresponding spectral types, ages, distances, and $V$\,magnitudes of all observed and detected stars. Figure\,\ref{fig:cmd} shows a \mbox{$V-K$} color-magnitude diagram (CMD) of the target stars, also with the 6\,GHz detections marked. It can already be seen in both Figs. that nearly all VLA-detected stars are young ($<$200\,Myr) and lie above the main-sequence, but these correlations are discussed in more detail in Sect.\,\ref{sec:dis:age}.
Stellar effective temperatures, $T_{\rm eff}$\ range from 3000\,K to 6700\,K (one outlier with 10,000\,K) with a median of 5315\,K. Spectral types range from F4 to M2, with a median of G8. 
Distances range from 8 to 400\,pc, with a median of 43\,pc and only six stars having distances $>$130\,pc. Visual magnitudes range from 3 to 13.8\,mag, with a median of 8.6\,mag.
Ages range from $\lesssim$5\,Myr\footnote{Because of the uncertainties involved, we treat the age of the TAU association, 1-2\,Myr \citep{KH1995,gagne2018} as $\lesssim$5\,Myr.} to 12\,Gyr with a median of 149\,Myr. Other than anticipated in our original target selection, our list now contains 25 stars with ages $>$500\,Myr, which actually serve as a nice comparison group (see discussion in Sect.\,\ref{sec:dis:age}).

Nearly all of our targets have IR (2.2\,$\mu$m\,--\,22\,$\mu$m; from 2MASS and WISE) spectral indices 
\mbox{$\alpha = \frac{d\rm{log}(\lambda F_{\lambda })}{d \rm{log}(\lambda )}$}
between $-$3.0 and $-$2.4 (mean $-2.8\pm-0.2$), i.e., they are Class\,III sources \citep{lada1987,ALS87} and correspond roughly to weak-line T\,Tauri stars which no longer possess protoplanetary disks. However, about 20 of our 169 observed targets have detected (mostly by {\it Herschel}) FIR excesses longward of 50\,$\mu$m, most likely indicative of cold debris disks.


\section{Observations and Data Reduction} \label{sec:obs}


\begin{deluxetable}{cccccc}[hb!]
\tablecaption{VLA observing sessions \label{tbl-obssessions}}
\tablecolumns{6}
\tablewidth{0pt}
\tablehead{
\colhead{Session}    	&
\colhead{No of} 		& 
\colhead{No of}		    & 
\colhead{Duration}		&
\colhead{Obs\,1}        & 
\colhead{Obs\,2}	  	\\
\colhead{}		    	&
\colhead{pointings} 	& 
\colhead{stars}		    & 
\colhead{}				&
\colhead{date}   	    & 
\colhead{date}	  	
}
\startdata
1 & 27 & 30 & 2:00 & 2013-10-29 & 2013-11-16 \\
2 & 28 & 30 & 2:00 & 2013-10-30 & 2013-11-06 \\
3 & 30 & 31 & 2:00 & 2013-11-17 & 2013-12-01 \\
4 & 20 & 21 & 1:45 & 2013-11-25 & 2014-01-03 \\
5 & 27 & 28 & 2:00 & 2013-11-27 & 2013-12-30 \\
6\tablenotemark{a} & 25 & 25 & \nodata & \nodata & \nodata \\
7 & 26 & 26 & 2:00 & 2013-10-29 & 2013-11-17 \\
8\tablenotemark{b} &  3  & 3 & 0:30 & 2014-01-10 & \nodata \\
\enddata
\tablenotetext{a}{Session\,6 was not scheduled due to oversubscription in the Galactic Centre R.A. range.}
\tablenotetext{b}{Session\,8 was a short filler to use the remaining granted time and could not be repeated.}
\end{deluxetable}


Observations were carried out with the VLA of the National Radio Astronomy Observatory (NRAO) in its B configuration between October 2013 and January 2014 (project 13B-111). The C-band receivers were used with dual polarisation and two sub-bands, each 1\,GHz wide and centred at 5.5 and 6.5\,GHz, respectively, thus providing a total bandwidth of 2\,GHz. The primary beam FWHM size (FoV) of the VLA antennas at 6\,GHz ($\lambda\approx5$\,cm) is 7\farcm5. The  mean synthesized FWHM beam width (angular resolution) of the VLA B-array at 6\,GHz is $\approx$2\farcs0 (major axis). For low elevation targets the major-axis is $\approx$3\farcs5.

In total we observed 169 target stars with 162 pointings (i.e., seven pointings had two target stars in one field), grouped in six 2-hr sessions with 25 to 30 pointings per session. The integration time per field was 165 seconds.
Each observing session started with a 10\,min integration on a suitable standard flux calibrator. Then followed a series of observations of typically 3-4 target fields located close together on the sky, bracketed by 1-2\,min integrations on a phase calibrator that was located within 10\degr\ from all target fields in that group.
To ensure that we also obtain a measure of the long-term flux variability, each session was repeated, with the two observations separated by one to a few weeks. One short extra session with three stars could be scheduled only once. Thus, we observed 166 stars twice, and three more stars could be observed only once. The individual observing session parameters are summarized in Table\,\ref{tbl-obssessions}.

The data were reduced using the VLA Calibration Pipeline as implemented in the Common Astronomy Software Applications package (CASA). 
In a nutshell, the pipeline determines the complex gains using the observations of the calibrators and transfers them to the target sources. It also automatically flags some radio frequency interferences (RFIs), but we did additional flagging by hand. In particular, one of the lower frequency spectral windows had to be flagged entirely as it contained strong RFIs that rendered it useless. After calibration, the data were imaged (also in CASA). The entire field of view (7\farcm5) was reconstructed in each field, and sources were searched systematically over the entire area mapped. The noise level in the final images was typically 15-20\,$\mu$Jy/beam. The probability that a noise peak exceeds 5\,$\sigma$\ is 3$\times$10$^{-7}$. There are roughly 16,500 independent resolution elements in each of our images. Thus, for each individual field, there is only a 0.5\% chance that a noise peak above 5$\sigma$\ exists. Accounting for the fact that we observed 162 fields in total, we expect at most one false positive. 

Source identification was done by visually inspecting the images. First, the regions near the known positions of targeted YSOs were examined. Then we looked for additional sources in the entire field. Positions, integrated and peak flux densities of each source were obtained by fitting an elliptical gaussian in a small region containing the source. The elliptical gaussian is adequate for both unresolved and only slightly resolved sources, as is the case for the expected emission of our targeted YSOs. Typical (median) positional uncertainties of the gaussian fit centers are 140\,mas in right ascension and 190\,mas in declination for the entire sample of VLA detections, and 70-80\,mas, respectively, for the detections associated to target stars (because these are on average brighter than the unrelated VLA detections; see Sect.\,\ref{ssec:res:overview} and Fig.\,\ref{fig:fluxdist}). 
The typical uncertainty in estimated flux densities is 20\% for the entire sample, and 7\% for the detections associated to target stars.


\section{Results} \label{sec:res}

\subsection{Overview of 6\,GHz detections} \label{ssec:res:overview}

\begin{figure}[ht!]
\epsscale{.95}
\plotone{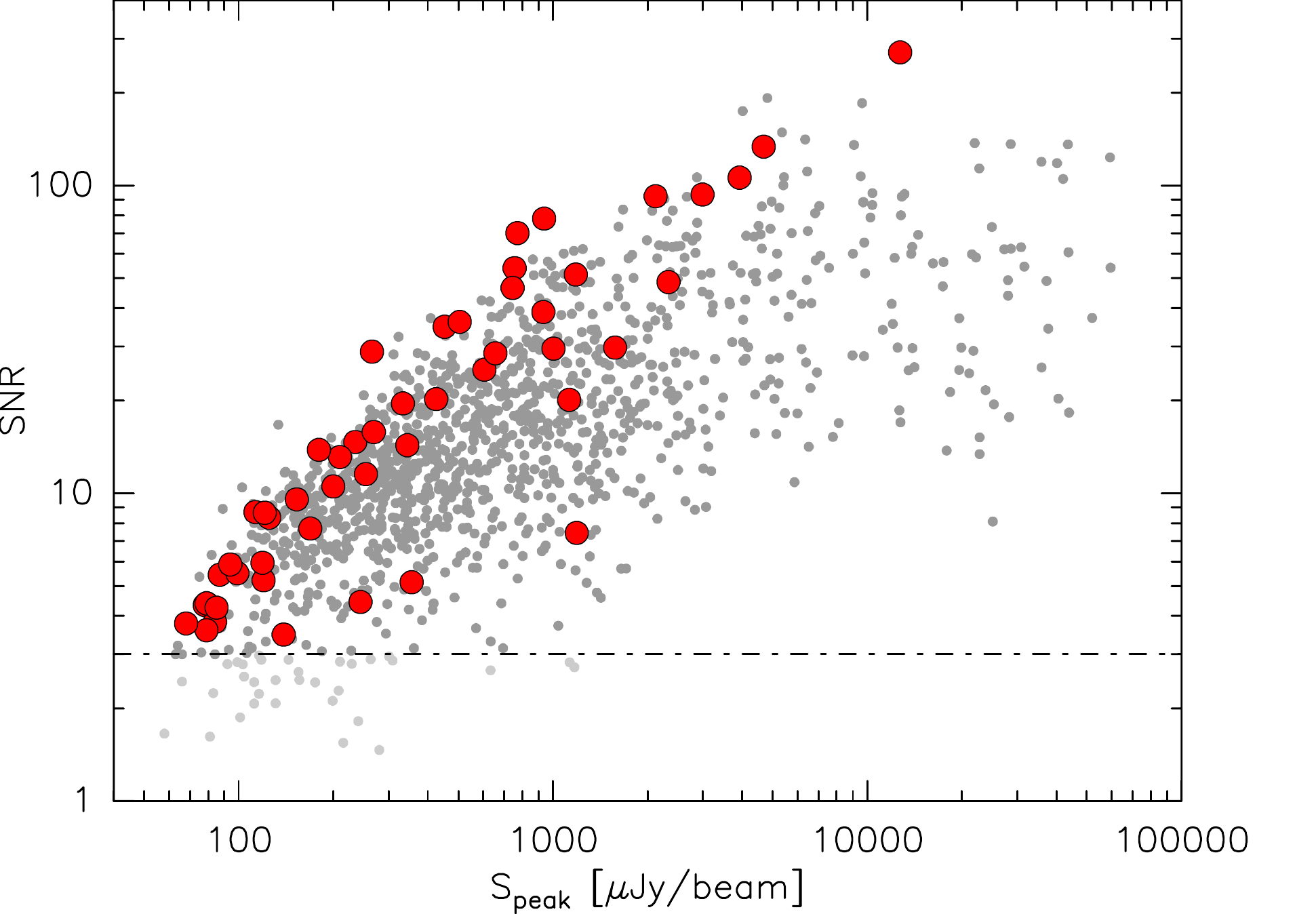}
\caption{\label{fig:fluxdist} Distribution of SNR vs. peak flux for all 1287 identified 6\,GHz radio sources (small grey dots). The horizontal dash-dotted line indicates SNR\,=\,3\,$\sigma$, the threshold above which we consider the radio detections significant (darker grey dots). Large red circles indicate those VLA detections with SNR\,$\ge$\,3 within $r\le6$\arcsec\ (cf. Fig.\,\ref{fig:sepdist1}) around the predicted positions of a target star that we consider as potential detection of the target star.}
\end{figure}

In total, we have identified 1287 6\,GHz radio sources in the 162 fields observed, of which 1252 sources (97.3\%) are significant with signal-to-noise ratios (SNR) $\ge$\,3 (1181 sources or 91.8\% with SNR\,$\ge$\,5).
Figure\,\ref{fig:fluxdist} shows the distribution of SNR vs. peak flux for all identified 6\,GHz radio sources. Note that many (but not all) sources may have been detected in both respective observing epochs. We did not check this epoch crossmatch for the full list of detections, but instead first identified associations between radio sources and target stars and then worked out the epoch crossmatch only on those radio sources that are likely associated with (known) stars. Thus, the actual number of physical radio sources in the observed fields is smaller than 1252, but still $\ge$\,626.

As a first step towards identifying possible radio counterparts to our target stars, we used the 2MASS positions \citep{cutri2003}, which are available for all our targets\footnote{At the time of this assessment, GDR2 was not yet released.}, together with the respective newest proper motion values we could obtain (see Sect.\,\ref{ssec:res:astrom}). For each target star, we then calculated the expected position and positional uncertainty at the mid-epoch of the VLA observations and determined the angular separation and its uncertainty to the nearest VLA source. The separation uncertainty is derived from the uncertainties of the predicted star position and of the VLA position. The first one is determined by both the formal uncertainty of the 2MASS positions and the uncertainty of the proper motion multiplied with the time between the respective 2MASS and VLA observing epochs. The latter one is derived by adding in quadrature the formal image fitting positional uncertainty of the VLA source (see Sect.\,\ref{sec:obs}), the formal uncertainty of the respective phase calibrator positions as listed in the VLA calibrator catalog, and a fixed term of 10\,mas (conservative estimate) for phase errors during the observations (see Sect.\,\ref{ssec:res:astrom}). The median separation uncertainty (between the predicted optical and the observed radio position) is $\approx$150\,mas with an rms scatter of $\approx$180\,mas.

\begin{figure}[ht!]
\epsscale{.95}
\plotone{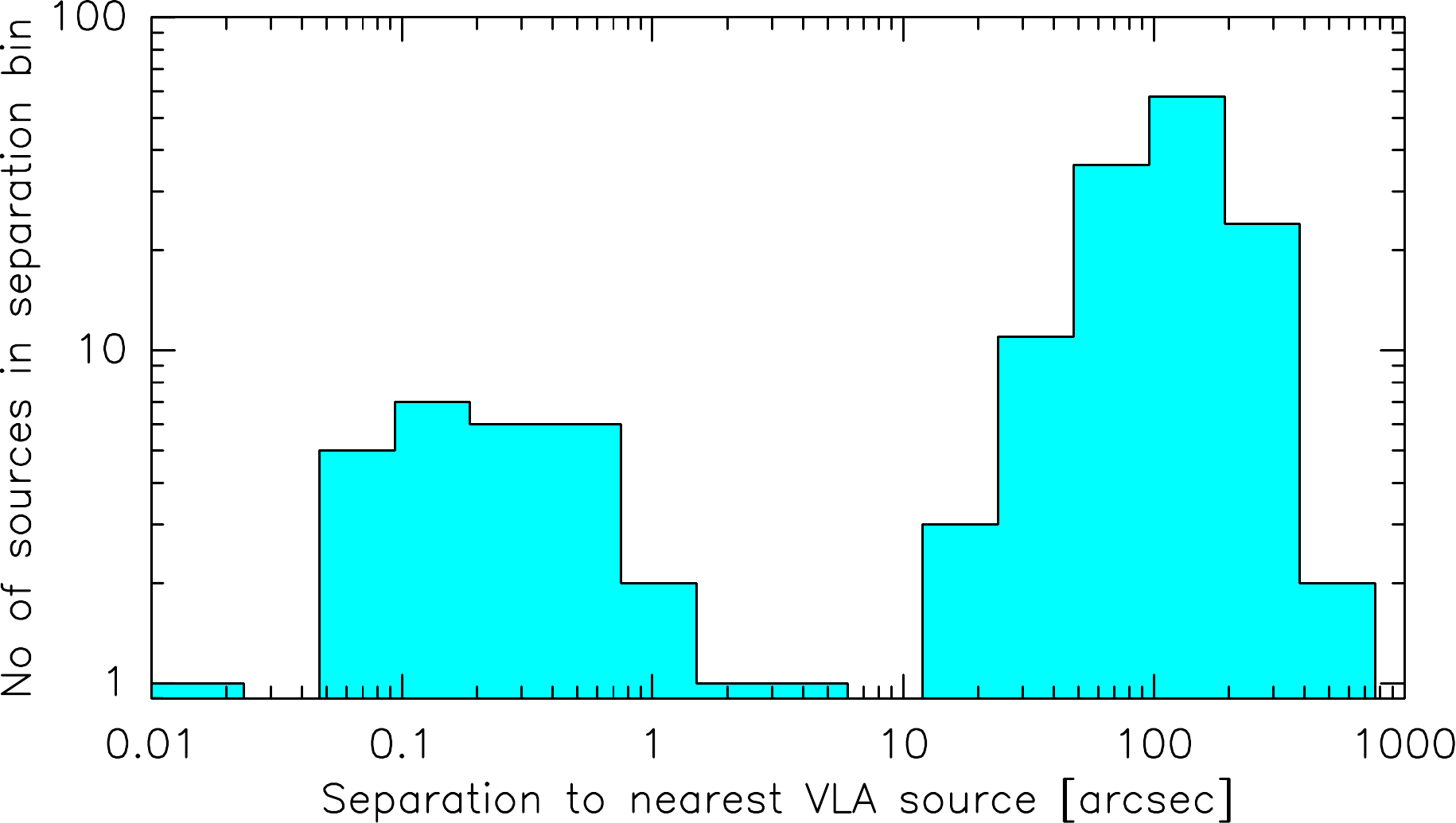}
\caption{\label{fig:sepdist1} Number of target stars as function of angular separation between the position predicted (from the 2MASS position and the proper motion) and the nearest VLA source (Sect.\,\ref{ssec:res:overview}). Note that the separation bins are logarithmic.}
\end{figure}

Figure\,\ref{fig:sepdist1} shows the resulting histogram of the number of target stars as function of angular separation between predicted star position and the nearest VLA source with logarithmic separation bins. The distribution of nearest neighbour separations is clearly bimodal, with the first peak located at a few tenth of an arcsecond and the second, much broader peak centered at a separation of $\approx$100\,--\,200\arcsec. The first peak encompasses 30 target stars, with 28 stars having VLA detections within $<1\arcsec$\ and only two stars having nearest VLA detections at separations 1\arcsec\,--\,5\arcsec. The second peak contains the remaining 139 target stars with nearest VLA detections at separations $>12\arcsec$. There is no target star with nearest VLA detection between 5\arcsec\ and 12\arcsec. Given the positional uncertainties and the typical angular separations of (potential) physically bound companions, plus the clear bimodality of the nearest neighbour separation distribution, it is safe to consider all 139 target stars with nearest neighbour VLA sources at separations $>10\arcsec$\ not to exhibit detectable radio emission (at the time of the observations). For the 30 target stars with nearest neighbour VLA detections within 5\arcsec, we also checked if a second-nearest VLA neighbour was located closer than $10\arcsec$, but we found none.

To evaluate if any of the $\approx$1200 VLA radio sources that are not related to our target stars might be related to other, not targeted stars, we carried out a Simbad query with search radius of 6\arcsec. This search radius was motivated by the 2MASS-based separation distribution between stars and nearest VLA source shown in Fig.\,\ref{fig:sepdist1}. The choice of an angular search radius, as opposed to an absolute one, is justified by the fact that we are looking for physical associations with discrepancies mainly originating from proper motion and pointing uncertainties and by the clear separation and large gap size between the two separation distribution groups (Fig.\,\ref{fig:sepdist1}). This search resulted in matches for only 175 out of the 1252 VLA sources. Of these, 117 matches are known galaxies, QSOs, or just listed radio sources with non-identified optical counterpart. The most frequent associations were found in the NRAO VLA Sky Survey \citep[NVSS:][]{condon1998} and in the Lyon-Meudon Extragalactic DAtabase \citep[LEDA:][]{LEDA1995}.  

Only 58 VLA sources are related to known stars, of which 53 are related to the 30 target detection candidates mentioned above and described in more detail in Sect.\,\ref{ssec:res:astrom}. Note that VLA detections in separate epochs or overlapping fields are all counted separately, such that the number of counted VLA sources is always larger than the number of physical sources to which they are related. 
Only five additional (out of the 58) VLA sources are located within 6\arcsec\ around two non-targeted stars. Three of these VLA detections (in two epochs and two fields) are located 5.6\arcsec\ away from HIP\,14807 (BD+21\,418B) at 03:11:12.334, +22:25:22.73. However, based on an astrometric analysis as described in Sect.\,\ref{ssec:res:astrom}, the VLA source does not seem to be associated with the star. It is most likely an unrelated background source.
The other two VLA detections are related to UCAC4\,832-014013 at 15:07:57.226, +76:13:59.15, which is a high proper motion M\,4.5 star at distance of $\sim30$\,pc (Tab.\,\ref{tbl-sourcelist}). A VLA radio source was detected in both epochs at 0\farcs19 from the predicted ICRS position of this star. The astrometric analysis confirmed that we have indeed detected radio emission from this star (Sect.\,\ref{ssec:res:astrom}). Thus, 55 out of our 1252 VLA detections are related to 31 (known) stars.

We conclude that 31 of the observed 169+1 target stars (not counting companions) are potentially associated with 6\,GHz radio emission above the mean 3\,$\sigma$\ detection limit of $\approx$45\,$\mu$Jy. 
This would correspond to a detection rate of $(18\pm3)$\% if all stars were single. Since this is not the case, we convert this system detection rate into a star detection rate in Sect\,\ref{sec:dis:bin}.
Of these, 22 sources were detected in two epochs, eight sources were detected in only one out of two epochs, and one detected object was observed only once. Figure\,\ref{fig:fluxdist} shows that the vast majority of the detection candidates line up at the upper envelope of the SNR\,--\,versus\,--\,peak flux distribution of detected radio sources, i.e., they are all significant. 
This is most likely related to the fact that the stars are nearly always located at the pointing centres, where the sensitivity is highest, while many of the other detections in the respective fields are located further out in the FoV of the synthesized beams where the sensitivity is lower.
%
To verify if the detection candidates are indeed associated with the target stars, or if they could possibly also be related to known or unknown (stellar) companions, we investigate all 31 detection candidates astrometrically in more detail in Sect.\,\ref{ssec:res:astrom}.

\begin{figure}[ht!]
\epsscale{.95}
\plotone{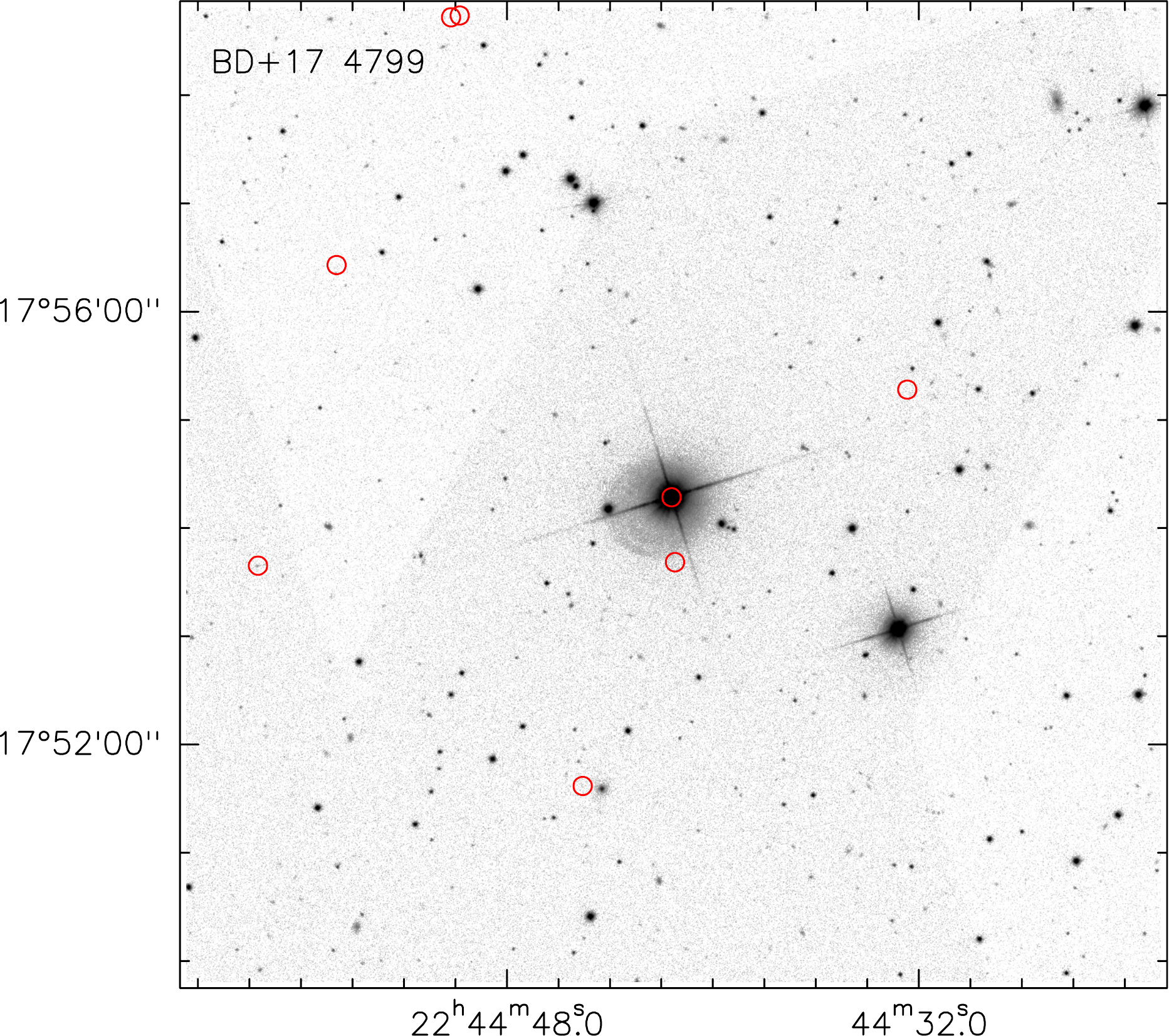}
\caption{\label{fig:BD+17.4799} Sloan Digitized Sky Survey (SDSS) r-band image of a field around the target star BD+17\,4799. Overlaid as red open circles are the positions of all eight VLA radio sources detected in this field.}
\end{figure}

Based on the analysis outlined above, we conclude that all 1197 (1252\,--\,55) radio sources not associated with known stars (96\%) are likely related to background sources (galaxies, quasars) at cosmological distances. This is nicely illustrated by Fig.\,\ref{fig:BD+17.4799}, which shows as one example an SDSS (red) image \citep{york2000} of the target star SAO\,108142 (=\,BD+17\,4799) overlaid with the positions of all eight 6\,GHz radio sources detected in this field. Apart from the VLA source at the target star position, none of the other radio sources is positionally associated with a visible star. Only one source at $\approx$\,22:44:57.7 +17:53:30 (Fig.\,\ref{fig:BD+17.4799}, center left) is associated with a very faint and fuzzy (i.e., not point-like) optical counterpart; but a Simbad search only lists the not further described radio source NVSS\,J224457+175338 at this position. Since we are interested in radio emission from nearby young stars only, we do not further follow up or discuss these apparently unrelated radio sources nor do we investigate which ones in the two epochs may be related to the same physical source.
However, we note that with $\geq$616 supposedly extragalactic background sources in 162 fields with 7\farcm5 diameter (Sect.\,\ref{sec:obs}), we observe a surface density of $\approx$0.1 sources per arcmin$^2$\ with a flux density at 6\,Ghz greater than $\approx$50\,$\mu$Jy. This compares well with the results of \citet{formalont1991}, who predict a surface density $0.2\pm0.1$\ background sources per arcmin$^2$\ at the 50\,$\mu$Jy detection threshold.


\begin{longrotatetable}
\begin{deluxetable}{rllrlrrrlll}
\tablecaption{Stars with 6\,GHz detections \label{tbl-det}}
\tabletypesize{\scriptsize}
\tablewidth{0pt}
\tablehead{
\colhead{No.\tablenotemark{a)}}         &
\colhead{Star\,ID (other name)}         & 
\colhead{date\,1} 				        &
\colhead{$S_\mathrm{6GHz}^1\pm \delta$} &
\colhead{date\,2} 				        &
\colhead{$S_\mathrm{6GHz}^2\pm \delta$} &
\colhead{$\Delta\,t$}			        &
\colhead{$S_1/S_2$}			            &
\colhead{$\Delta\,r$\tablenotemark{b)}} &
\colhead{Binarity\tablenotemark{c)}}    &
\colhead{Remarks\tablenotemark{d)}}    \\
& &  & [$\mu$Jy$\pm\mu$Jy] & & [$\mu$Jy$\pm\mu$Jy] & [d] & & [arcsec ($\sigma$)] & & 
}
\startdata
11 	& BD+17\,232\,A                &  2013-10-29 & 149$\pm$18  	 &  2013-11-16 	& 276$\pm$32   & 18 & 1.85    & 0.32 (2.75) & WDS\,(1\farcs7)          & D12 (1)\\
26  & HIP\,12545 (BD+05\,378)	   &  2013-10-29 & $<$52$\pm$17  &  2013-11-16  & 102$\pm$12   & 18 & $>$1.96 & 0.11 (0.41) & \nodata                  & D1 \\
27  & HIP\,12635 (HD\,16760\,B)    &  2013-10-29 & 261$\pm$69    &  2013-11-16  & $<$63$\pm$21 & 18 & $>$4.14 & 0.27 (1.10) & K19                      & D1 \\
33 	& V875\,Per                    &  2013-10-29 & 2394$\pm$86   &  2013-11-16 	& 1840$\pm$250 & 18 & 1.30    & 0.07 (2.48) & WDS\,(5\farcs8), RS\,CVn & D1  \\
35 	& TYC\,3301-2585-1             &  2013-10-30 & 154$\pm$35    &  2013-11-06 	& 392$\pm$34   &  7 & 2.55    & 0.68 (4.74) & WDS\,(2\farcs1)          & D12 (2) \\
48 	& HD\,22213 (BD-12\,674) 	   &  2013-11-17 & 623$\pm$24    &  2013-12-01 	& 108$\pm$20   & 14 & 5.77    & 0.11 (1.11) & WDS\,(1\farcs7)          & D1 \\
51  & HD\,23524 	               &  2013-10-30 & 2059$\pm$22   &  2013-11-06 	& 727$\pm$14   & 7  & 2.83    & 0.03 (0.22) & WDS\,(0\farcs3)          & D12 (2) \\
52 	& HD\,24681  (BD-02\,754) 	   &  2013-11-17 & 216$\pm$15  	 &  2013-12-01 	& $<$61$\pm$20 & 14 & $>$3.54 & 0.08 (0.84) & \nodata                  & D1 \\
53  & HD\,285281 (V1293\,Tau)       &  2013-10-30 & $<$112$\pm$37 &  2013-11-06 	& 314$\pm$80   & 7  & $>$2.80 & 0.19 (0.76) & WDS\,(0\farcs86)     & D1 \\
57 	& HD\,284135 (V1299\,Tau) 	   &  2013-10-30 & 386$\pm$27  	 &  2013-11-06 	& 254$\pm$13   & 7  & 1.52    & 0.30 (1.98) & WDS\,(0\farcs4)          & D12 (1, 4)  \\
74 	& HD\,31281 (V1349\,Tau) 	   &  2013-10-30 & 1089$\pm$37   &  2013-11-06 	& 2910$\pm$31  & 7  & 2.67    & 0.07 (2.46) & WDS\,(4\farcs4)          & D1 \\
77 	& BD-08\,995                   &  2013-11-17 & 197$\pm$37    &  2013-12-01 	& 206$\pm$27   & 14 & 1.05    & 0.25 (0.66) & \nodata                  & D1  \\
79  & HD\,286264 (V1841\,Ori) 	   &  2013-10-30 & $<$69$\pm$23  &  2013-11-06  & 222$\pm$37   & 7  & $>$3.22 & 0.05 (0.26) & \nodata                  & D1  \\
81 	& HD\,293857 (BD-04\,1063)	   &  2013-11-17 & 659$\pm$23 	 &  2013-12-01 	& 369$\pm$19   & 14 & 1.79    & 0.05 (0.51) & \nodata\tablenotemark{d)}& D1   \\
88 	& TYC\,713-661-1  	           &  2013-11-17 & 214$\pm$17 	 &  2013-12-01 	& $<$70$\pm$23 & 14 & $>$3.06 & 0.20 (2.34) & WDS\,(1\farcs8)          & D1  \\
92 	& TYC\,5925-1547-1 (CPD-19\,878) &  2013-11-17 & 274$\pm$27  &  2013-12-01 	&  57$\pm$16   & 14 & 4.81    & 0.90 (2.26) & \nodata                  & D1 (3) \\
93 	& SAO\,150676 (AI\,Lep)        &  2013-11-17 & 1280$\pm$250  &  2013-12-01 	& 709$\pm$203  & 14 & 1.81 	  & 0.64 (1.04) & WDS\,(8\farcs4), RS\,CVn & D1 (3) \\
98 	& HD\,62237 (BD-15\,1991)	   &  2013-11-25 & 91$\pm$24   	 &  2014-01-03 	& 168$\pm$12   & 39 & 1.85 	  & 0.16 (1.66) & \nodata                  & D1 \\
101 & SAO\,135659                  &  2013-11-25 & 4542$\pm$34   &  2014-01-03 	& 425$\pm$21   & 39 & 10.7 	  & 0.05 (1.08) & WDS\,(0\farcs14)         & D12 (2)  \\
106 & HD\,77407  	               &  2014-01-10 & 785$\pm$12    &  \nodata     & \nodata & \nodata  & \nodata  & 0.09 (1.89) & WDS\,(1\farcs7), K19   & D2 (WDS\,09035+3750\,B) \\
108 & HD\,82159 (GS\,Leo)          &  2013-11-25 & 3852$\pm$36   &  2014-01-03 	& 918$\pm$12   & 39 & 4.20 	  & 0.03 (0.74) & WDS\,(13\farcs7), D15    & D1 \\
109 & HD\,82558  (LQ\,Hya)         &  2013-11-25 &  112$\pm$76   &  2014-01-03 	& 716$\pm$15   & 39 & 6.39 	  & 0.07 (0.88) & \nodata                  & D1 \\
112 & GJ\,2079 (DK\,Leo)           &  2013-11-25 & 251$\pm$16    &  2014-01-03 	& 12784$\pm$47 & 39 & 51 	  & 0.05 (0.87) & WDS\,(0\farcs1), K19     & D12 (2) \\
144 & HD\,135363  (BD+76\,552) 	   &  2013-11-27 & 400$\pm$47    & 2013-12-30   & 167$\pm$39   & 33 & 2.40 	  & 0.49 (1.31) & WDS\,(0\farcs36), K19    & D12 \\
153 & HD\,199143\                  &  2013-10-29 & 1305$\pm$64   &  2013-11-17 	& 531$\pm$15   & 19 & 2.46 	  & 0.13 (0.62) & WDS\,(1\farcs05), K19    & D2 (HD\,199143\,B) \\
154 & HD\,358623 (AZ\,Cap)         &  2013-10-29 & 312$\pm$74    &  2013-11-17 	& 182$\pm$21   & 19 & 1.71 	  & 0.29 (1.65) & WDS\,(2\farcs2)          & D1 \\
155 & SAO\,50350 (BD+44\,3670)     &  2013-10-29 & 181$\pm$30 	 &  2013-11-17 	& $<$71$\pm$24 & 19 & $>$2.55 & 0.72 (5.26) & \nodata                  & D1 (5) \\
159 & GJ\,4199 (LO\,Peg)           &  2013-10-29 & 245$\pm$9  	 &  2013-11-17 	& 454$\pm$13   & 19 & 1.85 	  & 0.07 (0.97) & WDS\,(9\farcs2)          & D1 \\
164 & SAO\,51891 (V383\,Lac)       &  2013-10-29 & 1128$\pm$22 	 &  2013-11-17 	& $<$90$\pm$30 & 19 & $>$12.5 & 0.08 (1.66) & WDS\,(4\farcs0)          & D1 \\
165 & SAO\,108142 (BD+17\,4799)    &  2013-10-29 & 992$\pm$25 	 &  2013-11-17 	& 1610$\pm$950 & 19 & 1.62 	  & 0.21 (0.59) & \nodata                  & D1 \\
170 & UCAC4\,832-014013            &  2013-11-27 & 1070$\pm$120  &  2013-12-30  & 932$\pm$175  & 33 & 1.15    & 0.15 (0.96) & \nodata                  & D1 (6) \\
\enddata
\tablenotetext{a}{The No. refers to table\,\ref{tbl-sourcelist}, where also the coordinates and some basic stellar parameters are listed.}
\tablenotetext{b}{Absolute and relative positional discrepancy between VLA detection and predicted star position, given in arcsec and (in brackets) in units of the combined $1\,\sigma$\ positional uncertainties.}
\tablenotetext{c}{Companions listed in the Washington Visual Double Star Catalog \citep[WDS,][]{mason2020} with approximate projected separation given in brackets, or identified by \citet[][K19]{kervella2019} via PMa. RS\,CVn indicates RS-Canum-Venaticorum variability, which indicates the presence of a very close (interacting) companion. D15 indicates the presence of a  close companion with a 3.86-day period reported by \citet{desidera2015}. The star is also flagged with duplicity-induced variability in the {\it Hipparcos} catalogue.}
\tablenotetext{d}{D1: detection of target star (primary in case of visual binarity), 
  D2: detection of (visual) companion to target primary, 
  D12: VLA source centered in-between two binary components.}
\tablenotetext{e}{Large GDR2 astrometric excess noise (7.6\,mas) and $RUWE\approx59$ could hint at yet unknown companion.}
\tablenotetext{1}{VLA source centered closer to primary.}
\tablenotetext{2}{VLA source centered closer to secondary.}
\tablenotetext{3}{Bad calibrator (J\,0539-1550).}
\tablenotetext{4}{No valid GDR2 astrometry, HIP pm probably affected by secondary.}
\tablenotetext{5}{VLA source center off by 5.3\,$\sigma$, but star still within beam. Hint at unknown secondary?}
\tablenotetext{6}{UCAC4\,832-014013 was not an original target star, but was detected in the pointing on HD\,135363.}
\end{deluxetable}
\end{longrotatetable}


\subsection{Astrometric association of VLA sources with known stars} \label{ssec:res:astrom}

\figsetstart
\figsetnum{6}
\figsettitle{Astrometric charts for all 31 VLA sources associated with target stars.}

\figsetgrpstart
\figsetgrpnum{6.1}
\figsetgrptitle{BD+17\,232}
\figsetplot{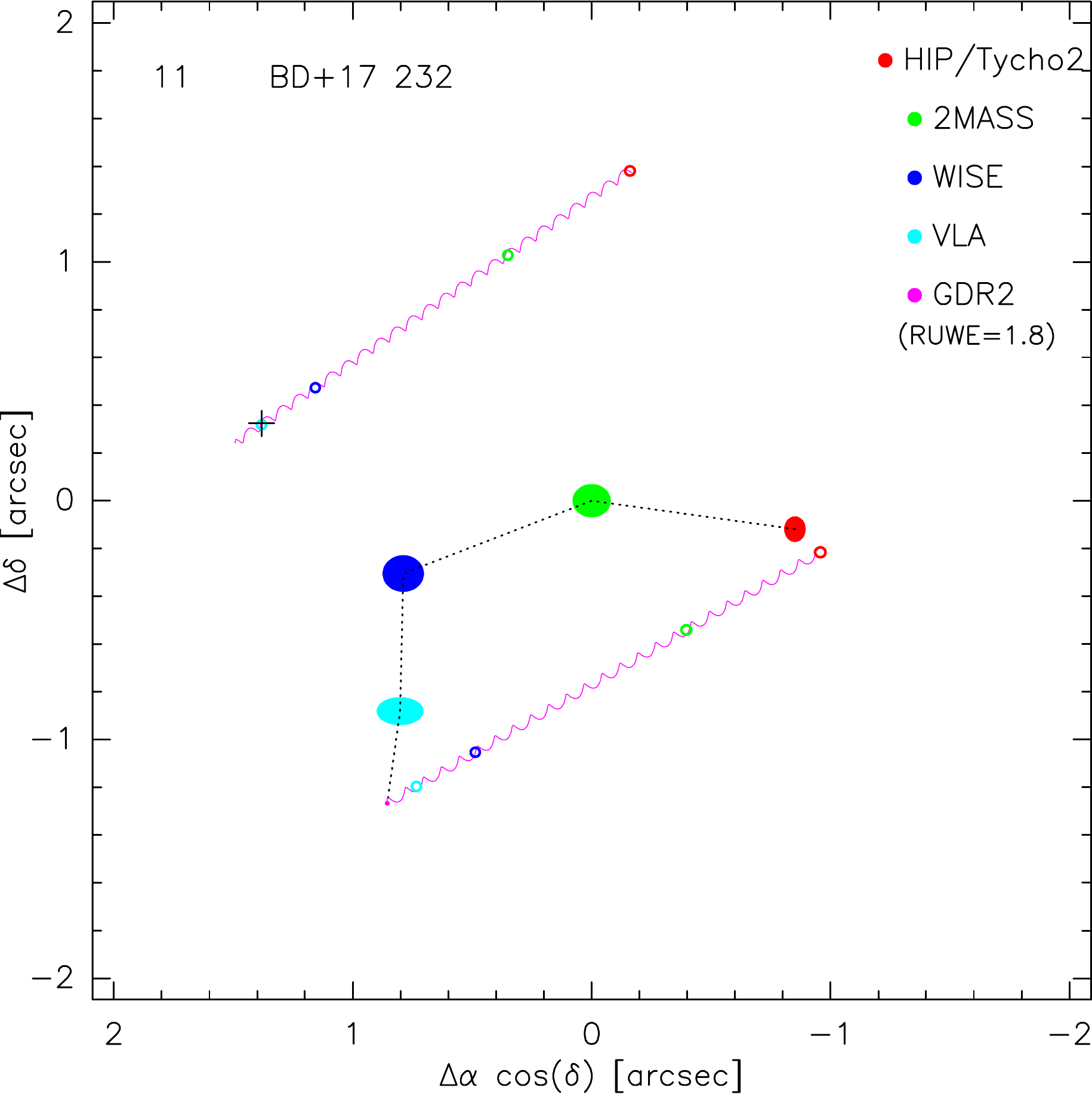}
\figsetgrpnote{Astrometric chart for BD+17\,232. 
The number in the upper left corner relates to the target number in Tables\,\ref{tbl-det} and \ref{tbl-sourcelist}. Measured positions (relative to the 2MASS) and 1\,$\sigma$\ uncertainties are drawn as filled ellipses with the colour coding for the respective missions/catalogs indicated in the upper right corner. Straight dotted lines connect the observed positions just to guide the eye. The lower wiggled line shows the combined proper motion and parallax prediction, starting at the GDR2 2015.5 position and projected backwards in time. Open ellipses show the position and uncertainty prediction from that starting point for the respective observing epochs. The black cross shows the position of the WDS companion BD+17\,232\,B \citep{mason2020} at the epoch of the VLA observations, i.e., relative to the center of the light blue open ellipse marking the predicted position of the primary. The upper wiggled line and open ellipses show the proper motion/parallax and position prediction for the secondary, which was also detected by GDR2.}
\figsetgrpend

\figsetgrpstart
\figsetgrpnum{6.2}
\figsetgrptitle{HIP\,12545}
\figsetplot{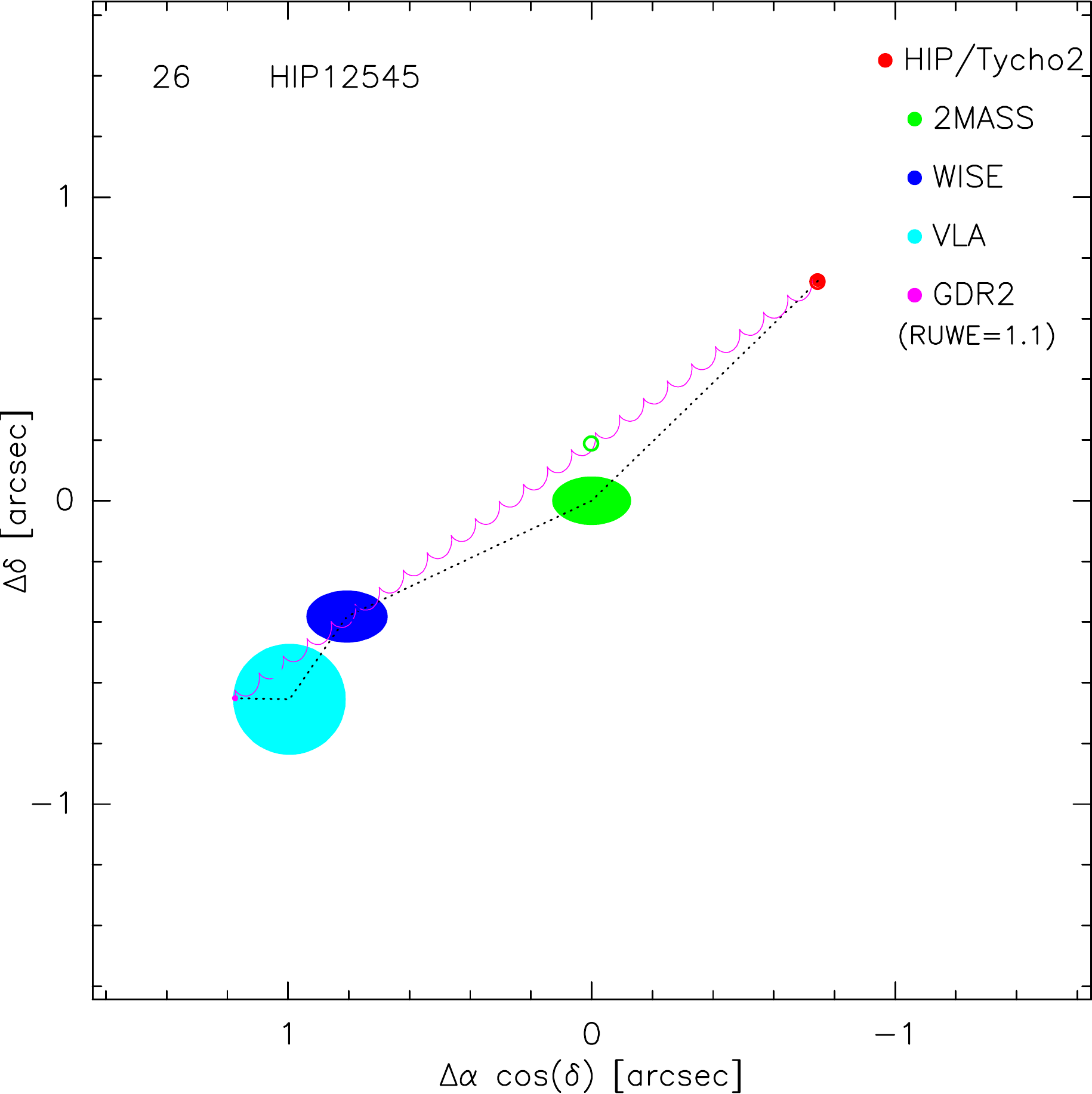}
\figsetgrpnote{Astrometric chart for HIP\,12545 (BD+05\,378). 
The number in the upper left corner relates to the target number in Tables\,\ref{tbl-det} and \ref{tbl-sourcelist}. Measured positions (relative to the 2MASS) and 1\,$\sigma$\ uncertainties are drawn as filled ellipses with the colour coding for the respective missions/catalogs indicated in the upper right corner. Straight dotted lines connect the observed positions just to guide the eye. The wiggled line shows the combined proper motion and parallax prediction, starting at the GDR2 2015.5 position and projected backwards in time. Open ellipses show the position and uncertainty prediction from that starting point for the respective observing epochs.}
\figsetgrpend

\figsetgrpstart
\figsetgrpnum{6.3}
\figsetgrptitle{HIP\,12635}
\figsetplot{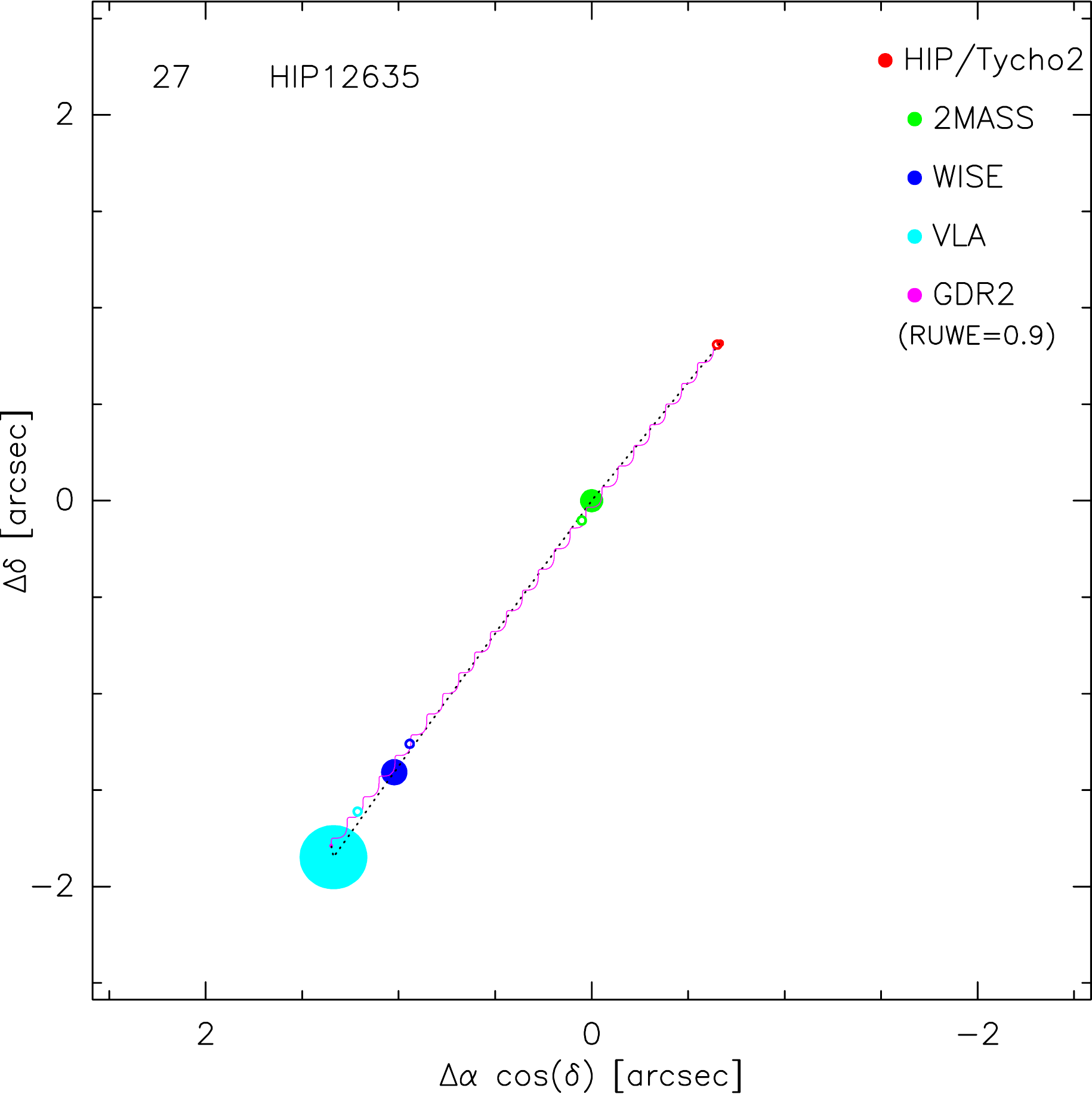}
\figsetgrpnote{Astrometric chart for HIP\,12635 (HD\,16760\,B).
The number in the upper left corner relates to the target number in Tables\,\ref{tbl-det} and\ref{tbl-sourcelist}. Measured positions (relative to the 2MASS) and 1\,$\sigma$\ uncertainties are drawn as filled ellipses with the colour coding for the respective missions/catalogs indicated in the upper right corner. Straight dotted lines connect the observed positions just to guide the eye. The wiggled line shows the combined proper motion and parallax prediction, starting at the GDR2 2015.5 position and projected backwards in time. Open ellipses show the position and uncertainty prediction from that starting point for the respective observing epochs.}
\figsetgrpend

\figsetgrpstart
\figsetgrpnum{6.4}
\figsetgrptitle{V875\,Per}
\figsetplot{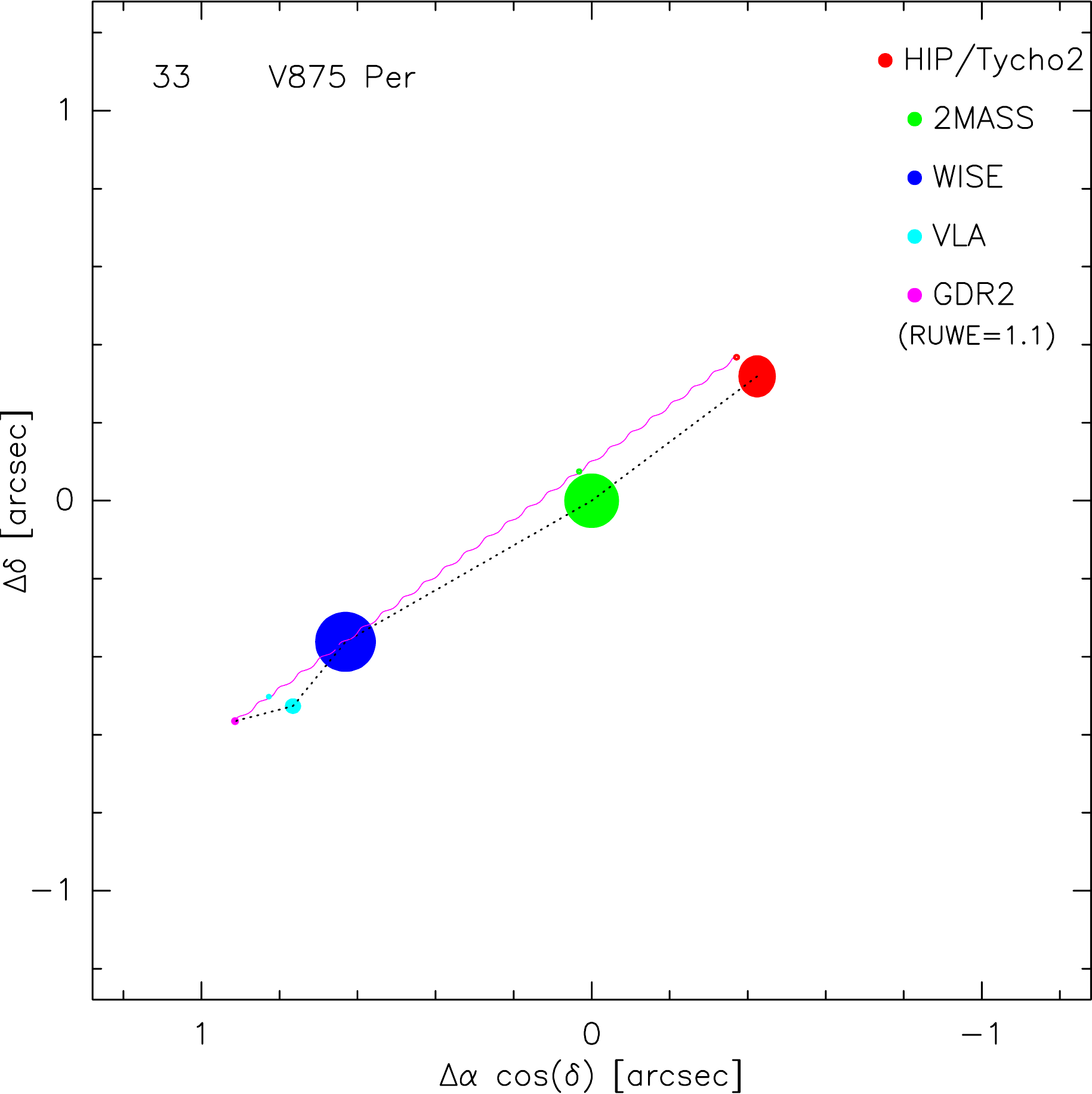}
\figsetgrpnote{Astrometric chart for V875\,Per.
The number in the upper left corner relates to the target number in Tables\,\ref{tbl-det} and \ref{tbl-sourcelist}. Measured positions (relative to the 2MASS) and 1\,$\sigma$\ uncertainties are drawn as filled ellipses with the colour coding for the respective missions/catalogs indicated in the upper right corner. Straight dotted lines connect the observed positions just to guide the eye. The wiggled line shows the combined proper motion and parallax prediction, starting at the GDR2 2015.5 position and projected backwards in time. Open ellipses show the position and uncertainty prediction from that starting point for the respective observing epochs.}
\figsetgrpend

\figsetgrpstart
\figsetgrpnum{6.5}
\figsetgrptitle{TYC\,3301-2585-1}
\figsetplot{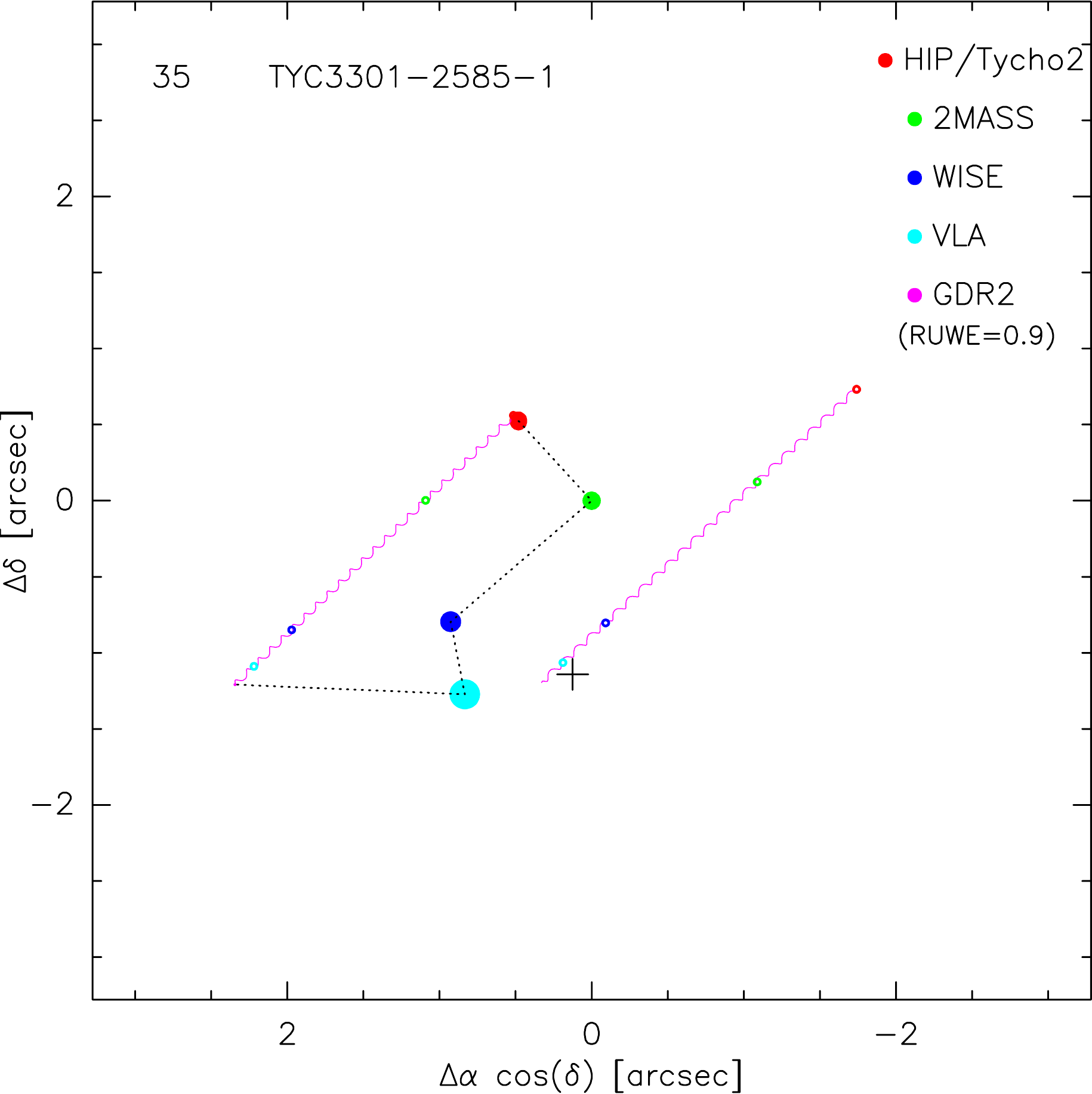}
\figsetgrpnote{Astrometric chart for TYC\,3301-2585-1.
The number in the upper left corner relates to the target number in Tables\,\ref{tbl-det} and \ref{tbl-sourcelist}. Measured positions (relative to the 2MASS) and 1\,$\sigma$\ uncertainties are drawn as filled ellipses with the colour coding for the respective missions/catalogs indicated in the upper right corner. Straight dotted lines connect the observed positions just to guide the eye. The left wiggled line shows the combined proper motion and parallax prediction, starting at the GDR2 2015.5 position and projected backwards in time. Open ellipses show the position and uncertainty prediction from that starting point for the respective observing epochs. The black cross shows the position of the companion WDS\,02557+4746\,B \citep{mason2020} at the epoch of the VLA observations, i.e., relative to the center of the left light blue open ellipse marking the predicted position of the primary. The right wiggled line and open ellipses show the proper motion/parallax and position prediction for the secondary, which was also detected by GDR2.}
\figsetgrpend

\figsetgrpstart
\figsetgrpnum{6.6}
\figsetgrptitle{HD\,22213}
\figsetplot{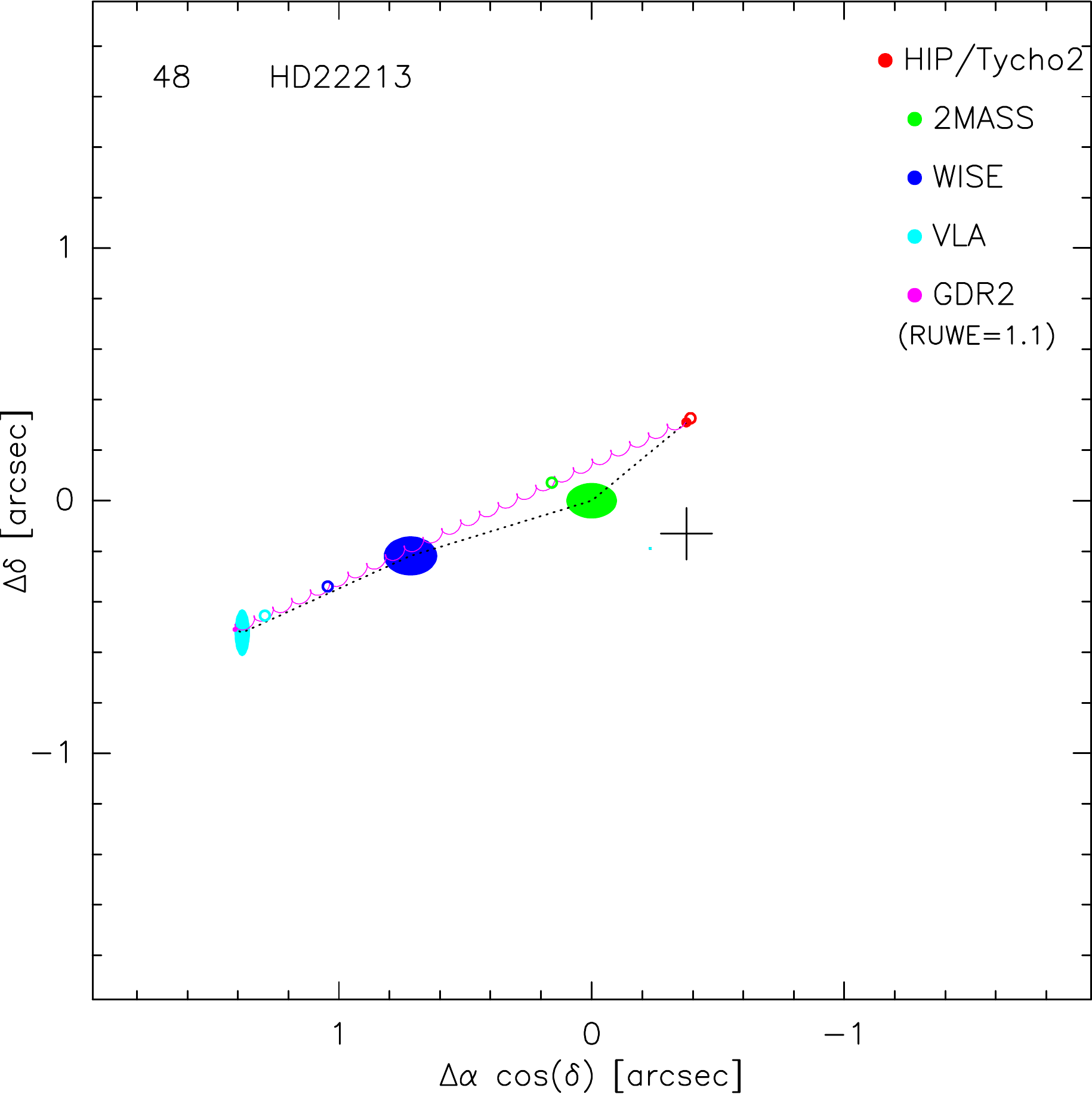}
\figsetgrpnote{Astrometric chart for HD\,22213 (BD-12\,674).
The number in the upper left corner relates to the target number in Tables\,\ref{tbl-det} and \ref{tbl-sourcelist}. Measured positions (relative to the 2MASS) and 1\,$\sigma$\ uncertainties are drawn as filled ellipses with the colour coding for the respective missions/catalogs indicated in the upper right corner. Straight dotted lines connect the observed positions just to guide the eye. The wiggled line shows the combined proper motion and parallax prediction, starting at the GDR2 2015.5 position and projecting backwards in time. Open ellipses show the position and uncertainty prediction from that starting point for the respective observing epochs. The black cross shows the position of the companion WDS\,03343-1204\,B \citep{mason2020} at the epoch of the VLA observations, i.e., relative to the center of the light blue open ellipse marking the predicted position of the primary. No independent astrometric solution exists for this secondary since it was neither detected by Gaia, nor by Hipparcos.}
\figsetgrpend

\figsetgrpstart
\figsetgrpnum{6.7}
\figsetgrptitle{HD\,23524}
\figsetplot{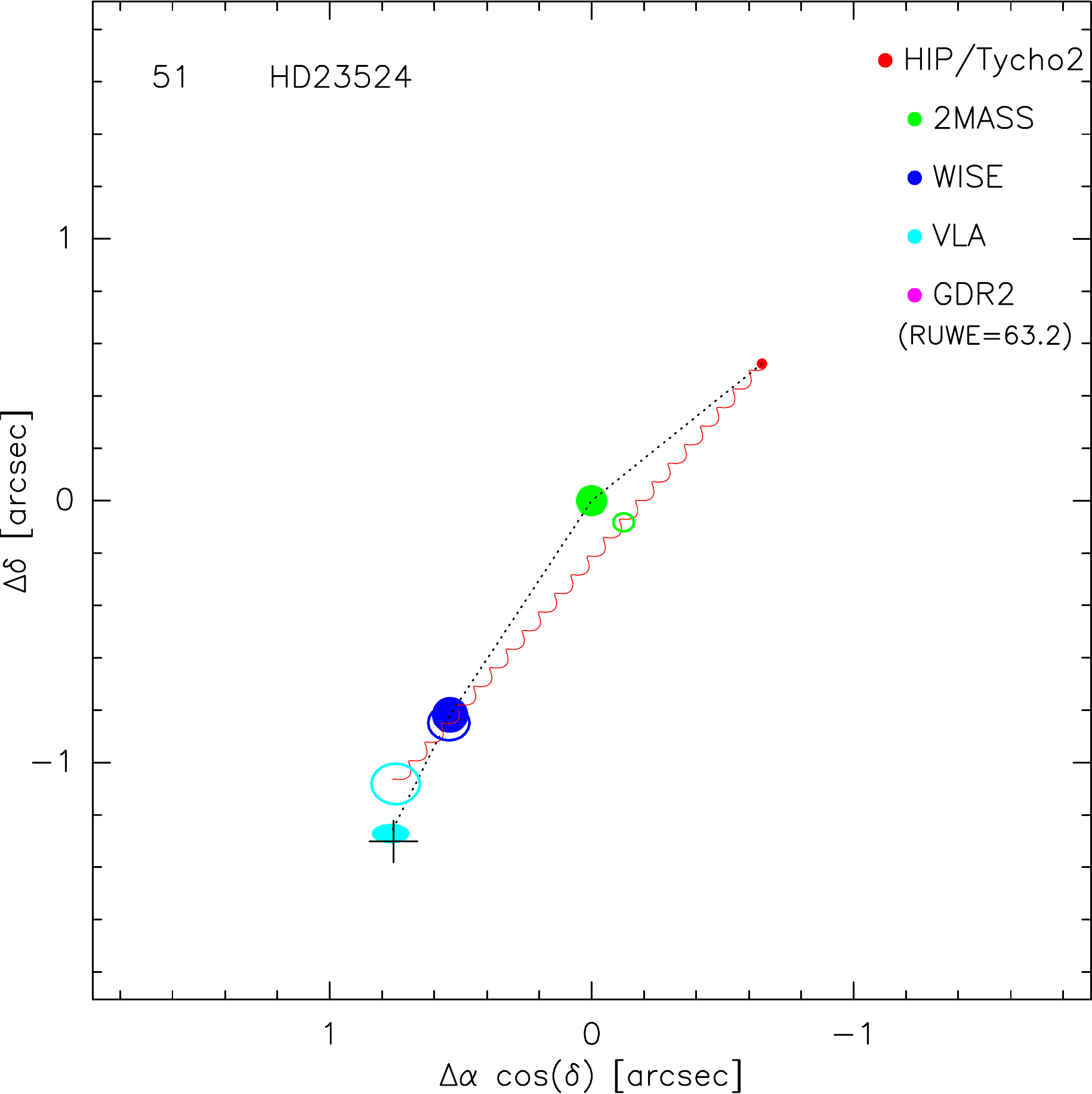}
\figsetgrpnote{Astrometric chart for HD\,23524.
The number in the upper left corner relates to the target number in Tables\,\ref{tbl-det} and \ref{tbl-sourcelist}. Measured positions (relative to the 2MASS) and 1\,$\sigma$\ uncertainties are drawn as filled ellipses with the colour coding for the respective missions/catalogs indicated in the upper right corner. Straight dotted lines connect the observed positions just to guide the eye. The red wiggled line shows the combined proper motion and parallax prediction, starting at the HIP/Tycho2 position and projected forward in time. The black cross shows the position of the WDS companion HD\,23524\,B \citep{mason2020} at the epoch of the VLA observations, i.e., relative to the center of the light blue open ellipse marking the predicted position of the primary. No GDR2 (nor GeDR3) astrometric solution exists for this 0\farcs3 binary star since it was most likely not spatially resolved by Gaia.}
\figsetgrpend

\figsetgrpstart
\figsetgrpnum{6.8}
\figsetgrptitle{HD\,24681}
\figsetplot{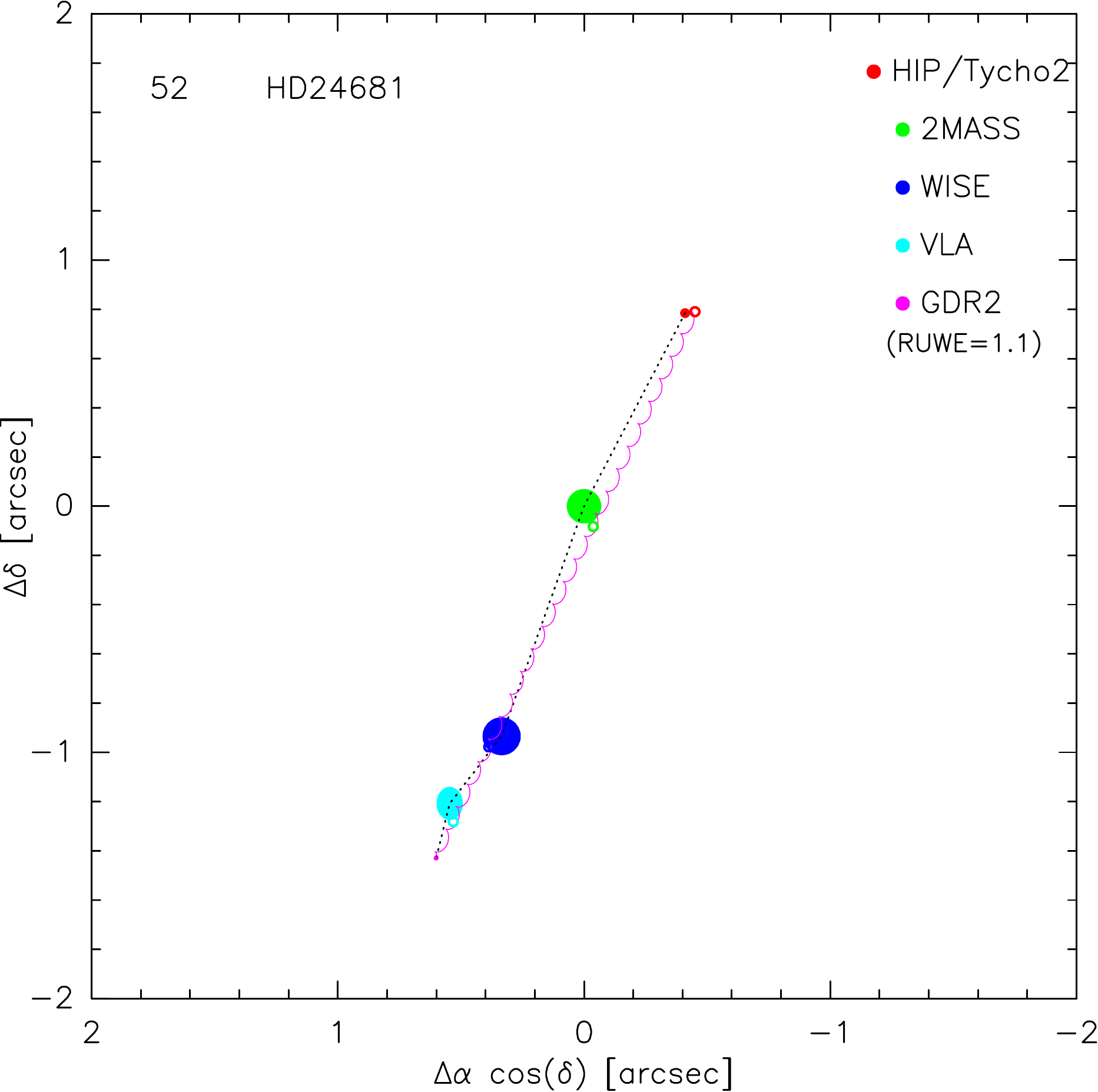}
\figsetgrpnote{Astrometric chart for HD\,24681 (BD-02\,754).
The number in the upper left corner relates to the target number in Tables\,\ref{tbl-det} and \ref{tbl-sourcelist}. Measured positions (relative to the 2MASS) and 1\,$\sigma$\ uncertainties are drawn as filled ellipses with the colour coding for the respective missions/catalogs indicated in the upper right corner. Straight dotted lines connect the observed positions just to guide the eye. The wiggled line shows the combined proper motion and parallax prediction, starting at the GDR2 2015.5 position and projected backwards in time. Open ellipses show the position and uncertainty prediction from that starting point for the respective observing epochs.}
\figsetgrpend

\figsetgrpstart
\figsetgrpnum{6.9}
\figsetgrptitle{HD\,285281}
\figsetplot{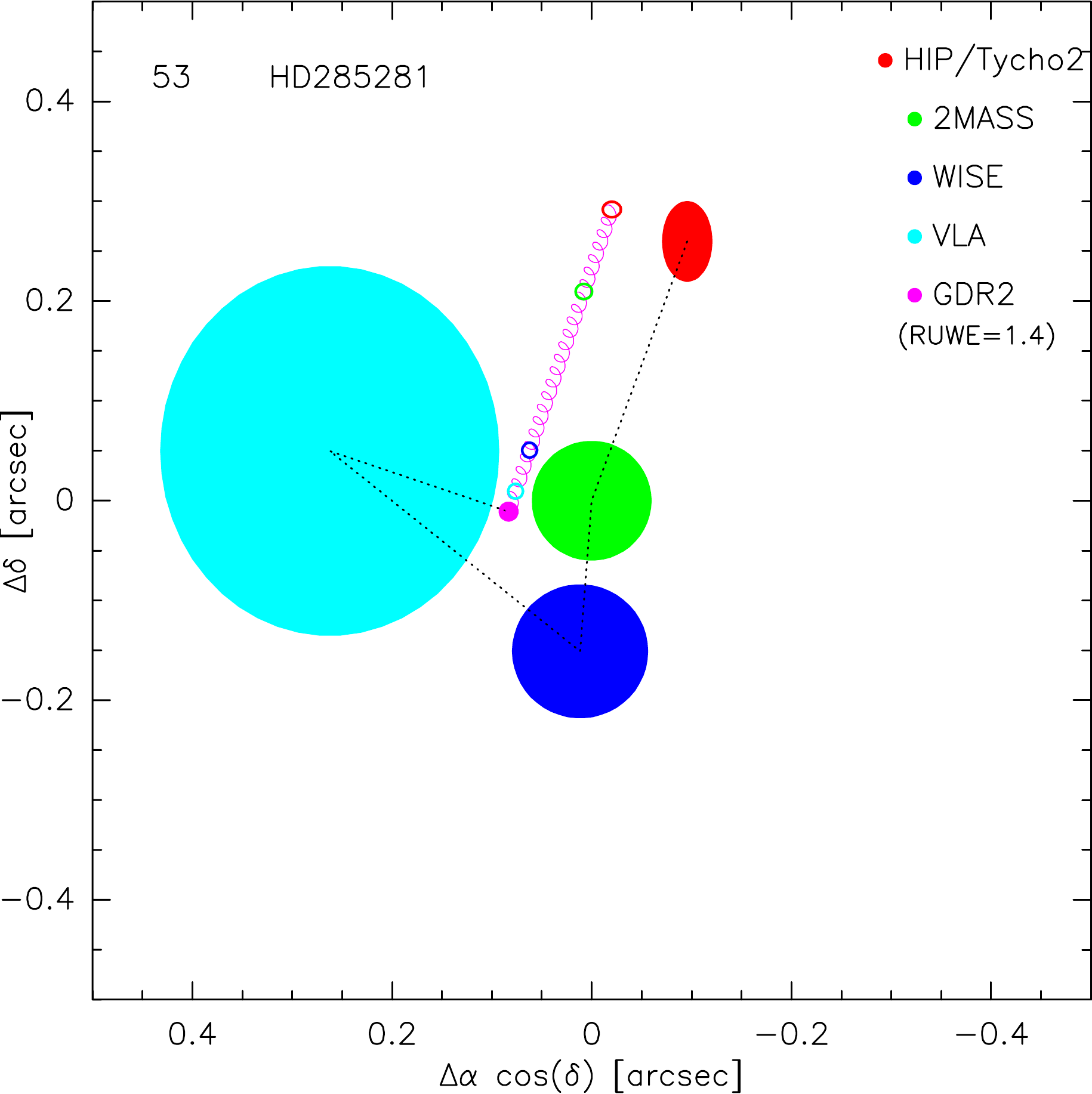}
\figsetgrpnote{Astrometric chart for HD\,285281 (V1293\,Tau).
The number in the upper left corner relates to the target number in Tables\,\ref{tbl-det} and \ref{tbl-sourcelist}. Measured positions (relative to the 2MASS) and 1\,$\sigma$\ uncertainties are drawn as filled ellipses with the colour coding for the respective missions/catalogs indicated in the upper right corner. Straight dotted lines connect the observed positions just to guide the eye. The wiggled line shows the combined proper motion and parallax prediction, starting at the GDR2 2015.5 position and projected backwards in time. Open ellipses show the position and uncertainty prediction from that starting point for the respective observing epochs.}
\figsetgrpend

\figsetgrpstart
\figsetgrpnum{6.10}
\figsetgrptitle{HD\,284135}
\figsetplot{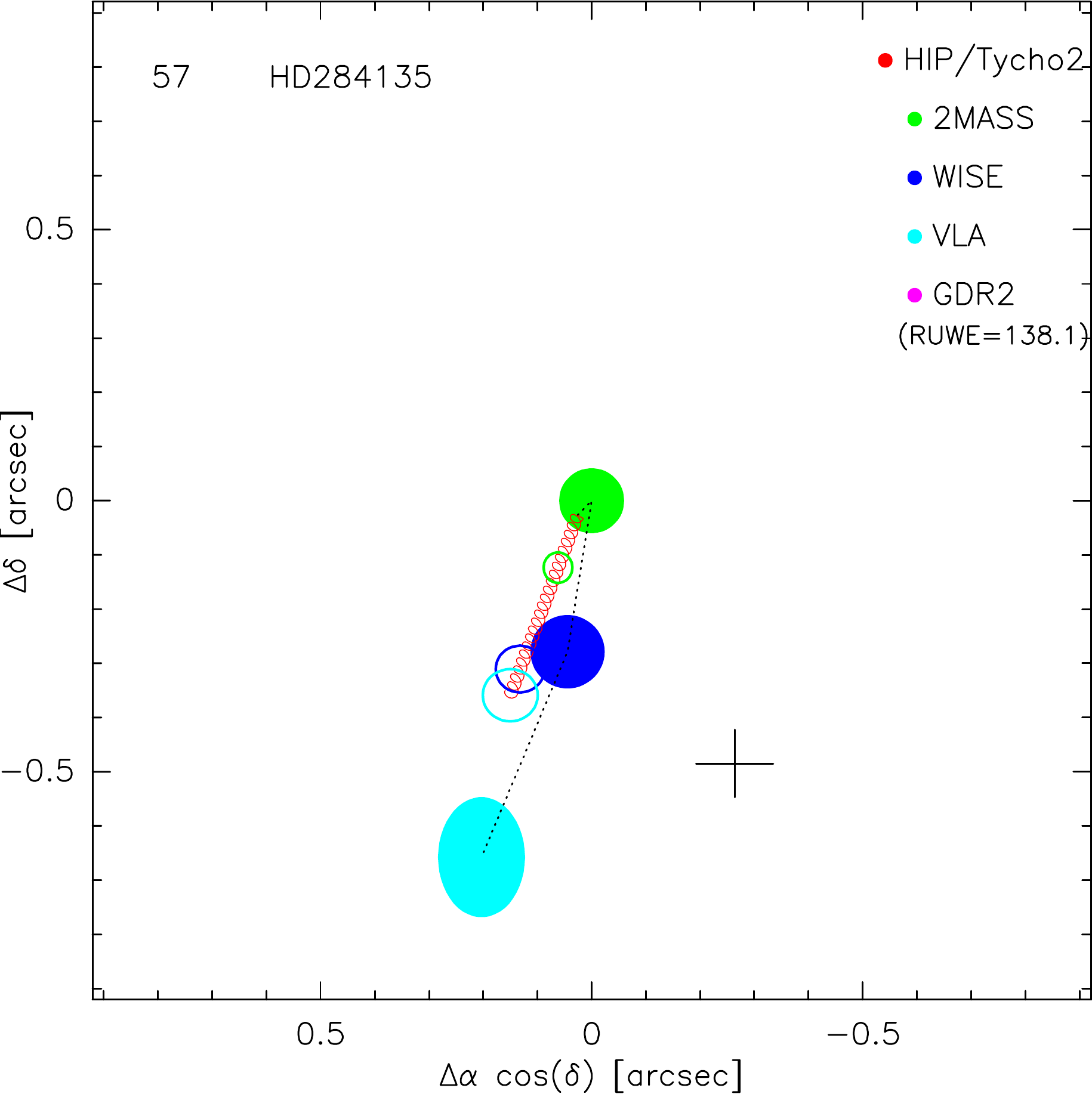}
\figsetgrpnote{Astrometric chart for HD\,284135 (V1299\,Tau). 
The number in the upper left corner relates  to the target number in Tables\,\ref{tbl-det} and \ref{tbl-sourcelist}. Measured positions (relative to the 2MASS) and 1\,$\sigma$\ uncertainties are drawn as filled ellipses with the colour coding for the respective missions/catalogs indicated in the upper right corner. Straight dotted lines connect the observed positions just to guide the eye. The red wiggled line shows the combined proper motion and parallax prediction, starting at the HIP/Tycho2 position and projected forward in time. The black cross shows the position of the companion WDS\,J04057+2248\,B \citep{mason2020} at the epoch of the VLA observations, i.e., relative to the center of the light blue open ellipse marking the predicted position of the primary. No GDR2 (nor GeDR3) astrometric solution exists for this 0\farcs4 binary star since it was most likely not spatially resolved by Gaia.}
\figsetgrpend

\figsetgrpstart
\figsetgrpnum{6.11}
\figsetgrptitle{HD\,31281}
\figsetplot{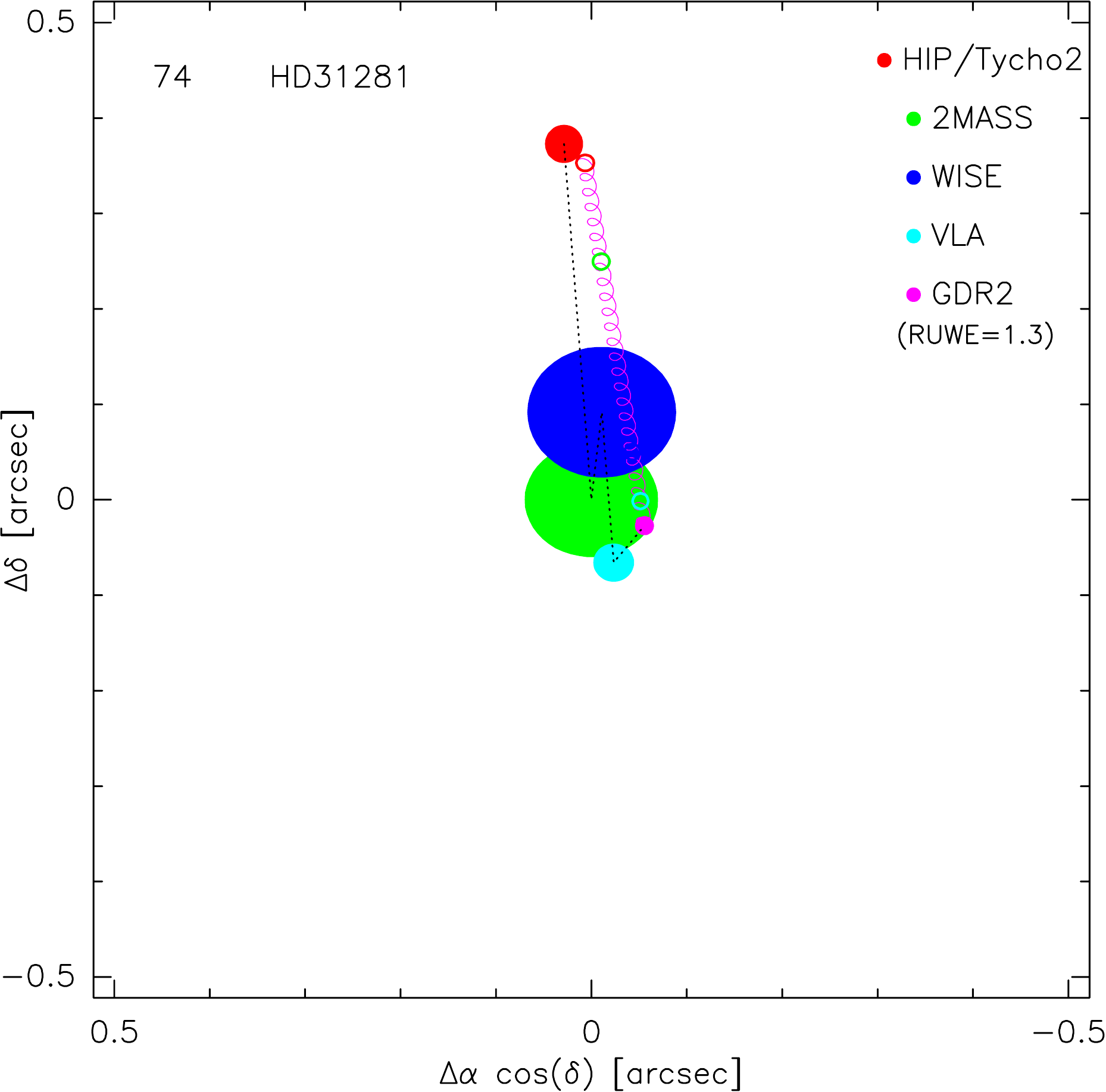}
\figsetgrpnote{Astrometric chart for HD\,31281 (V1349\,Tau).
The number in the upper left corner relates to the target number in Tables\,\ref{tbl-det} and \ref{tbl-sourcelist}. Measured positions (relative to the 2MASS) and 1\,$\sigma$\ uncertainties are drawn as filled ellipses with the colour coding for the respective missions/catalogs indicated in the upper right corner. Straight dotted lines connect the observed positions just to guide the eye. The wiggled line shows the combined proper motion and parallax prediction, starting at the GDR2 2015.5 position and projected backwards in time. Open ellipses show the position and uncertainty prediction from that starting point for the respective observing epochs.}
\figsetgrpend

\figsetgrpstart
\figsetgrpnum{6.12}
\figsetgrptitle{BD-08\,995}
\figsetplot{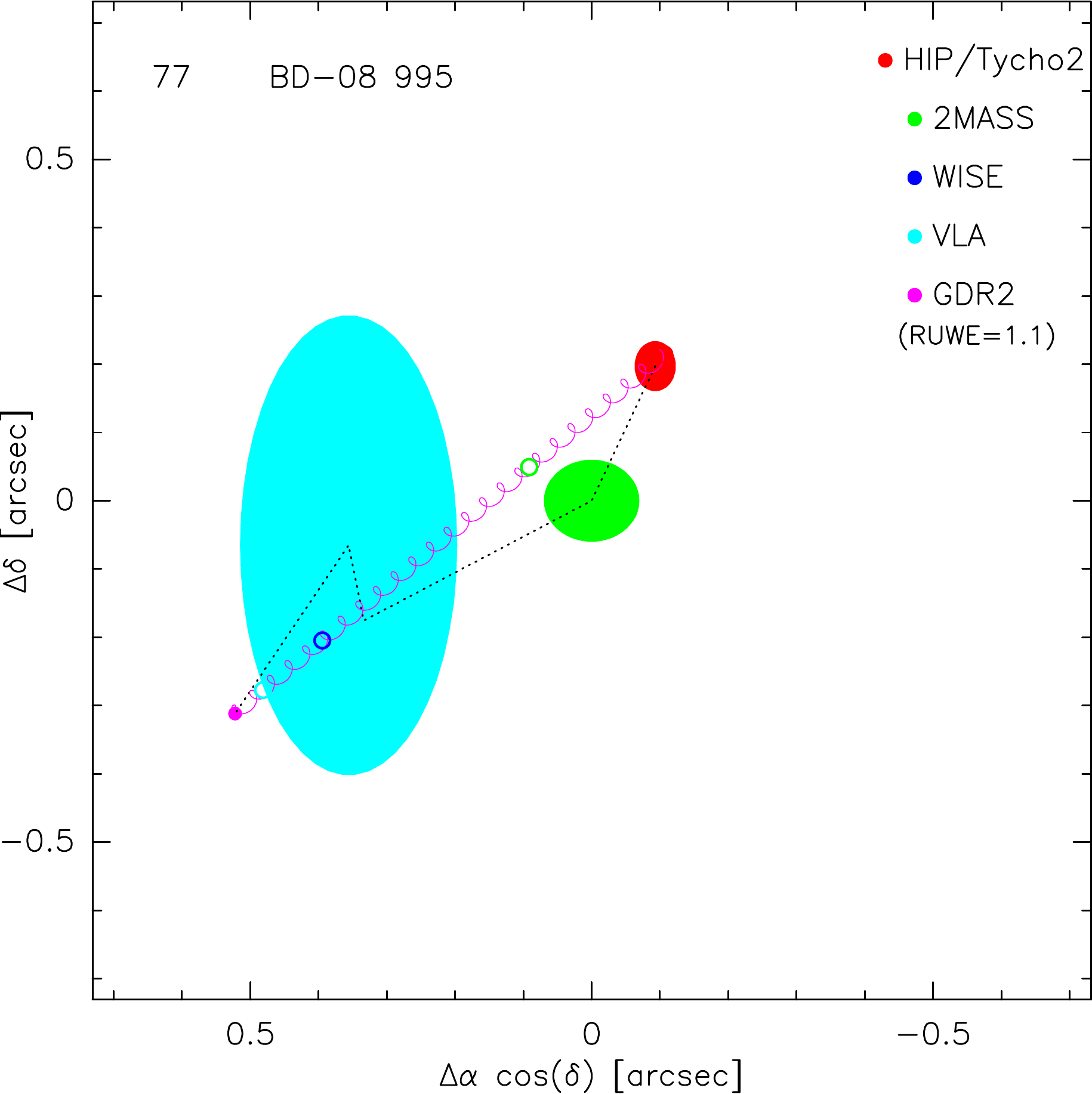}
\figsetgrpnote{Astrometric chart for BD-08\,995.
The number in the upper left corner relates to the target number in Tables\,\ref{tbl-det} and \ref{tbl-sourcelist}. Measured positions (relative to the 2MASS) and 1\,$\sigma$\ uncertainties are drawn as filled ellipses with the colour coding for the respective missions/catalogs indicated in the upper right corner. Straight dotted lines connect the observed positions just to guide the eye. The wiggled line shows the combined proper motion and parallax prediction, starting at the GDR2 2015.5 position and projected backwards in time. Open ellipses show the position and uncertainty prediction from that starting point for the respective observing epochs.}
\figsetgrpend

\figsetgrpstart
\figsetgrpnum{6.13}
\figsetgrptitle{HD\,286264}
\figsetplot{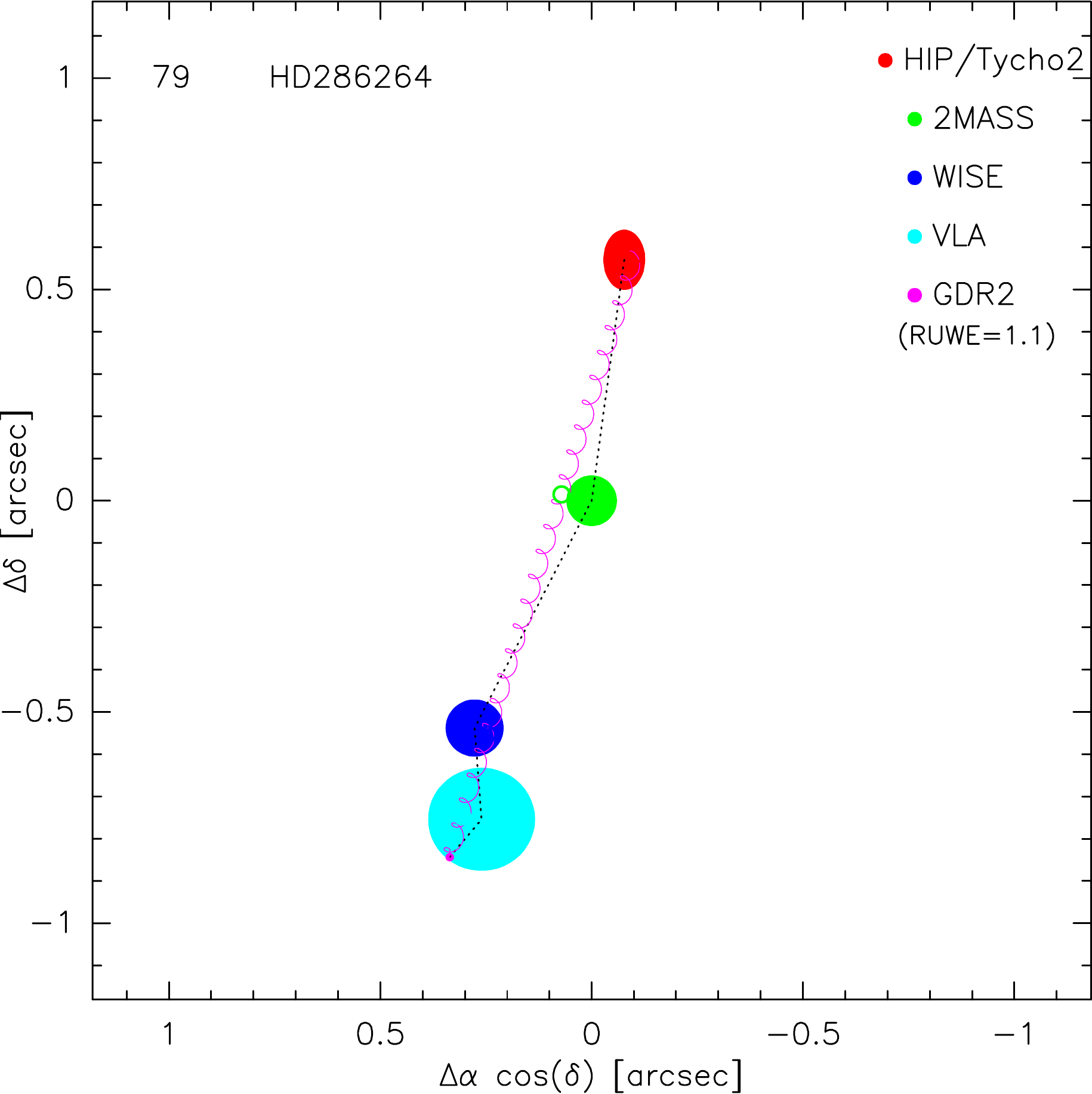}
\figsetgrpnote{Astrometric chart for HD\,286264 (V1841\,Ori).
The number in the upper left corner relates to the target number in Tables\,\ref{tbl-det} and \ref{tbl-sourcelist}. Measured positions (relative to the 2MASS) and 1\,$\sigma$\ uncertainties are drawn as filled ellipses with the colour coding for the respective missions/catalogs indicated in the upper right corner. Straight dotted lines connect the observed positions just to guide the eye. The wiggled line shows the combined proper motion and parallax prediction, starting at the GDR2 2015.5 position and projected backwards in time. Open ellipses show the position and uncertainty prediction from that starting point for the respective observing epochs.}
\figsetgrpend

\figsetgrpstart
\figsetgrpnum{6.14}
\figsetgrptitle{HD\,293857}
\figsetplot{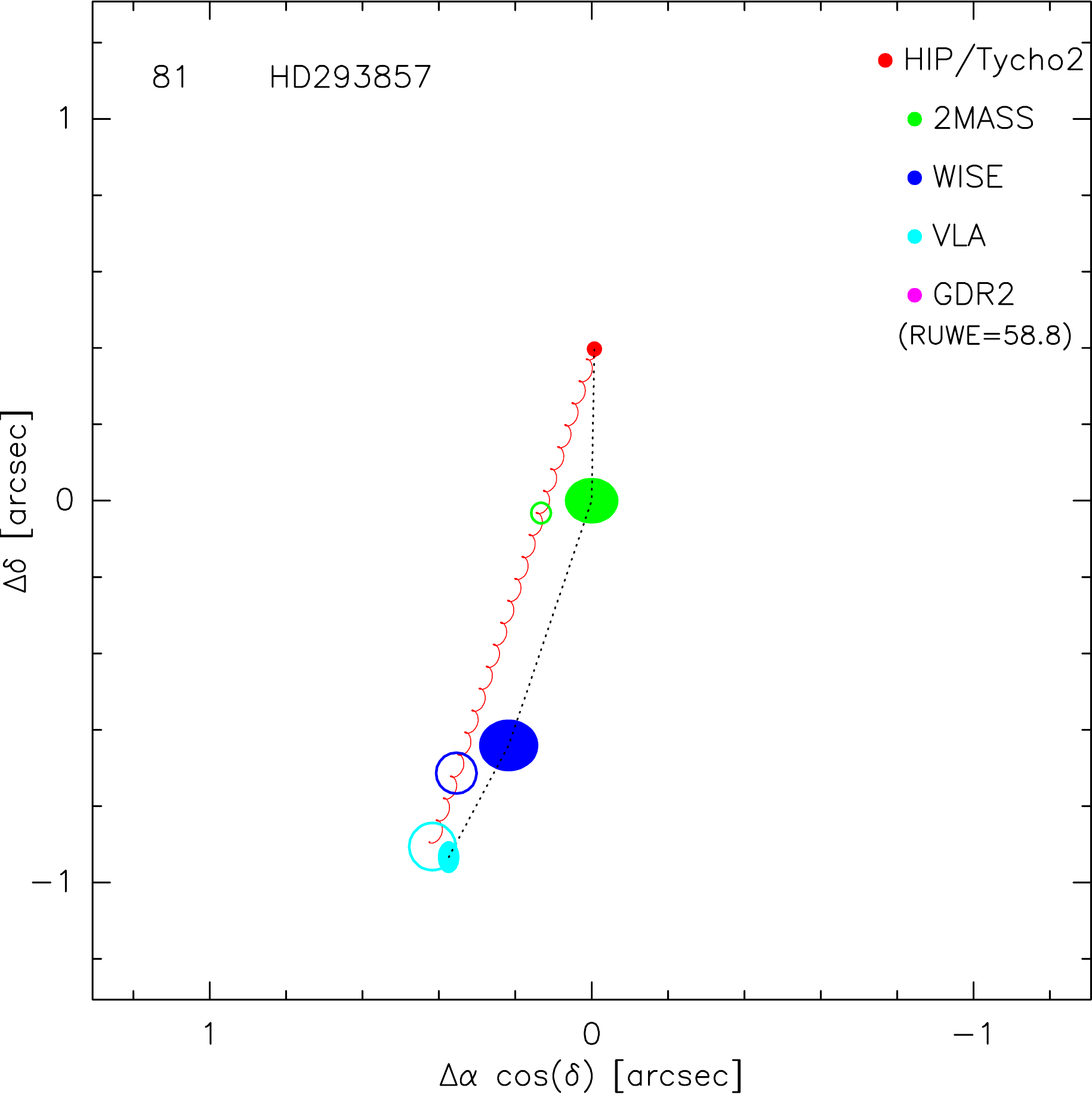}
\figsetgrpnote{Astrometric chart for HD\,293857 (BD-04\,1063).
The number in the upper left corner relates to the target number in Tables\,\ref{tbl-det} and \ref{tbl-sourcelist}. Measured positions (relative to the 2MASS) and 1\,$\sigma$\ uncertainties are drawn as filled ellipses with the colour coding for the respective missions/catalogs indicated in the upper right corner. Straight dotted lines connect the observed positions just to guide the eye. The red wiggled line shows the combined proper motion and parallax prediction, starting at the HIP/Tycho2 position and projected forward in time. Open ellipses show the position and uncertainty prediction from that starting point for the respective observing epochs. No GDR2 (nor GeDR3) astrometric solution exists for this star.}
\figsetgrpend

\figsetgrpstart
\figsetgrpnum{6.15}
\figsetgrptitle{TYC\,713-661-1}
\figsetplot{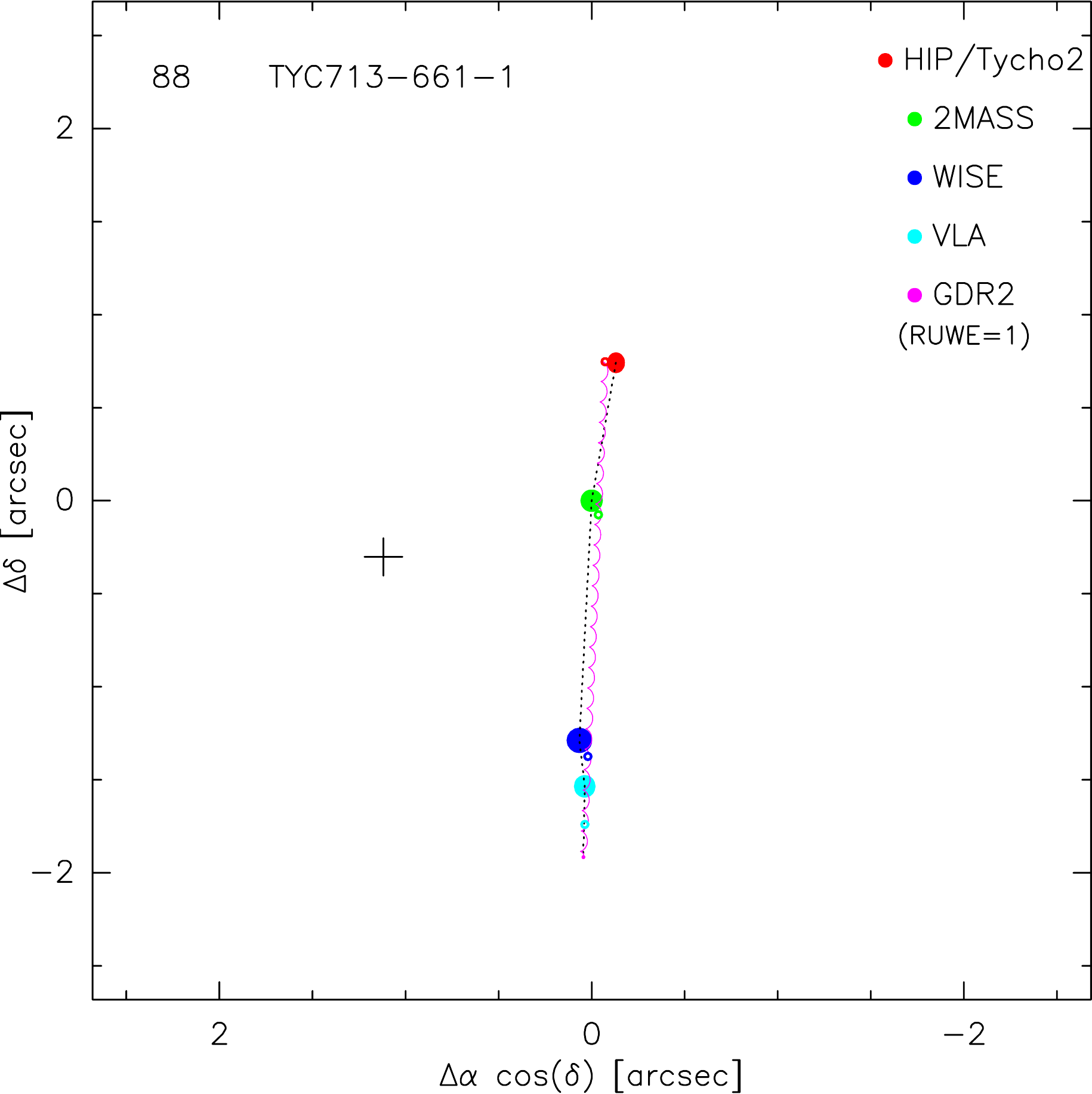}
\figsetgrpnote{Astrometric chart for TYC\,713-661-1.
The number in the upper left corner relates to the target number in Tables\,\ref{tbl-det} and \ref{tbl-sourcelist}. Measured positions (relative to the 2MASS) and 1\,$\sigma$\ uncertainties are drawn as filled ellipses with the colour coding for the respective missions/catalogs indicated in the upper right corner. Straight dotted lines connect the observed positions just to guide the eye. The wiggled line shows the combined proper motion and parallax prediction, starting at the GDR2 2015.5 position and projected backwards in time. Open ellipses show the position and uncertainty prediction from that starting point for the respective observing epochs.}
\figsetgrpend

\figsetgrpstart
\figsetgrpnum{6.16}
\figsetgrptitle{TYC\,5925-1547-1}
\figsetplot{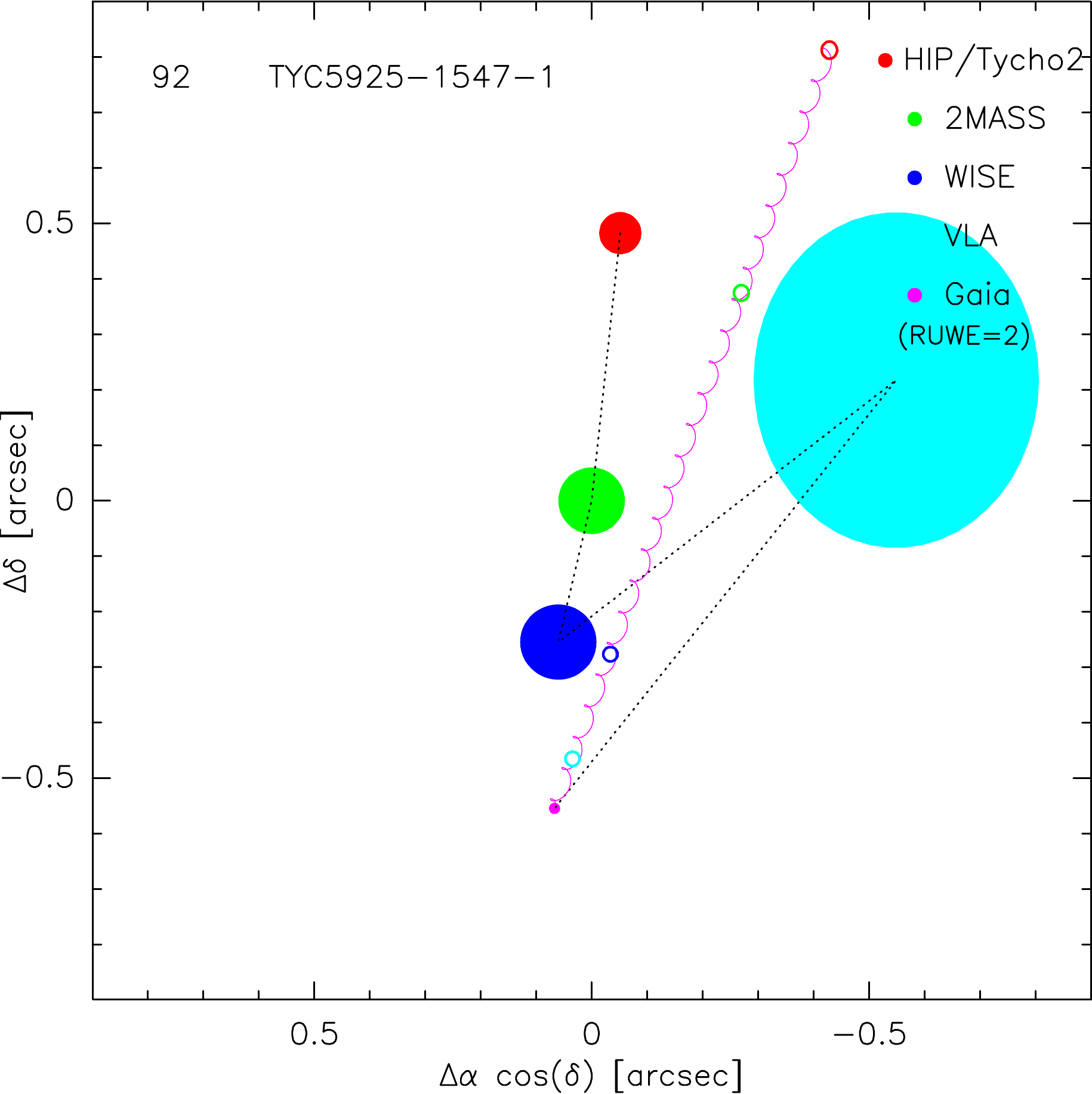}
\figsetgrpnote{Astrometric chart for TYC\,5925-1547-1 (CPD-19\,878).
The number in the upper left corner relates to the target number in Tables\,\ref{tbl-det} and \ref{tbl-sourcelist}. Measured positions (relative to the 2MASS) and 1\,$\sigma$\ uncertainties are drawn as filled ellipses with the colour coding for the respective missions/catalogs indicated in the upper right corner. Straight dotted lines connect the observed positions just to guide the eye. The wiggled line shows the combined proper motion and parallax prediction, starting at the GDR2 2015.5 position and projected backwards in time. Open ellipses show the position and uncertainty prediction from that starting point for the respective observing epochs.}
\figsetgrpend

\figsetgrpstart
\figsetgrpnum{6.17}
\figsetgrptitle{SAO\,150676}
\figsetplot{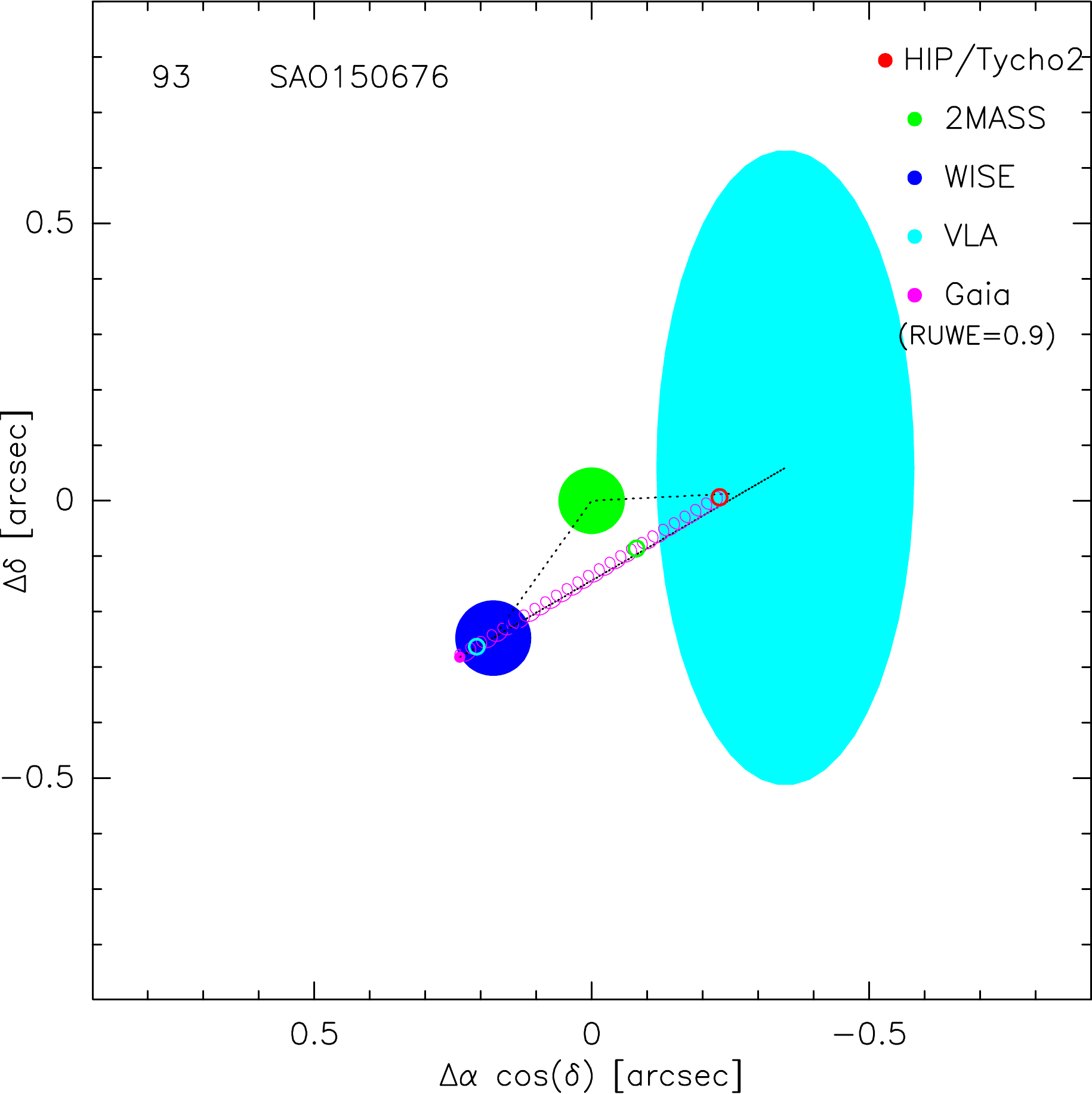}
\figsetgrpnote{Astrometric chart for SAO\,150676 (AI\,Lep).
The number in the upper left corner relates to the target number in Tables\,\ref{tbl-det} and \ref{tbl-sourcelist}. Measured positions (relative to the 2MASS) and 1\,$\sigma$\ uncertainties are drawn as filled ellipses with the colour coding for the respective missions/catalogs indicated in the upper right corner. Straight dotted lines connect the observed positions just to guide the eye. The wiggled line shows the combined proper motion and parallax prediction, starting at the GDR2 2015.5 position and projected backwards in time. Open ellipses show the position and uncertainty prediction from that starting point for the respective observing epochs.}
\figsetgrpend

\figsetgrpstart
\figsetgrpnum{6.18}
\figsetgrptitle{HD\,62237}
\figsetplot{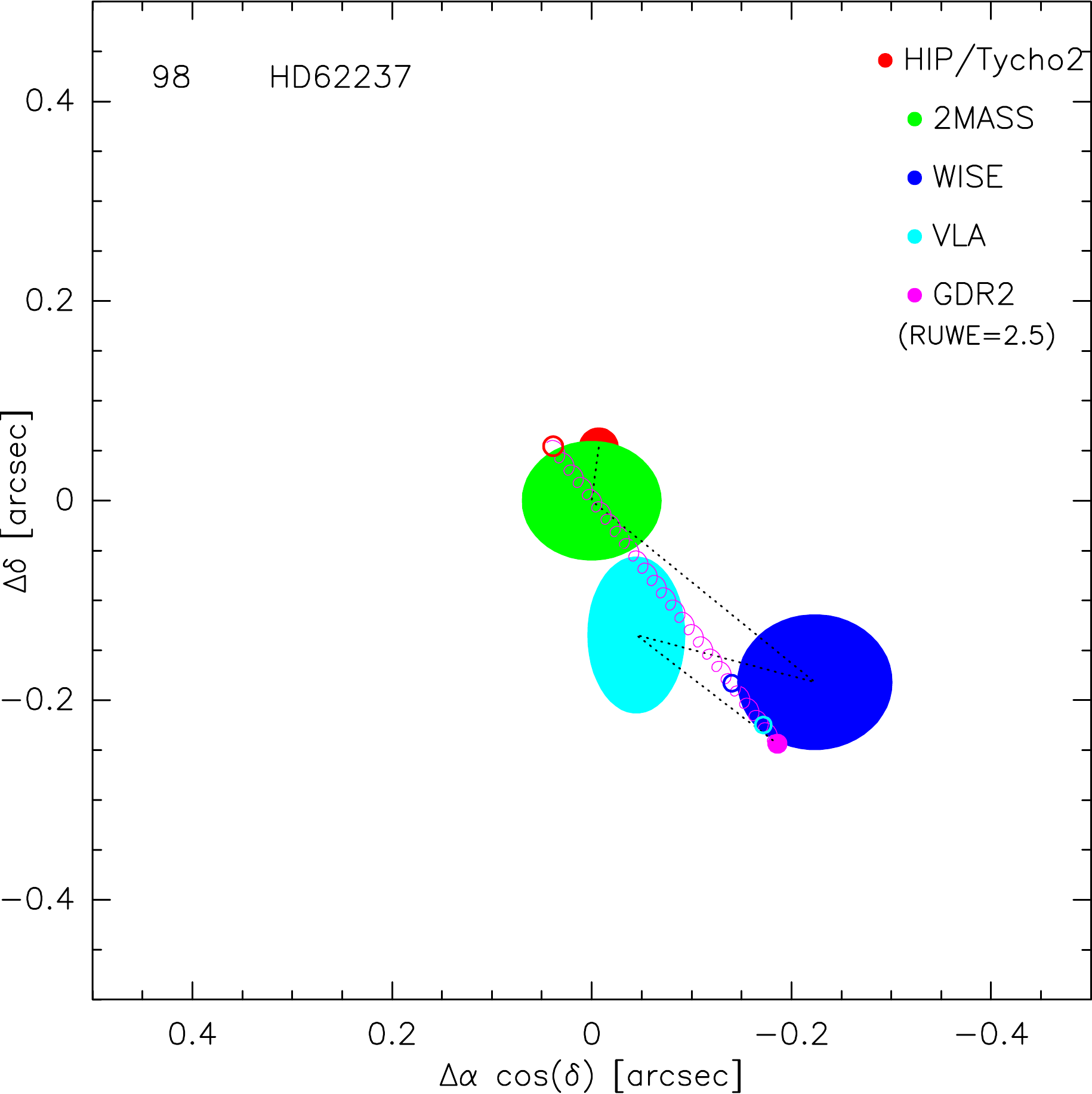}
\figsetgrpnote{Astrometric chart for HD\,62237 (BD-15\,1991).
The number in the upper left corner relates to the target number in Tables\,\ref{tbl-det} and \ref{tbl-sourcelist}. Measured positions (relative to the 2MASS) and 1\,$\sigma$\ uncertainties are drawn as filled ellipses with the colour coding for the respective missions/catalogs indicated in the upper right corner. Straight dotted lines connect the observed positions just to guide the eye. The wiggled line shows the combined proper motion and parallax prediction, starting at the GDR2 2015.5 position and projected backwards in time. Open ellipses show the position and uncertainty prediction from that starting point for the respective observing epochs.}
\figsetgrpend

\figsetgrpstart
\figsetgrpnum{6.19}
\figsetgrptitle{SAO\,135659}
\figsetplot{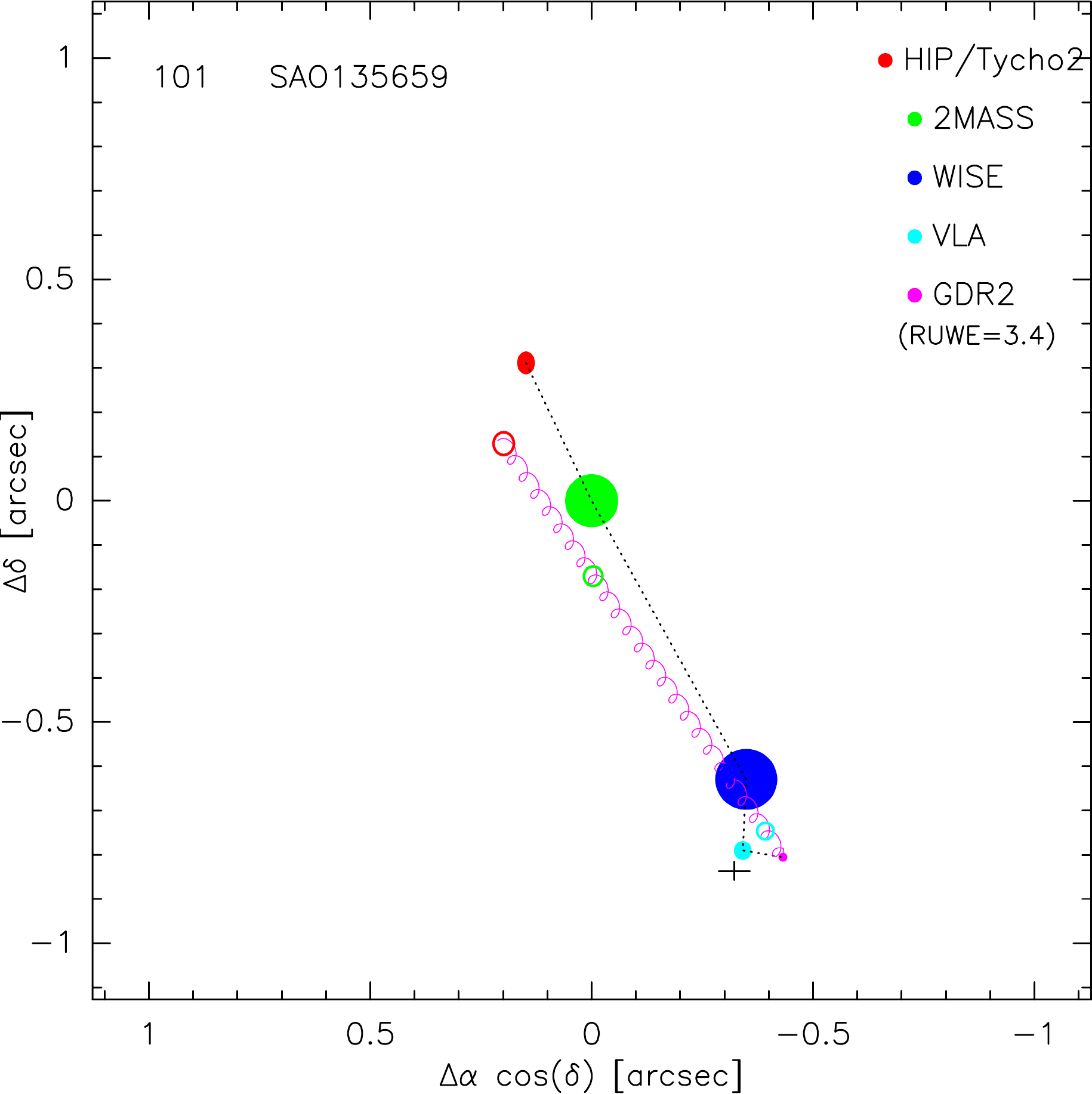}
\figsetgrpnote{Astrometric chart for SAO\,135659.
The number in the upper left corner relates to the target number in Tables\,\ref{tbl-det} and \ref{tbl-sourcelist}. Measured positions (relative to the 2MASS) and 1\,$\sigma$\ uncertainties are drawn as filled ellipses with the colour coding for the respective missions/catalogs indicated in the upper right corner. Straight dotted lines connect the observed positions just to guide the eye. The wiggled line shows the combined proper motion and parallax prediction, starting at the GDR2 2015.5 position and projected backwards in time. Open ellipses show the position and uncertainty prediction from that starting point for the respective observing epochs. The black cross shows the position of the companion WDS\,08138-0738\,B \citep{mason2020} at the epoch of the VLA observations, i.e., relative to the center of the light blue open ellipse marking the predicted position of the primary. No independent astrometric solution exists for this secondary since it was neither detected by Gaia, nor by Hipparcos.}
\figsetgrpend

\figsetgrpstart
\figsetgrpnum{6.20}
\figsetgrptitle{HD\,77407}
\figsetplot{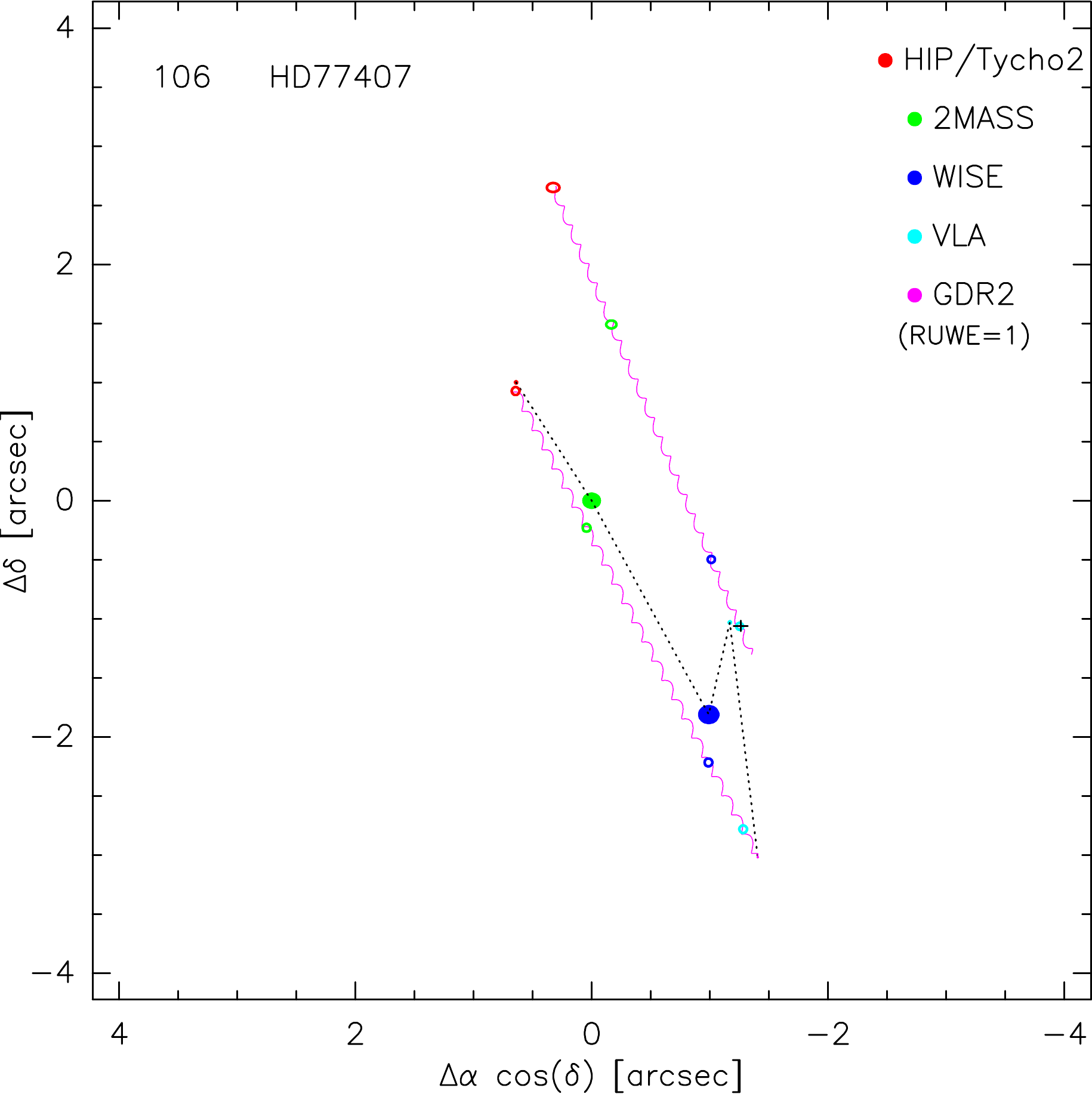}
\figsetgrpnote{Astrometric chart for HD\,77407.
The number in the upper left corner relates to the target number in Tables\,\ref{tbl-det} and \ref{tbl-sourcelist}. Measured positions (relative to the 2MASS) and 1\,$\sigma$\ uncertainties are drawn as filled ellipses with the colour coding for the respective missions/catalogs indicated in the upper right corner. Straight dotted lines connect the observed positions just to guide the eye. The lower wiggled line shows the combined proper motion and parallax prediction, starting at the GDR2 2015.5 position and projected backwards in time. Open ellipses show the position and uncertainty prediction from that starting point for the respective observing epochs. The black cross shows the position of the companion WDS\,09035+3750\,B \citep{mason2020} at the epoch of the VLA observations, i.e., relative to the center of the light blue open ellipse marking the predicted position of the primary. The upper wiggled line and open ellipses show the proper motion/parallax and position prediction for the secondary, which was also detected by GDR2.}
\figsetgrpend

\figsetgrpstart
\figsetgrpnum{6.21}
\figsetgrptitle{HD\,82159}
\figsetplot{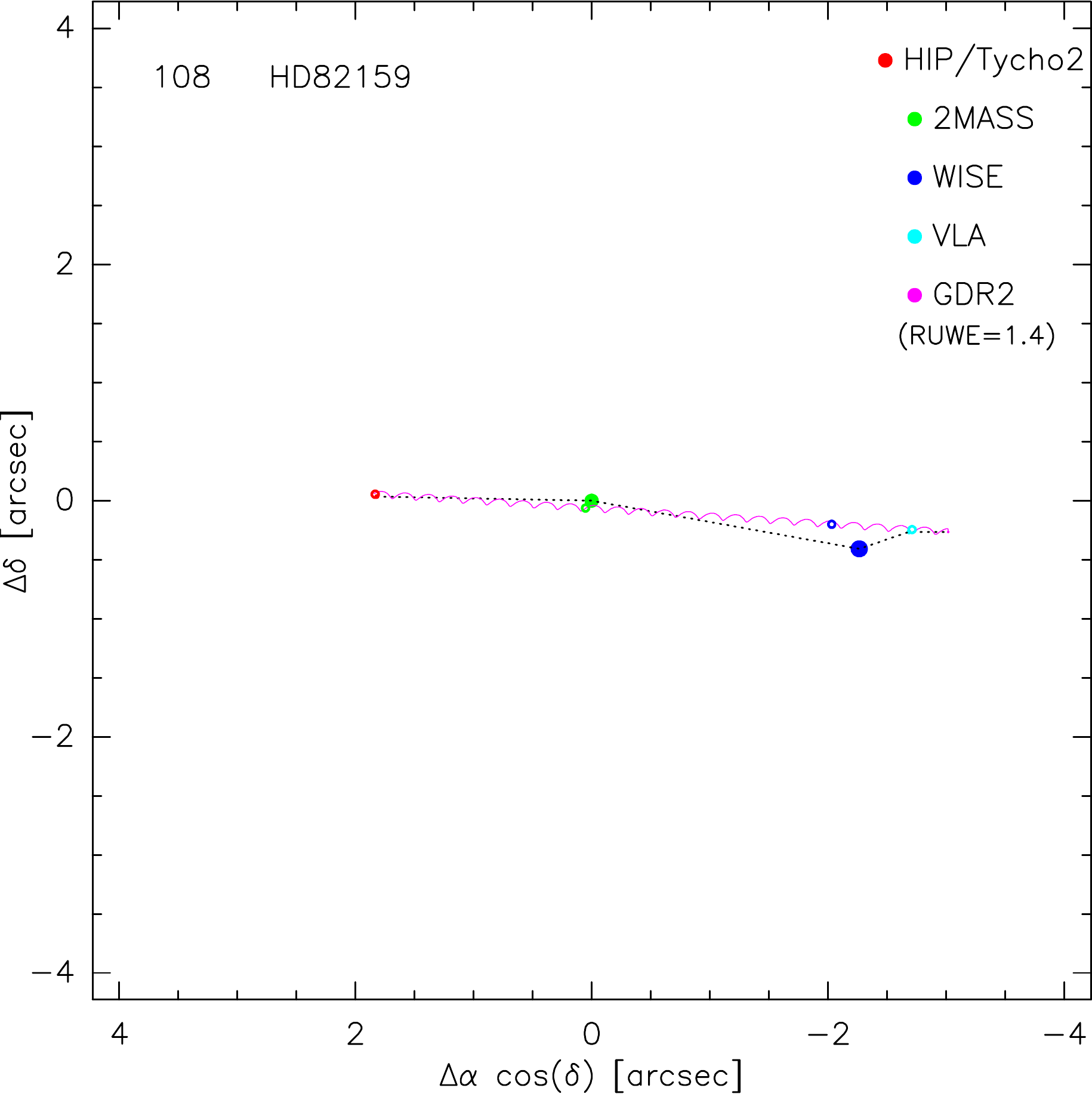}
\figsetgrpnote{Astrometric chart for HD\,82159 (GS\,Leo).
The number in the upper left corner relates to the target number in Tables\,\ref{tbl-det} and \ref{tbl-sourcelist}. Measured positions (relative to the 2MASS) and 1\,$\sigma$\ uncertainties are drawn as filled ellipses with the colour coding for the respective missions/catalogs indicated in the upper right corner. Straight dotted lines connect the observed positions just to guide the eye. The wiggled line shows the combined proper motion and parallax prediction, starting at the GDR2 2015.5 position and projected backwards in time. Open ellipses show the position and uncertainty prediction from that starting point for the respective observing epochs.}
\figsetgrpend

\figsetgrpstart
\figsetgrpnum{6.22}
\figsetgrptitle{HD\,82558}
\figsetplot{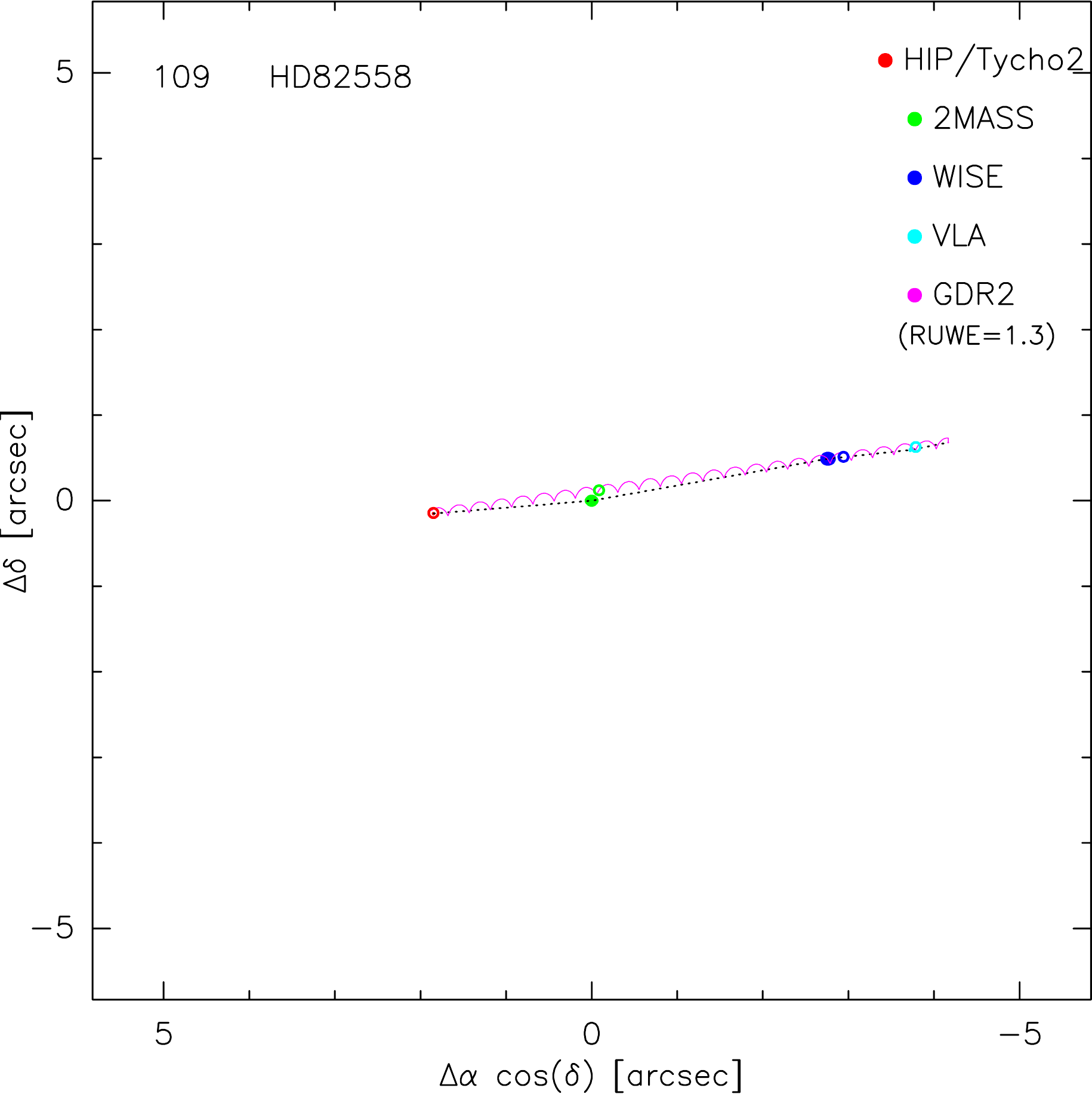}
\figsetgrpnote{Astrometric chart for HD\,82558 (LQ\,Hya).
The number in the upper left corner relates to the target number in Tables\,\ref{tbl-det} and \ref{tbl-sourcelist}. Measured positions (relative to the 2MASS) and 1\,$\sigma$\ uncertainties are drawn as filled ellipses with the colour coding for the respective missions/catalogs indicated in the upper right corner. Straight dotted lines connect the observed positions just to guide the eye. The wiggled line shows the combined proper motion and parallax prediction, starting at the GDR2 2015.5 position and projected backwards in time. Open ellipses show the position and uncertainty prediction from that starting point for the respective observing epochs.}
\figsetgrpend

\figsetgrpstart
\figsetgrpnum{6.23}
\figsetgrptitle{GJ\,2079}
\figsetplot{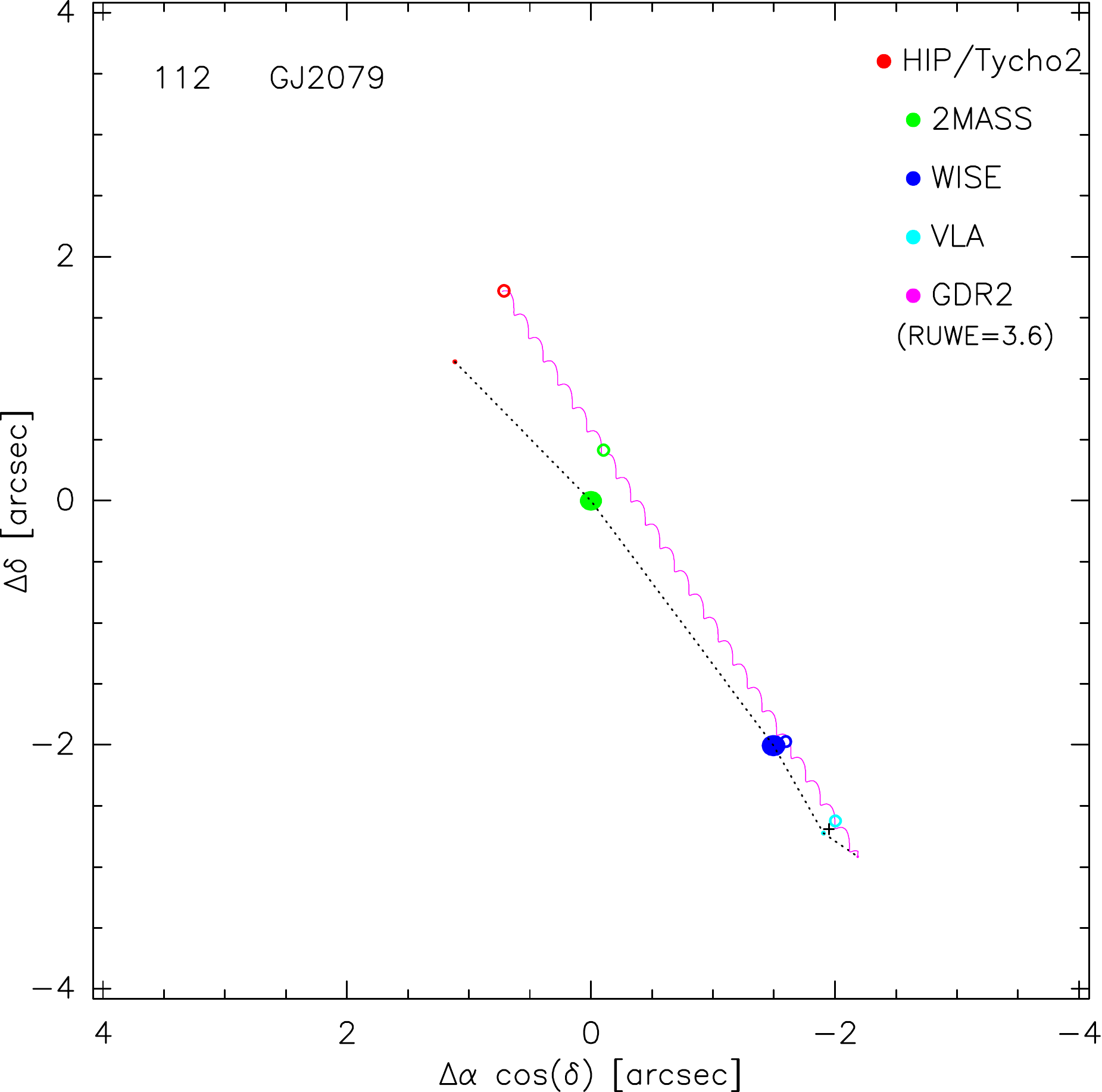}
\figsetgrpnote{Astrometric chart for GJ\,2079 (DK\,Leo).
The number in the upper left corner relates to the target number in Tables\,\ref{tbl-det} and \ref{tbl-sourcelist}. Measured positions (relative to the 2MASS) and 1\,$\sigma$\ uncertainties are drawn as filled ellipses with the colour coding for the respective missions/catalogs indicated in the upper right corner. Straight dotted lines connect the observed positions just to guide the eye. The wiggled line shows the combined proper motion and parallax prediction, starting at the GDR2 2015.5 position and projected backwards in time. Open ellipses show the position and uncertainty prediction from that starting point for the respective observing epochs. The black cross shows the position of the companion WDS\,08138-0738\,B \citep{mason2020} at the epoch of the VLA observations, i.e., relative to the center of the light blue open ellipse marking the predicted position of the primary. No independent astrometric solution exists for this secondary since it was neither detected by Gaia, nor by Hipparcos.}
\figsetgrpend

\figsetgrpstart
\figsetgrpnum{6.24}
\figsetgrptitle{HD\,135363}
\figsetplot{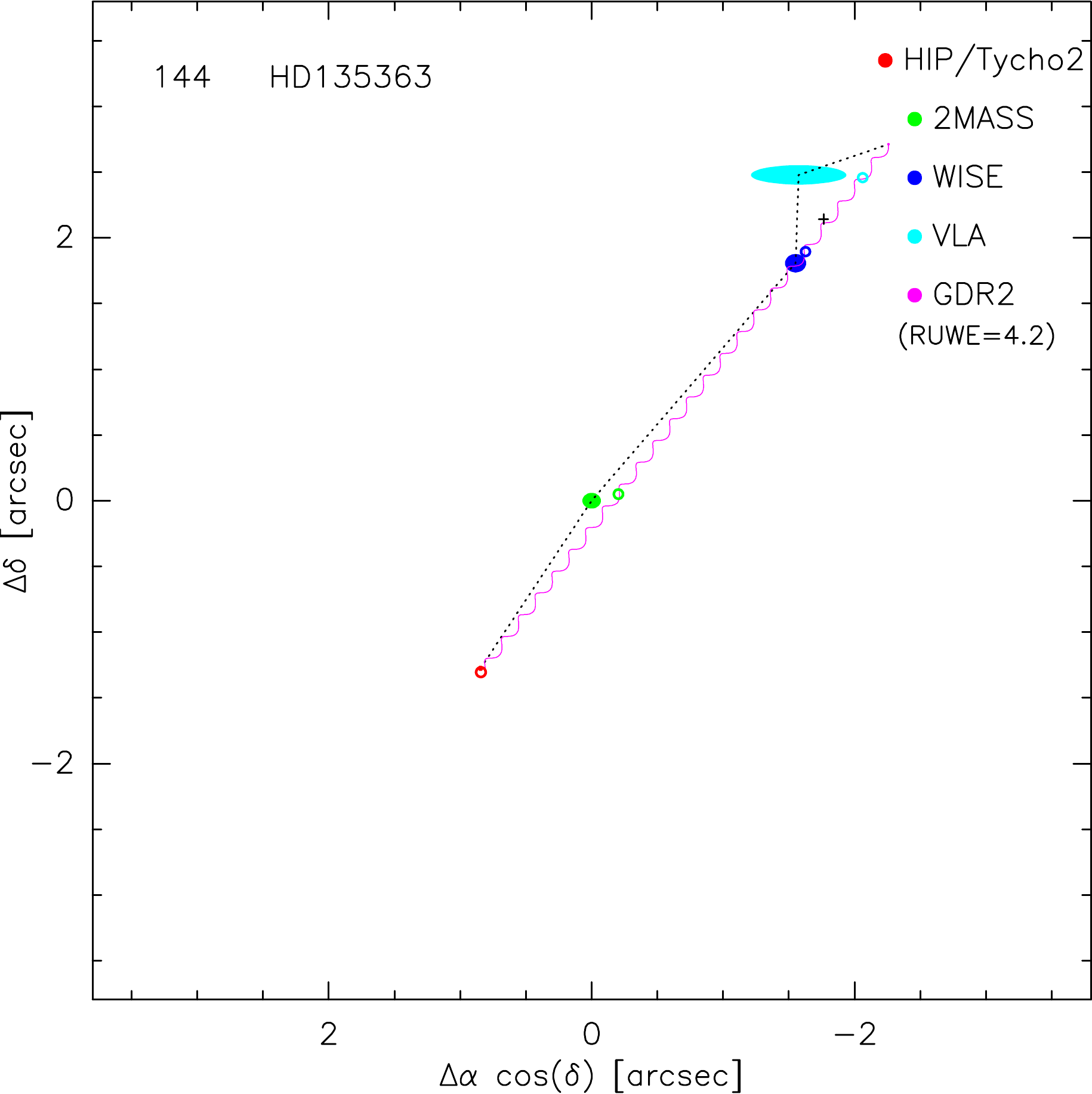}
\figsetgrpnote{Astrometric chart for HD\,135363 (BD+76\,552).
The number in the upper left corner relates to the target number in Tables\,\ref{tbl-det} and \ref{tbl-sourcelist}. Measured positions (relative to the 2MASS) and 1\,$\sigma$\ uncertainties are drawn as filled ellipses with the colour coding for the respective missions/catalogs indicated in the upper right corner. Straight dotted lines connect the observed positions just to guide the eye. The wiggled line shows the combined proper motion and parallax prediction, starting at the GDR2 2015.5 position and projected backwards in time. Open ellipses show the position and uncertainty prediction from that starting point for the respective observing epochs.}
\figsetgrpend

\figsetgrpstart
\figsetgrpnum{6.25}
\figsetgrptitle{HD\,199143}
\figsetplot{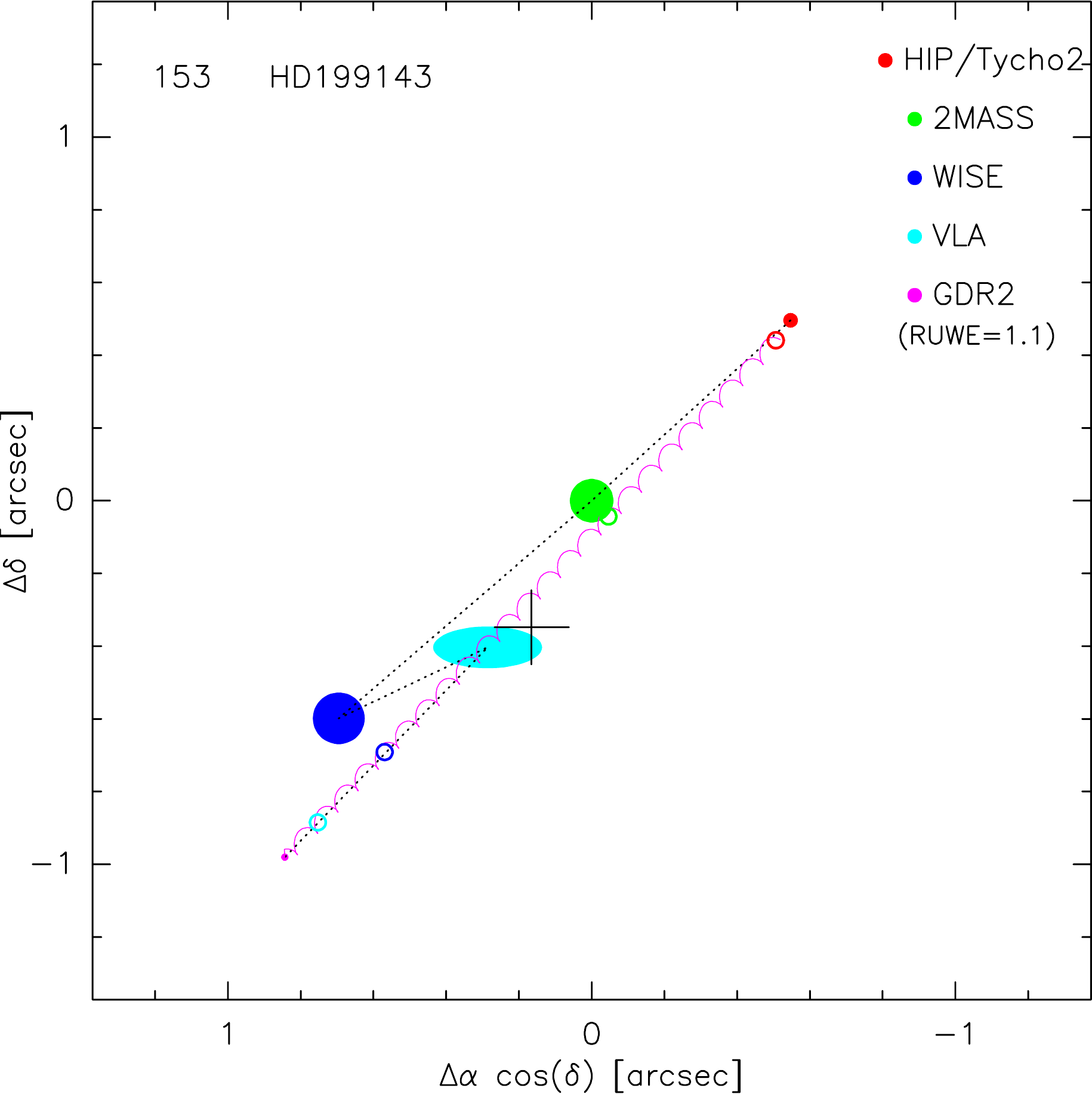}
\figsetgrpnote{Astrometric chart for HD\,199143.
The number in the upper left corner relates to the target number in Tables\,\ref{tbl-det} and \ref{tbl-sourcelist}. Measured positions (relative to the 2MASS) and 1\,$\sigma$\ uncertainties are drawn as filled ellipses with the colour coding for the respective missions/catalogs indicated in the upper right corner. Straight dotted lines connect the observed positions just to guide the eye. The wiggled line shows the combined proper motion and parallax prediction, starting at the GDR2 2015.5 position and projected backwards in time. Open ellipses show the position and uncertainty prediction from that starting point for the respective observing epochs. The black cross shows the position of the WDS companion HD\,199143\,B \citep{mason2020} at the epoch of the VLA observations, i.e., relative to the center of the light blue open ellipse marking the predicted position of the primary. No independent astrometric solution exists for this secondary since it was neither detected by Gaia, nor by Hipparcos.}
\figsetgrpend

\figsetgrpstart
\figsetgrpnum{6.26}
\figsetgrptitle{HD\,358623}
\figsetplot{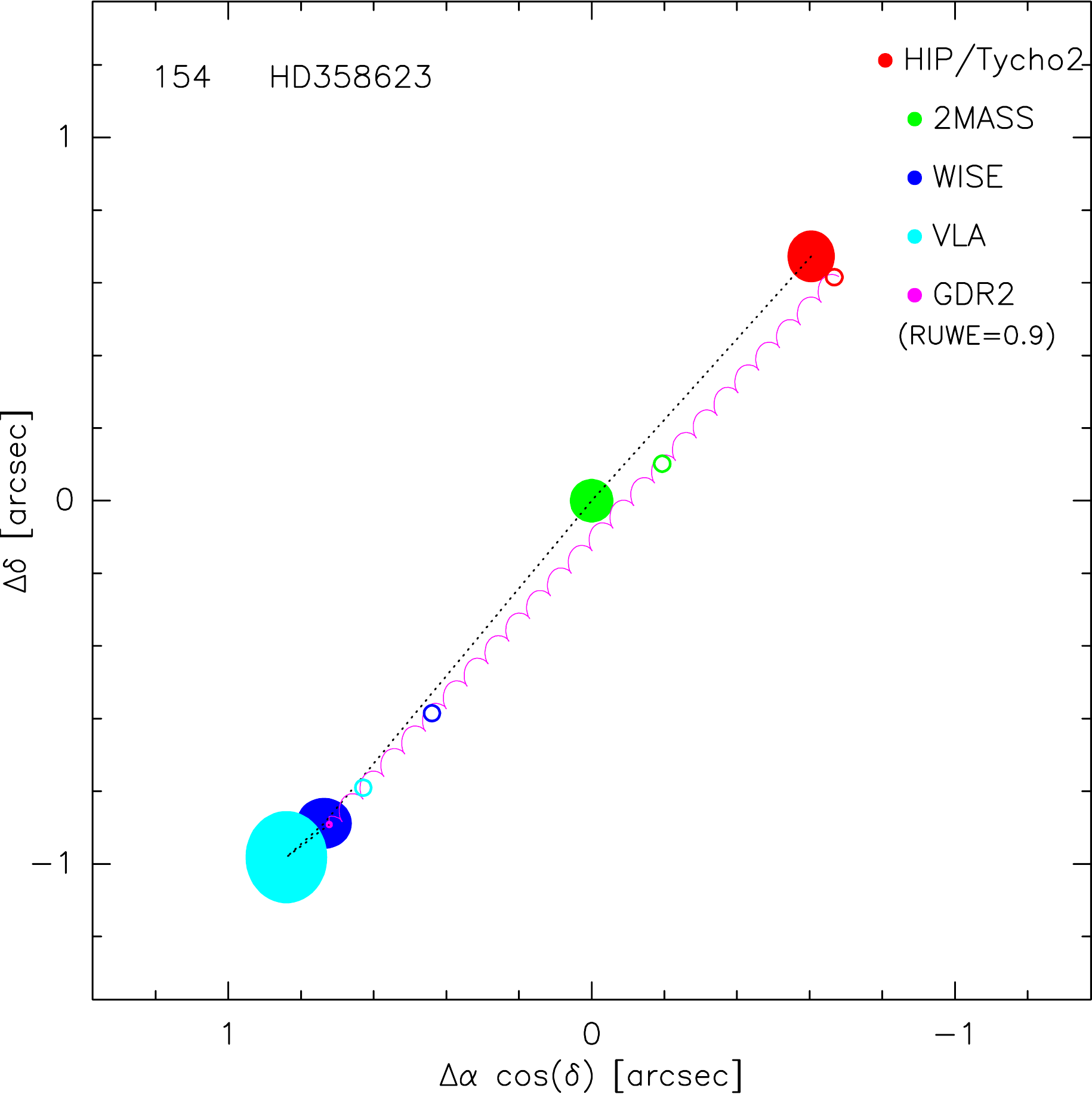}
\figsetgrpnote{Astrometric chart for HD\,358623 (AZ\,Cap).
The number in the upper left corner relates to the target number in Tables\,\ref{tbl-det} and \ref{tbl-sourcelist}. Measured positions (relative to the 2MASS) and 1\,$\sigma$\ uncertainties are drawn as filled ellipses with the colour coding for the respective missions/catalogs indicated in the upper right corner. Straight dotted lines connect the observed positions just to guide the eye. The wiggled line shows the combined proper motion and parallax prediction, starting at the GDR2 2015.5 position and projected backwards in time. Open ellipses show the position and uncertainty prediction from that starting point for the respective observing epochs.}
\figsetgrpend

\figsetgrpstart
\figsetgrpnum{6.27}
\figsetgrptitle{SAO\,50350}
\figsetplot{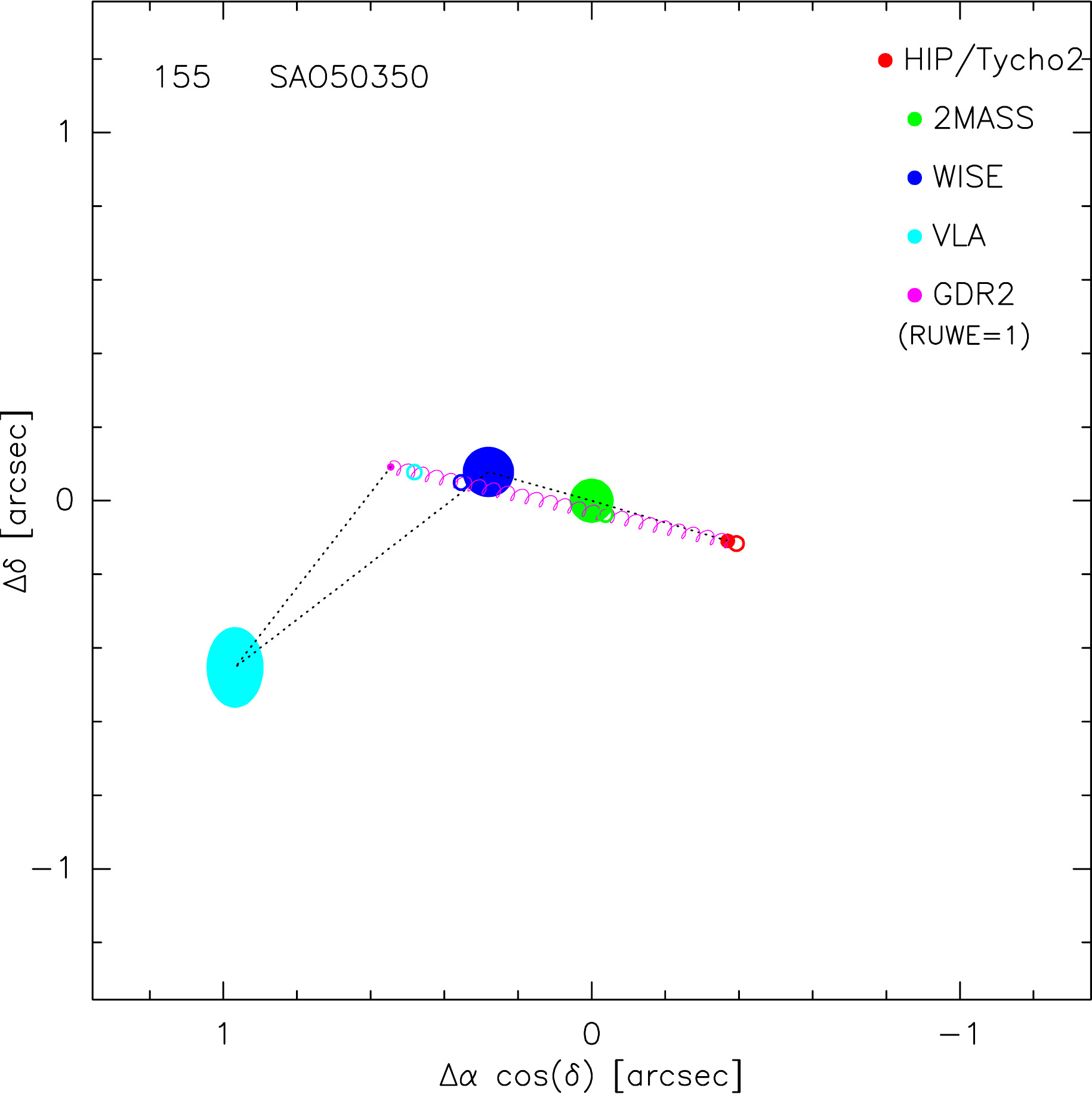}
\figsetgrpnote{Astrometric chart for SAO\,50350 (BD+44\,3670).
The number in the upper left corner relates to the target number in Tables\,\ref{tbl-det} and \ref{tbl-sourcelist}. Measured positions (relative to the 2MASS) and 1\,$\sigma$\ uncertainties are drawn as filled ellipses with the colour coding for the respective missions/catalogs indicated in the upper right corner. Straight dotted lines connect the observed positions just to guide the eye. The wiggled line shows the combined proper motion and parallax prediction, starting at the GDR2 2015.5 position and projected backwards in time. Open ellipses show the position and uncertainty prediction from that starting point for the respective observing epochs.}
\figsetgrpend

\figsetgrpstart
\figsetgrpnum{6.28}
\figsetgrptitle{GJ\,4199}
\figsetplot{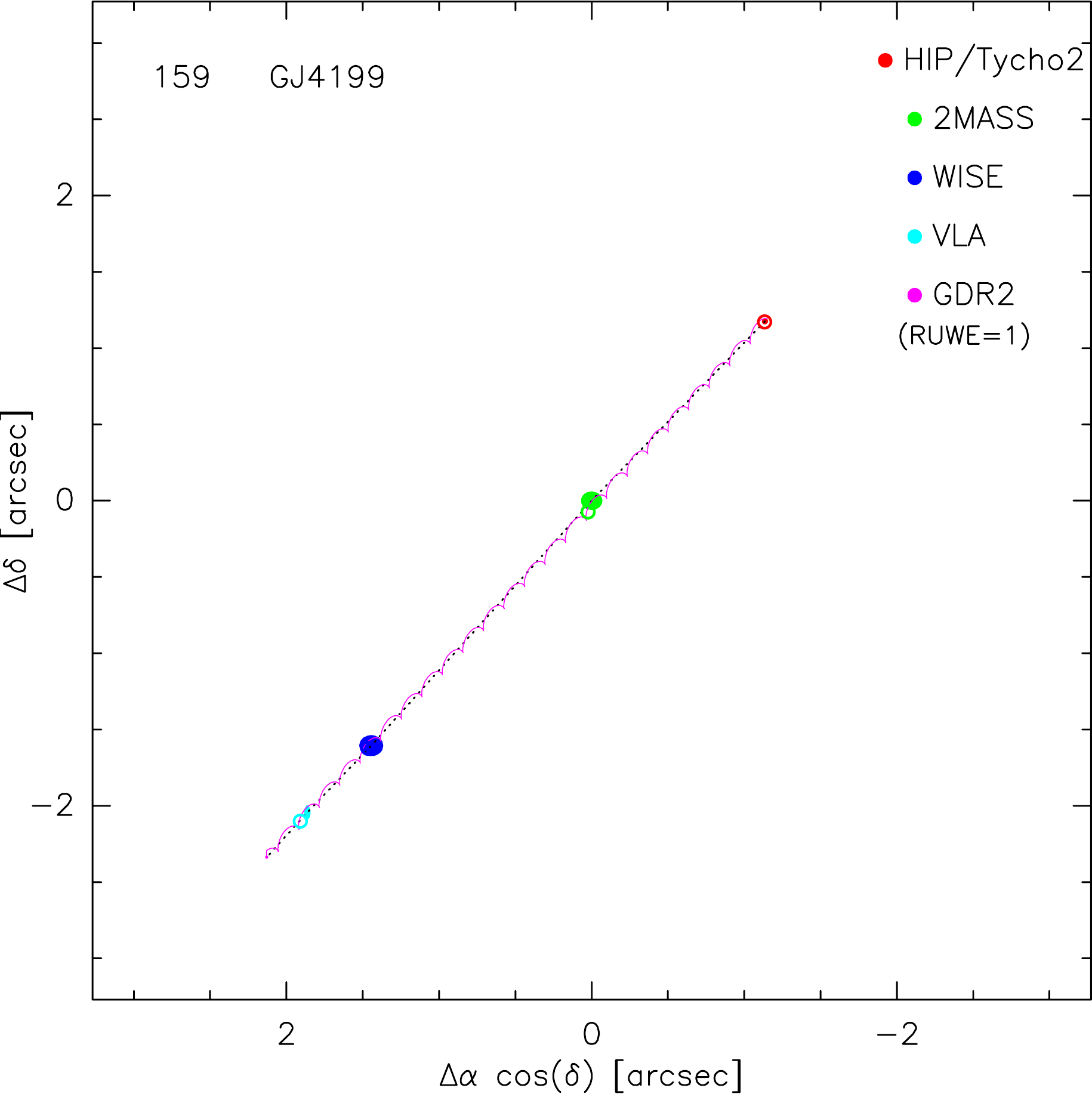}
\figsetgrpnote{Astrometric chart for GJ\,4199 (LO\,Peg).
The number in the upper left corner relates to the target number in Tables\,\ref{tbl-det} and \ref{tbl-sourcelist}. Measured positions (relative to the 2MASS) and 1\,$\sigma$\ uncertainties are drawn as filled ellipses with the colour coding for the respective missions/catalogs indicated in the upper right corner. Straight dotted lines connect the observed positions just to guide the eye. The wiggled line shows the combined proper motion and parallax prediction, starting at the GDR2 2015.5 position and projected backwards in time. Open ellipses show the position and uncertainty prediction from that starting point for the respective observing epochs.}
\figsetgrpend

\figsetgrpstart
\figsetgrpnum{6.29}
\figsetgrptitle{SAO\,51891}
\figsetplot{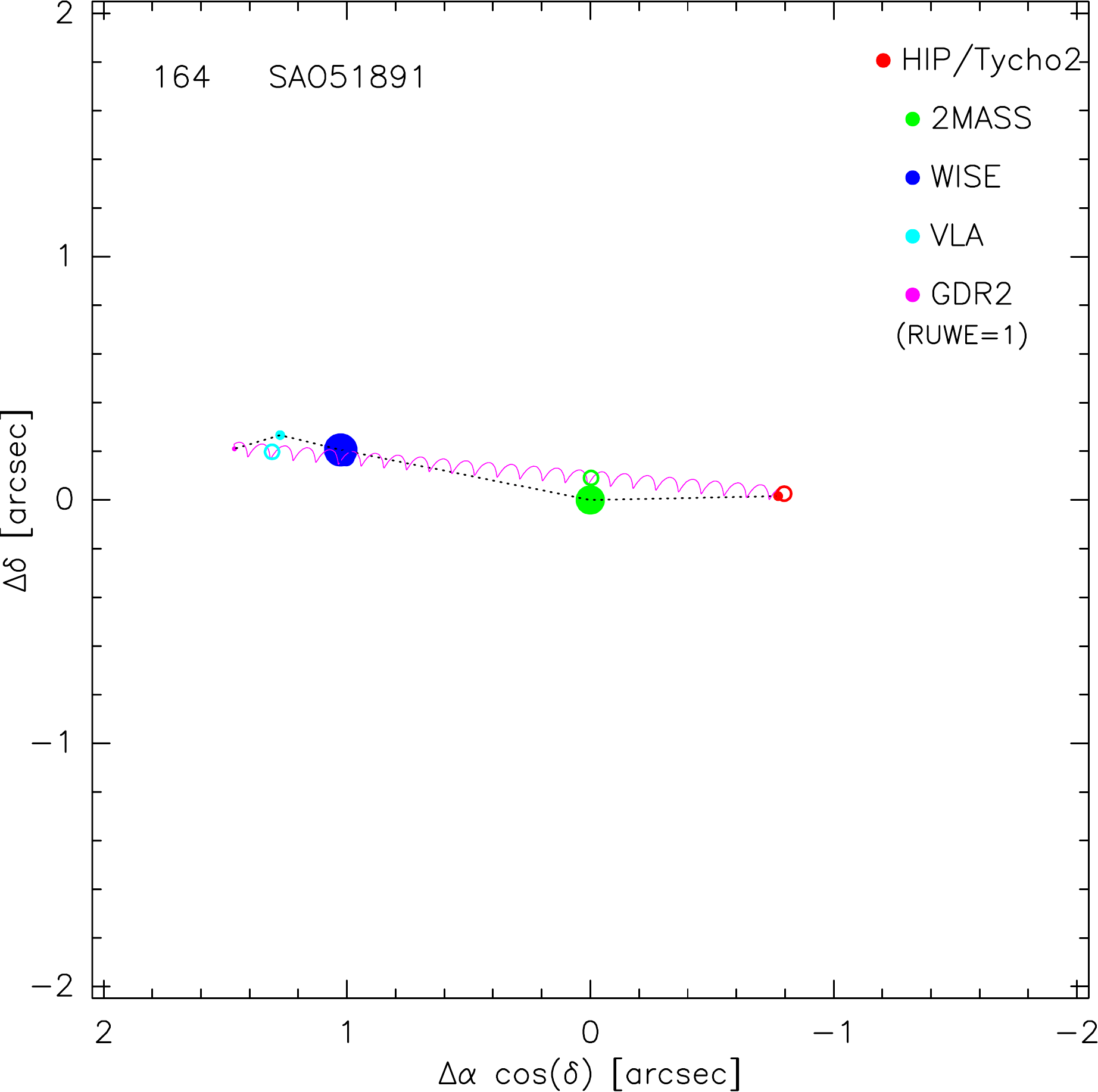}
\figsetgrpnote{Astrometric chart for SAO\,51891 (V383\,Lac).
The number in the upper left corner relates to the target number in Tables\,\ref{tbl-det} and \ref{tbl-sourcelist}. Measured positions (relative to the 2MASS) and 1\,$\sigma$\ uncertainties are drawn as filled ellipses with the colour coding for the respective missions/catalogs indicated in the upper right corner. Straight dotted lines connect the observed positions just to guide the eye. The wiggled line shows the combined proper motion and parallax prediction, starting at the GDR2 2015.5 position and projected backwards in time. Open ellipses show the position and uncertainty prediction from that starting point for the respective observing epochs.}
\figsetgrpend

\figsetgrpstart
\figsetgrpnum{6.30}
\figsetgrptitle{SAO\,108142}
\figsetplot{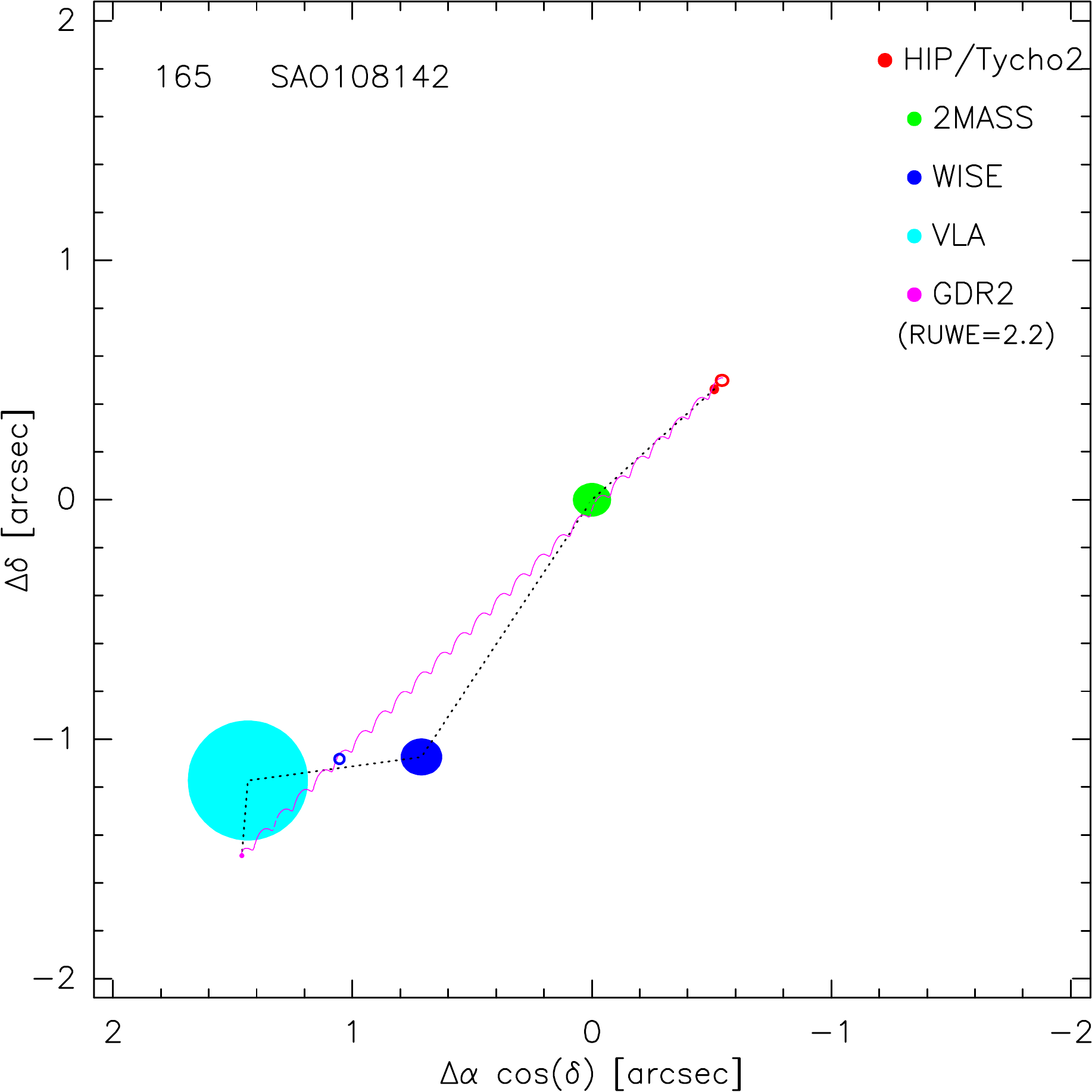}
\figsetgrpnote{Astrometric chart for SAO\,108142 (BD+17\,4799).
The number in the upper left corner relates to the target number in Tables\,\ref{tbl-det} and \ref{tbl-sourcelist}. Measured positions (relative to the 2MASS) and 1\,$\sigma$\ uncertainties are drawn as filled ellipses with the colour coding for the respective missions/catalogs indicated in the upper right corner. Straight dotted lines connect the observed positions just to guide the eye. The wiggled line shows the combined proper motion and parallax prediction, starting at the GDR2 2015.5 position and projected backwards in time. Open ellipses show the position and uncertainty prediction from that starting point for the respective observing epochs.}
\figsetgrpend

\figsetgrpstart
\figsetgrpnum{6.31}
\figsetgrptitle{UCAC4\,832-014013}
\figsetplot{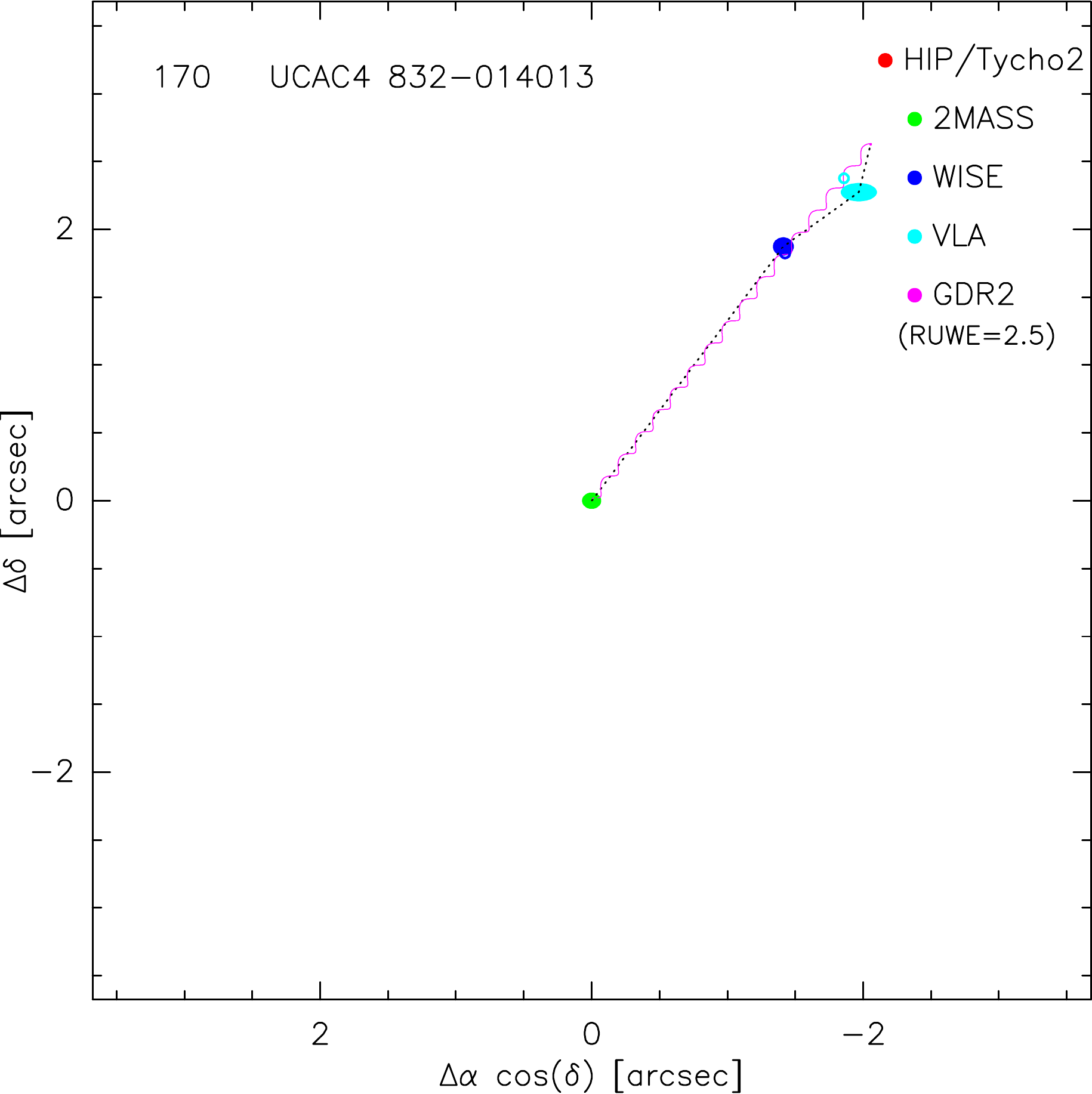}
\figsetgrpnote{Astrometric chart for UCAC4\,832-014013.
The number in the upper left corner relates to the target number in Tables\,\ref{tbl-det} and \ref{tbl-sourcelist}. Measured positions (relative to the 2MASS) and 1\,$\sigma$\ uncertainties are drawn as filled ellipses with the colour coding for the respective missions/catalogs indicated in the upper right corner. Straight dotted lines connect the observed positions just to guide the eye. The wiggled line shows the combined proper motion and parallax prediction, starting at the GDR2 2015.5 position and projected backwards in time. Open ellipses show the position and uncertainty prediction from that starting point for the respective observing epochs.}
\figsetgrpend

\figsetend

\begin{figure}
\plotone{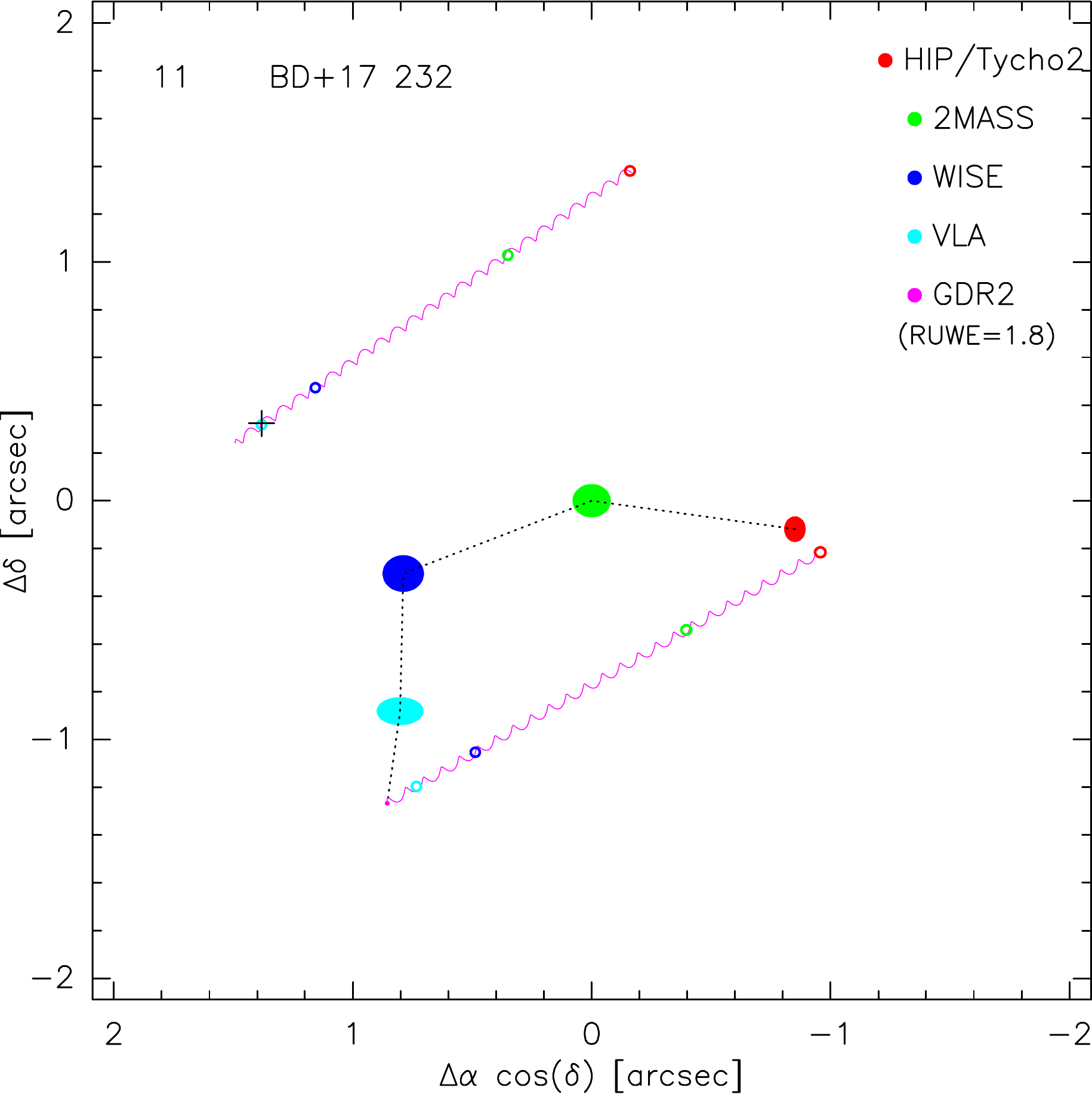}
\caption{Astrometric chart for BD+17\,232.  
The number in the upper left corner relates to the target number in Tables\,\ref{tbl-det} and \ref{tbl-sourcelist}. Measured positions (relative to the 2MASS) and 1\,$\sigma$\ uncertainties are drawn as filled ellipses with the colour coding for the respective missions/catalogs indicated in the upper right corner. Straight dotted lines connect the observed positions just to guide the eye. The lower wiggled line shows the combined proper motion and parallax prediction, starting at the GDR2 2015.5 position and projected backwards in time. Open ellipses show the position and uncertainty prediction from that starting point for the respective observing epochs. The black cross shows the position of the WDS companion BD+17\,232\,B \citep{mason2020} at the epoch of the VLA observations, i.e., relative to the center of the light blue open ellipse marking the predicted position of the primary. The upper wiggled line and open ellipses show the proper motion/parallax and position prediction for the secondary, which was also detected by GDR2. The complete figure set\label{fig:astrom} (31 images) is available in the online journal.}
\end{figure}

To verify the physical association of the 31 VLA detection candidates with the target stars, we evaluate position measurements and proper motion predictions from up to four additional catalogues (besides the VLA measurements), thus covering observing epochs from 1991.25 ({\it Hipparcos}) to 2015.5 (GDR2). In particular, we evaluate the following catalogues, for which we also list the epoch (or epoch range) and the typical (mean) positional uncertainty, $\sigma_{\mathrm pos}$:
\begin{enumerate}
\item {\it Hipparcos}, the New Reduction \citep{leeuwen2010}, Ep\,=\,1991.25 (JD\,2448349.0625), \mbox{$\sigma_{\mathrm pos}\approx 1$\,mas}. For targets with no entry in this catalog, we use the positions, proper motions, and individual epochs from the Tycho2 catalogue  \citep{hog2000} instead.
\item 2MASS \citep{cutri2003}, epochs for individual data sets 1997\,--\,2000 (JD\,2450727\,--\,2451875 for our targets), \mbox{$\sigma_{\mathrm pos}\approx 80$\,mas}.
\item WISE \citep{cutri2012}, default mid epoch of WISE coordinates = 2010.5589 (JD\,2455400.63889), \mbox{$\sigma_{\mathrm pos}\approx 37$\,mas}.
\item VLA observations between Oct. 2013 and Jan. 2014 (see Tab.\,\ref{tbl-obssessions}, JD\,2456595\,--\,2456668), \mbox{$\sigma_{\mathrm pos}\approx$\,80\,mas}.
\item GDR2 \citep{gaia_dr2}, Ep\,=\,2015.5 (JD\,2457206), \mbox{$\sigma_{\mathrm pos}\approx 0.03$\,mas}.
\end{enumerate}

In Sect.\,\ref{sec:dis:bin}, we also evaluate various indicators that could hint at the existence of hitherto overlooked companions.
%
All 31 target stars with VLA detection candidates have 2MASS, WISE, and {\it Gaia} astrometry, although we do not use the GDR2 positions for three of the sources because of the astrometric fit quality issues mentioned above (RUWE$\,>50$: HD\,23524, HD\,284135, and HD\,293857). The first two of these are actually known close visual double stars. Nine stars also have {\it Hipparcos} \citep{leeuwen2010} astrometry and for 21 stars there are at least Tycho2 positions available. Only one star has neither {\it Hipparcos} nor Tycho(2) astrometry (UCAC4\,832-014013). Twenty target stars have known WDS companions, of which 14 are closer than 2.5\arcsec\ to the primary (see Table\,\ref{tbl-det} and discussion in Sect\,\ref{sec:dis:bin}). Five stars have significant ($>3\,\sigma$) proper motion anomalies (PMa) between the long-term HIP\,--\,{\it Gaia} proper motion vector and the GDR2 measurements \citep[PMaG2;][]{kervella2019}. However, four of these five stars have known WDS companions with separations $<$2.5\arcsec\ \citep{mason2020}, and the PMaG2 are consistent with being caused by the presence of these companions (Figs.\,\ref{fig:pma_hd199143} through \ref{fig:pma_hip12635}). The implications of these PMaG2 on the association between VLA and stellar positions are further discussed in Sects.\,\ref{ssec:res:individ} and \ref{sec:dis:bin}.

Figure\,\ref{fig:astrom} and the associated figure set show the astrometric charts for all 31 VLA detection candidates with the measured positions relative to the 2MASS position and the respective 1\,$\sigma$\ uncertainty ellipses. Positional uncertainties of the VLA detections are derived as described in Sect.\,\ref{ssec:res:overview}. In addition to the observed positions, we also show the most recent and accurate estimate of the combined proper and parallactic motion along with the positions and uncertainties predicted by this motion for the epochs of the respective other catalog positions. If a star has a valid GDR2 solution, the motion vector and position predictions are projected backwards in time from the mean GDR2 position. If a star has a {\it Hipparcos}/Tycho2 entry, but no valid GDR2 astrometry, the motion vector and position predictions are projected forward in time starting from the observed {\it Hipparcos}/Tycho2 position. Of prime interest are of course the position predictions for the stars in question at the time of the respective VLA observations, which we compare in the following to the actually observed VLA detections. The position predictions for the other epochs/missions are only used to assess the consistency of the multi-epoch astrometry for the respective star and to identify stars with generally troublesome astrometry, which could indicate, e.g., unresolved multiplicity \citep[see, e.g.,][]{dzib2021,forbrich2021}.


With the now more precise astrometry, the positional discrepancies between the 31 VLA detection candidates and the respective nearest known star are all well below 1\arcsec.
%
We find that the radio sources related to nineteen target stars are securely associated with the target stars. Three targets have somewhat problematic astrometry, but we still consider the VLA detections very likely associated with the respective stars. In the following subsection, we discuss all those stars individually for which the astrometry or the association between VLA source and star had some issues.


\subsection{Notes on individual targets} \label{ssec:res:individ}

For two stars (TYC\,5925-1547-1 and SAO\,150676), which used the same VLA phase calibrator and which have the same large discrepancy of $\approx$0\farcs7 between predicted and observed VLA position, we conclude that the problem is most likely related to an unfortunate combination of calibrator choice and observing schedule. None of the two stars has a known close ($<$\,5\arcsec) companion, but the respective calibrator J\,0539-1550 was on the faint side (0.5\,Jy) and about 5\degr\ away from the two targets. While this would have been acceptable under good conditions, the observations were carried out with the calibrator at an elevation of 21\degr\ and the two targets at 18\degr. This, together with the faintness of the calibrator, may be the reason for the large phase calibration error. 

The third star with large positional discrepancy is SAO\,50350, which is not a known binary \citep{mason2020}, and for which the detections by Tycho2, 2MASS, WISE, and {\it Gaia} all align very well. The VLA detection is 0\farcs72 ($\approx5.3\sigma$) offset from the {\it Gaia}-predicted optical position of the star, which has a well-behaved GDR2 astrometric solution (RUWE $\approx$1.0), and does not show a PMa. In this case we cannot say whether the VLA measurement is just an outlier with some astrometry issues not accounted for, or whether we have actually detected an hitherto unknown companion or an unrelated background source (but see discussion in Sect.\,\ref{ssec:res:overview}).

For two sources, the radio emission is clearly associated with the known secondary components and no emission is detected from the respective primaries. HD\,77407 has a visual secondary at 1\farcs8, which is a M3-6V physical companion to the G0V primary and which was also detected by {\it Gaia} with its own astrometric solution. Our VLA detection is clearly associated with this secondary and no radio emission is detected from the primary. The primary exhibits a large PMa (PMaG2\,=\,24.5\,$\sigma$). Figure \,\ref{fig:pma_hd77407} shows that this visual companion is compatible with having caused this PMa. 
%
The other star, HD\,199143, has an about 3\,mag (V) fainter visual secondary located at 1\farcs1 north-west of the primary, which is not detected by {\it Gaia}. Our VLA detection is clearly associated with this visual secondary and no radio emission was detected from the primary. The primary also exhibits a significant (24\,$\sigma$) PMaG2 \citep{kervella2019}. Figure\,\ref{fig:pma_hd199143} shows that this visual companion is compatible with having caused this PMa. 

For two sources with known visual companions at 1\farcs7 and 2\farcs1, the VLA detection is located in-between the two binary components and the data do not allow us to decide from which component the radio emission arises or whether possibly both stars contribute.  
TYC\,3301-2585-1 has an approximately equally bright (at {\it V}-band) companion at an angular separation of 2\farcs1 (**ES\,1217\,AB). While both components are detected separately by GDR2 (with 'good' astrometric solutions), only the primary has a Tycho2 entry, and both 2MASS and WISE did obviously not resolve the source and list a position in-between the two components. The VLA detection is also located in-between the two components, albeit closer to the secondary. It is thus possible that the radio emission arises from both components, but is not resolved by the $>$2\arcsec\ VLA beam.
BD+17\,232 has an approximately equally bright ({\it V}-band) companion at $\sim$1\farcs7 projected separation (WDS\,J01377+1836\,B). The primary is well-detected by both Tycho2 and GDR2 (see Fig.\,\ref{fig:astrom}), although with moderate excess noise/RUWE for the GDR2 solution, such that the predicted position at the time of the VLA observations may have a larger (systematic) uncertainty than formally adopted. 2MASS and WISE did obviously not resolve the source and list a position in-between the two components. The VLA position is also located in-between the two stars, albeit only 2.8\,$\sigma$\ away from the primary. It is this thus likely that the radio emission arises from the primary only.

For five sources with known visual companions at 0\farcs1\,--\,0\farcs4, i.e., significantly smaller than the synthesized $>$2\arcsec\ VLA beam, the data also do not allow us to decide from which component the radio emission arises or whether possibly both stars contribute. Statistically, one would expect that the lower-mass, i.e., cooler component, which is in most cases the secondary, is the more likely radio emitter (assuming coevality, see Sect.\,\ref{sec:dis:spt} and Fig.\,\ref{fig:detratios}).
HD\,23524 has a secondary component (WDS) of approximately equal ({\it V}) brightness at 0\farcs3 south of the primary. Our astrometry indicates that the detected radio emission is centered on the secondary component, although the primary is also covered by the synthesized beam. However, GDR2 indicates a large value of RUWE (9.7) for the primary and does not list solutions for parallax and proper motion. {\it Hipparcos} resolved the system as a binary, but the proper motion solution might still be affected by the binarity. 

HD\,284135 also has an approximately equally bright ({\it V}-band) companion at $\sim$0\farcs4 (WDS\,J04057+2248\,B). The primary does not have a valid {\it Hipparcos} solution, but is detected by both Tycho2 and GDR2, albeit in both cases with flags that indicate bad quality solutions, such that the astrometry may be affected by the not separately detected secondary and therefore unreliable. The VLA detection is $\approx$2.5\,$\sigma$ from the Tycho2-predicted position of the primary along the proper motion vector (but further away from the secondary). 
SAO\,135659 has a secondary component (WDS) of approximately equal ({\it V}) brightness at 0\farcs1 south-east of the primary. The VLA detection is centered in-between the two stars. 

GJ\,2079 has a secondary component of approximately equal ({\it V}-band) brightness at a projected separation of $\leq$0\farcs1 ($\approx$2.3\,au). \citet{bowler2015} estimate an orbital period of 4.8\,yrs and \citet{kammerer2019} report the latest relative projected separation from mid 2016 as $\rho\approx77$\,mas at P.A.$\approx$339\degr, but a good orbit solution is not yet known. The VLA detection is within 1\,$\sigma$ of both components, and the positional uncertainty is comparable to the separation of the two components. The primary also exhibits a large PMaG2 of $\approx 67\sigma$ and the curvature of its motion can be clearly seen in the astrometric chart (Fig.\,\ref{fig:astrom}). Figure\,\ref{fig:pma_gj2079} shows that the observed PMa is compatible with this visual companion. 

HD\,135363 has a visual secondary at 0\farcs36 south-east, which is a M2/4V physical companion to the G5V primary and which was not independently detected by {\it Gaia}. The primary also has a significant (9\,$\sigma$) PMaG2, which could be  consistent with the visual secondary (Fig.\,\ref{fig:pma_hd135363}). Although our VLA detection is centered slightly closer to the primary, the M-type secondary is the more likely radio emitter.

One more target star with 6\,GHz detection, HIP\,12635 (HD\,16760\,B), exhibits a 4\,$\sigma$\ PMaG2 \citep[][Fig.\,\ref{fig:pma_hip12635}]{kervella2019}, but has no known close companion. The WDS primary, HD\,16760, at a projected separation of 14\farcs3, which has itself a close ($\approx$0\farcs27) companion \citep{evans2012} and exhibits a 5.6\,$\sigma$ PMa, is actually not physically associated with HIP\,12635 since it has a significantly different parallax with good-quality solution (Tab.\,\ref{tbl-sourcelist}). No radio emission was detected from HD\,16760. Hence, the PMa of HIP\,12635 could hint at a hitherto undisclosed close companion. Our VLA detection is only 0\farcs27 (1.1\,$\sigma$) offset from the {\it Gaia}-predicted position of HIP\,12635, such that we consider this a secure detection.

In summary, we have thus detected radio emission from 22 target stars or their respective primaries. For seven close binary systems, the data do not allow us to decide from which of the two components the detected radio emission arises, and for two targets, the radio emission is clearly associated with the secondary components only.


\subsection{Statistical assessment of positional mismatches} \label{ssec:res:stat}

\begin{figure}[ht!]
\epsscale{.95}
\plotone{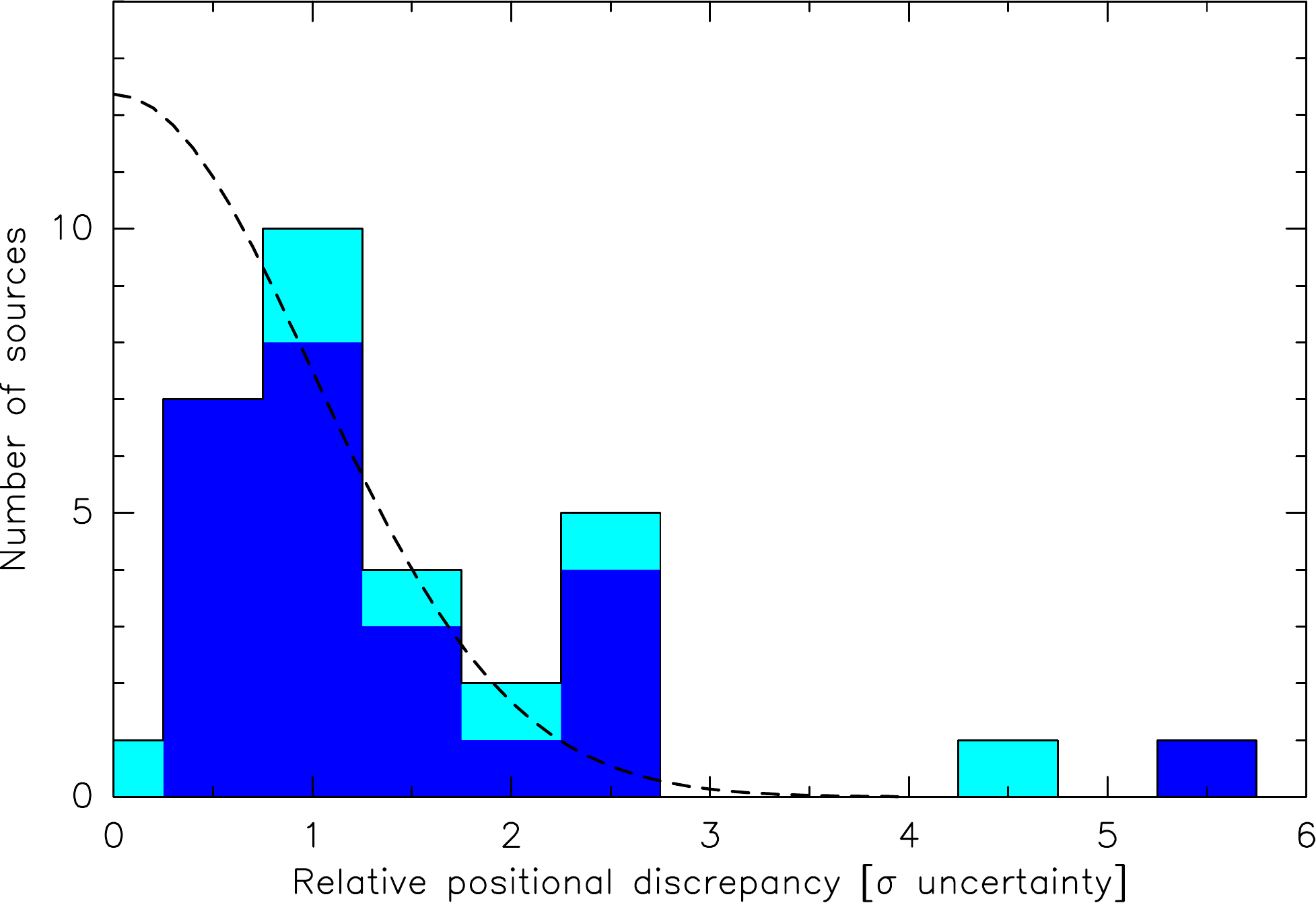}
\caption{\label{fig:sepdist3} Number of target stars as function of angular separation between predicted position and the nearest VLA source in units of the combined 1\,$\sigma$\ uncertainties of the two positions. The dark blue histogram accounts for all targets in which the VLA source could be associated with one star (primary or secondary in binaries). 
The light blue histogram on top indicates those sources in which two binary components are likely to contribute to the radio emission. The dashed curve shows the normal distribution for all 31 detections, i.e., the expected shape of the histogram for purely random and uncorrelated astrometric measurement errors.
}
\end{figure}

As discussed above and shown in Fig.\,\ref{fig:sepdist1}, the association of radio emission with stars emitting mostly at optical and NIR wavelengths is by far not straight-forward. Even with such detailed astrometric investigations, there are remaining positional discrepancies between the radio position and the predicted optical position of the star in question, which may not only result from measurement uncertainties (both accounted and unaccounted for), but could also indicate that the radio emission does not arise from the star itself, but e.g. from a yet unknown companion. To investigate this, we show in Fig.\,\ref{fig:sepdist3} the distribution of the relative positional discrepancies (angular separation) between the predicted and the observed VLA positions, and compare it to the normal distribution for all detections which we would expect if the positional discrepancies would be related entirely to purely random and uncorrelated measurement errors. We find that the distribution is roughly consistent with the expected normal distribution for random errors, but that there is a clear lack of exact matches (discrepancy $<$1\,$\sigma$) and a slight excess of at relative discrepancies between 2 and 3\,$\sigma$. The figure also shows that this mismatch is unlikely to be caused entirely by the contribution of radio emission from the known, but unresolved secondary components. We discuss the possible reasons in Sect.\,\ref{sec:dis:bin}.


\subsection{Radio flux variability} \label{ssec:res:var}


As described in Sect.\,\ref{sec:obs}, all but three fields were observed in two epochs separated by 7 to 39 days. Of the 31 stars from which radio emission was detected, 22 sources were detected in two epochs, eight sources were detected in only one epoch, and one detected object was observed only once. 
Flux ratios range from 1.05 up to about 11, plus one clear outlier with the strongest peak flux (12.8\,mJy) and a flux ratio of 51 (GJ\,2079). 
It is interesting to note that GJ\,2079 is the closest known visual binary in our sample (projected separation $\leq100$\,mas or 2.3\,au) and the emission is most likely arising from both equally bright (in $V$-band) components. GJ\,2079 has the second-lowest X-ray luminosity in our radio-bright sample ($9\times10^{-6}$\,L$_{\odot}$, see Fig.\,\ref{fig:rx-ir}), but is otherwise not significantly distinguished in any property from the other radio-detected young stars. This star thus resembles the group of 13 extremely radio-variable young stars in the Orion nebula cluster identified by \citet{forbrich2017}. However, since we only have 2 epochs of radio observations separated by 39\,d, we have no further information on the time scale of this extreme variability. 
Lower limit flux ratios for stars detected in only one epoch range from 2 to 12, with a median of 3.1. Ignoring the one outlier, the median flux ratio between the two epochs is $\approx2.6$. There is no significant correlation between the variability amplitude and time lag between the two epochs, nor with stellar age or spectral type.


\section{Discussion} \label{sec:dis}


\subsection{Correlation between radio emission and spectral type} \label{sec:dis:spt}
%
\begin{figure}[htb!]
\epsscale{.95}
\plotone{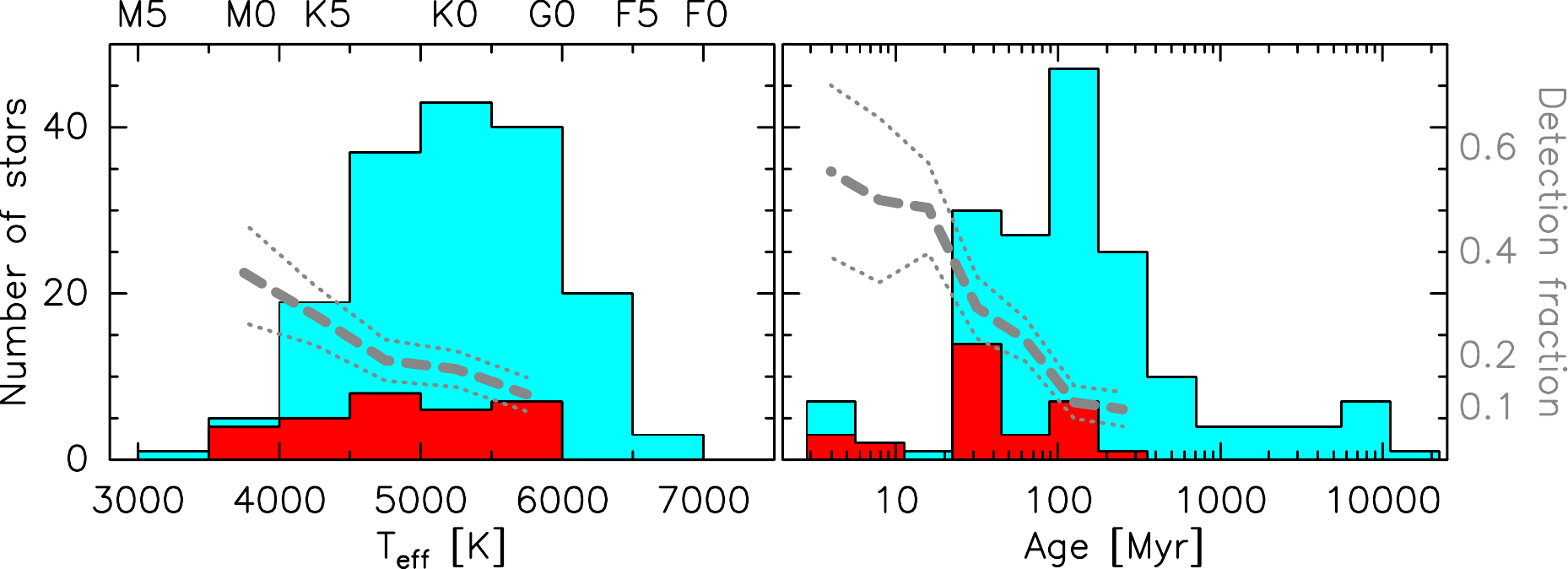}
\caption{\label{fig:detratios} 
Distribution of stellar effective temperatures (corresponding spectral types for main-sequence stars indicated on top), and ages for the observed (light blue histograms) and detected (at 6\,GHz) stars (red histograms, cf. Fig.\,\ref{fig:sourcedist}). Shown in grey are the detection fractions (thick dashed lines; smoothed over 3 bins) with 1\,$\sigma$\ confidence intervals (dotted lines). The scale is indicated on the right side.
}
\end{figure}

The left panel of Fig.\,\ref{fig:detratios} shows the distribution of effective temperatures and spectral types of the stars with detected 6\,GHz emission, along with that of all 170 observed stars. The latest and earliest spectral types in the observed sample are M2 ($T_{\rm eff}\approx3500$\,K) and F4 ($T_{\rm eff}\approx6700$\,K (with one exception of spectral type B9.5, see Tab.\,\ref{tbl-sourcelist}).
The latest and earliest spectral types with detected radio emission are M0 ($T_{\rm eff}\approx3800$\,K) and F8 ($T_{\rm eff}\approx5920$\,K). 
Figure\,\ref{fig:detratios} also shows that the detection fraction (smoothed over three bins) decreases from 36\,$\pm$\,10\%\footnote{The 1\,$\sigma$\ confidence intervals are derived from the standard deviation of the respective binomial distributions.} at 3800\,K to 12.6\,$\pm$\,3.3\% at 5900\,K, i.e., by a factor of three. No star with $T_{\rm eff}>6000$\,K was detected, although 22 such stars were observed.
On the other hand, we find no significant correlation between $T_{\rm eff}$\ (or spT) and the strength of the 6\,GHz emission if it is detected.


\subsection{Correlation between radio emission and age} \label{sec:dis:age}
The right panel of Fig.\,\ref{fig:detratios} shows the distribution of ages of the stars with detected 6\,GHz emission, along with that of all 170 observed stars. The youngest and oldest stars in our sample have ages of $\lesssim$5\,Myr and 12\,Gyr, respectively. The youngest and oldest stars with detected radio emission have ages of $\lesssim$5\,Myr and 200\,Myr, respectively, i.e., gyro-synchrotron radio emission occurs at all ages between $\lesssim$5 and 200\,Myr\footnote{Note that no valid age estimate could be obtained for UCAC4\,832-014013.}. 
Figure\,\ref{fig:detratios} also shows that the detection fraction (smoothed over three bins) decreases from 56\,$\pm$\,20\% at $\leq$10\,Myr to 10\,$\pm$\,3\% at 100-200\,Myr, i.e., by a factor of nearly six. No star older than 200\,Myr was detected, although 45 such stars were observed. This suggests a significant decline of coronal gyro-synchrotron emission with age from a few Myr to about 200\,Myr and a complete termination afterwards. 
A general decline of stellar magnetic activity and high-energy output due to the rotational spin-down with age has been observed and modeled for the Sun and other stars by many authors \citep[e.g.,][]{bouvier2014,guedel2004,guedel2020}. Our survey results confirm, for the first time, that strong (observable) coronal gyro-synchrotron emission indeed declines on the same timescale, or even a bit faster, as the pre-main-sequence spin-down of the stellar rotation period. 
On the other hand, we find no significant correlation between age and the strength of the 6\,GHz emission (if it is detected) within the age range 1-200\,Myr.


\subsection{Correlation between radio emission and infrared excess and X-ray activity} \label{sec:dis:irxray}
%
\begin{figure}[htb!]
\epsscale{.95}
\plotone{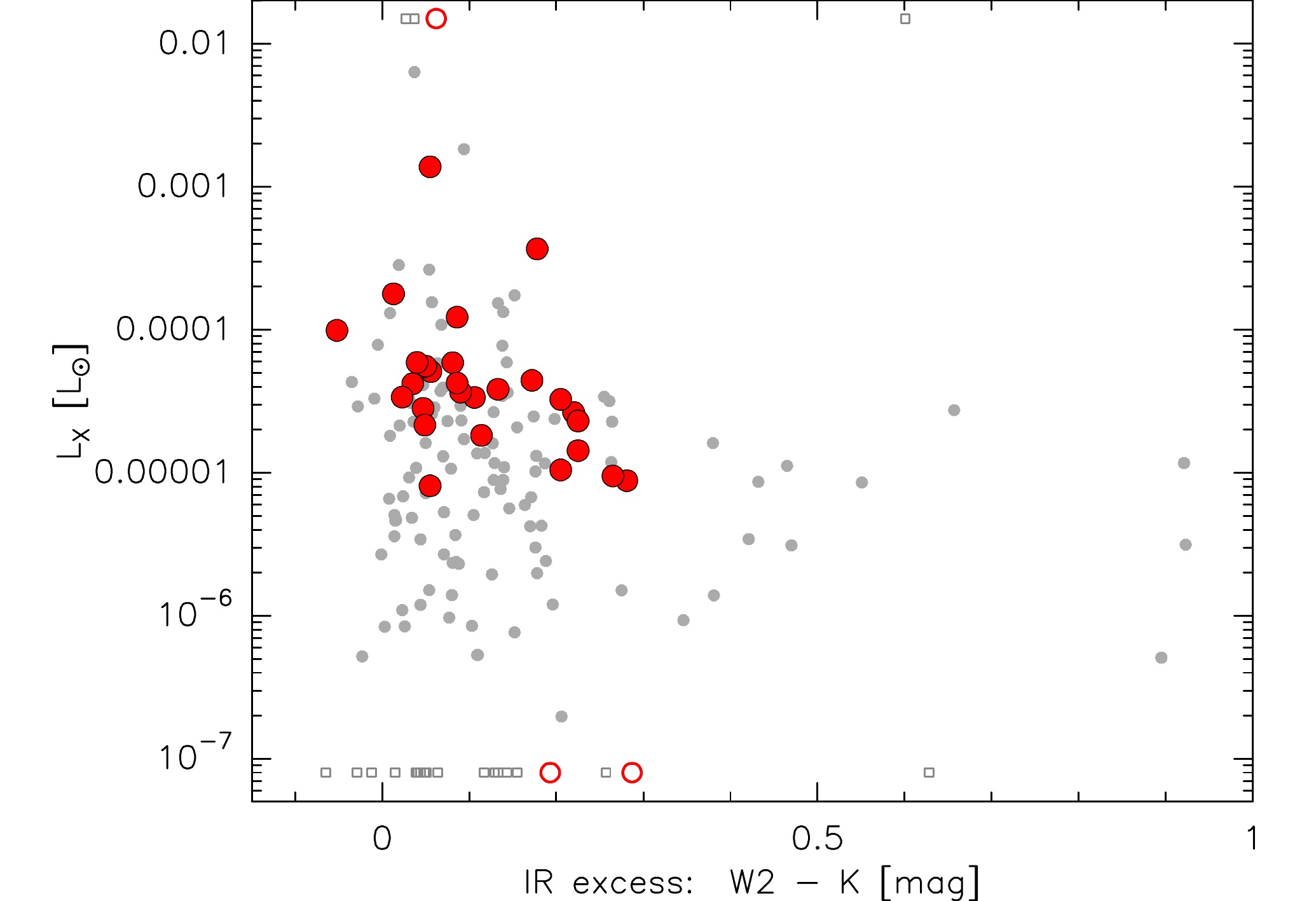}
\caption{\label{fig:rx-ir} X-ray luminosity \citep[2RXS absorption-corrected flux from a power-law fit;][]{boller2016} vs. IR excess \citep[WISE W2(4.618\,$\mu$m) minus K(2.2\,$\mu$m);][]{cutri2012,cutri2003} of observed non-detected (grey dots) and radio-detected (filled red circles) stars. Targets with X-ray non-detections or X-ray luminosities above the plot limit are shown as open squares and circles lined up at the lower and upper boundaries of the plot, respectively. 
}
\end{figure}

All but 21 out of the 170 VLA-observed stars exhibit ROSAT-detected X-ray emission. All 21 X-ray non-detected stars are older than 20\,Myr.
Figure\,\ref{fig:rx-ir} shows the absorption-corrected X-ray luminosity \citep[derived from the 2RXS flux obtained from a power-law fit;][]{boller2016} vs. IR excess \citep[WISE W2/4.618\,$\mu$m minus K/2.2\,$\mu$m;][]{cutri2012,cutri2003} of the stars with detected 6\,GHz emission, along with that of all 170 observed stars. All but three of the radio-detected stars have X-ray luminosities between $8\times10^{-6}$\ and $2\times10^{-3}$\,L$_{\odot}$. 
Only one star (HD\,31281, $<5$\,Myr) has an X-ray luminosity which is 5 orders of magnitude higher than that of the other radio-bright stars, i.e. ROSAT probably caught this star (and three other radio-non-detected stars) during an X-ray super-flare \citep{getman2021}.
Apart from two X-ray non-detections, no star with $L_{\rm X}<8\times10^{-6}$\,L$_{\odot}$\ was detected at 6\,GHz, although more than 70 such stars were observed.
The only two radio-bright stars without an X-ray counterpart are HD\,217014 and HD\,358623. 

Figure\,\ref{fig:rx-ir} also shows that, owing to their youth and the likely existence of debris disks, nearly all of our target stars have $4.6-2.2\,\mu$m excesses of $>$0\,mag, but $<$1\,mag. Yet, only two out of the 31 radio-bright stars have significant debris disk excess at longer wavelengths, albeit with fractional disk luminosities $\le10^{-4}$ \citep[see Fig.\,6 in][]{launhardt2020}. Nevertheless, the corresponding color excess of main-sequence stars without debris disks would be very close to zero.
It is also evident that all radio-bright sources have relatively small $4.6-2.2\,\mu$m excesses of $<0\ldots 0.4$\,mag, and no star with IR excess $>0.4$\,mag was detected at 6\,GHz, although the number of such stars is small (13). The stars with larger IR excess have ages of 8\,--170\,Myr, i.e., they are all in the young age range where the radio detection rate is relatively high as compared to older stars (Sect.\,\ref{sec:dis:age} and Fig.\,\ref{fig:detratios}). Since with a mean detection rate of $\sim30$\% in this younger age range, we should have detected about 3-4 out of the 13 stars observed, it is likely that the non-detection of stars with large IR excess is primarily related to the mass of the disks and not to the age.
We thus conclude that future searches for stars with strong gyro-synchrotron radio emission, e.g., for identifying suitable targets for VLBA astrometry, could boost their efficiency by pre-selecting stars with 
\mbox{$L_{\rm X}>5\times10^{-6}\,{\rm L}_{\odot}$} and IR excess \mbox{W2--K\,$<0.5$\,mag}.

\begin{figure}[htb!]
\epsscale{.95}
\plotone{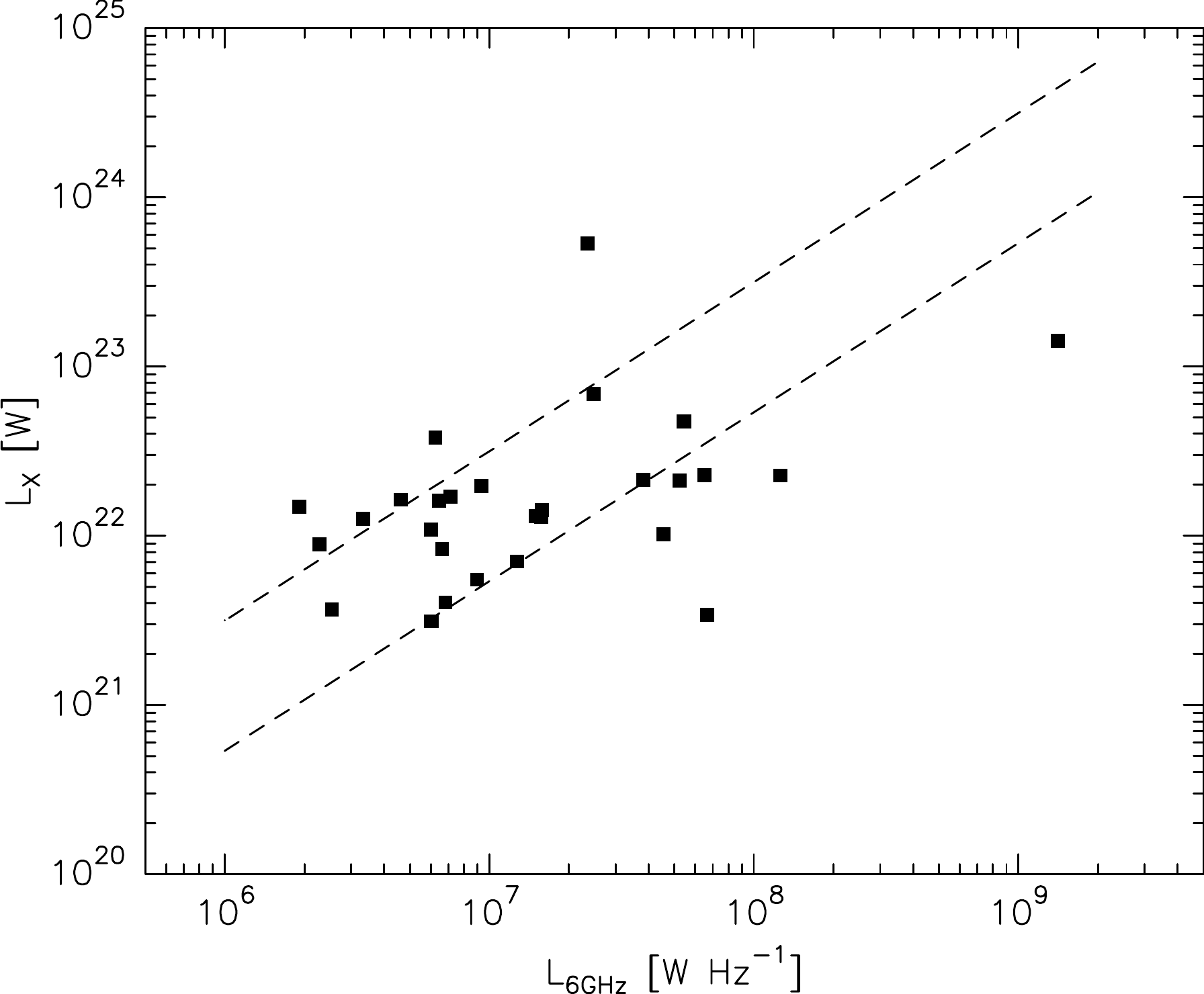}
\caption{\label{fig:rx-6cm} X-ray luminosity vs. 6\,GHz radio luminosity (max of the two epochs) of observed and radio-detected stars. Dashed lines show the empirical \citet{gb1993} relation between 'quiescent' (i.e., not accounting for flares) X-ray and radio luminosity for different types of magnetically active stars \citep[see][]{bg1994}. 
}
\end{figure}

Figure\,\ref{fig:rx-6cm} shows the relation between the 6\,GHz radio luminosity (maximum of the two epochs) of observed and radio-detected stars and their respective X-ray luminosity. Most of the radio-detected stars have X-ray luminosities between $3e21$\ and $2e23$\,W (one outlier with $5.5e23$\,W). While the lower boundary might be related to ROSAT's detection limit, the upper boundary might not be significant due to low-number statistics. Most of our radio-detected stars thus roughly follow the 'G{\"u}del-Benz' relation \citep{gb1993} between 'quiescent' (i.e., not accounting for flares) X-ray and radio luminosity for magnetically active stars. The few sources to the lower right of the relation might have being caught during a radio flare. 


\subsection{Correlation between radio emission and binarity} \label{sec:dis:bin}
To verify if and to what degree binarity is related to the occurrence of gyro-synchrotron emission, we evaluate for all detection candidates the ninth catalogue of spectroscopic binary orbits \citep[SB9;][]{pourbaix2004} and the Washington Visual Double Star Catalog \citep[WDS;][]{mason2020} and obtained the individual measurements were necessary to determine the positional offset of the components at the time of the VLA observations. Although these two catalogs are by far not complete, they provide the most complete binarity data base we have available and the only useful means of comparison with binarity fractions of other surveys. In addition, we check the NASA Exoplanet Archive for known planetary companions to our radio-bright target stars, but we find none. 
We also look for PMaG2 \citep[][]{kervella2019}, which could be indicative of the presence of a perturbing secondary object, but would also lead to a wrong prediction of the star position at the time of the VLA observations. In addition, we search the literature for reported companion detections.
For the GDR2 astrometry, we also evaluate the re-normalized "unit weight error" \citep[RUWE;][]{lindegren2018}, which should be close to 1.0 for well-behaved solutions of single stars. Excessively large values may indicate that not every {\it Gaia} observation was consistent with the single-star model. 
 
We find that twenty out of the 31 radio-bright stars identified in our sample (65\%) have known companions of some type, of which 14 (45\%$\pm$10\%) have at least one component located within 2\farcs5 of the primary. 
Two more stars (V875\,Per and AI\,Lep) are supposed to have close (interacting) companions revealed by their photometric variability (Sect.\,\ref{sec:dis:act}), which are not directly detected. One more star without known close companion exhibits a significant PMa (HIP\,12635), which could hint at the presence of a yet unknown close companion. One other star (HD\,293857) exhibits a large GDR2 astrometric excess noise (7.6\,mas) and $RUWE\approx59$, which could also hint at yet unknown companion. These indirect detections would increase the close companion fraction to 58\%$\pm$10\% (18 stars).
The Gould's Belt Distances Survey  \citep[GOBELINS;][]{ortiz-leon2017a,ortiz-leon2017b,kounkel2017} reported 59 out of 156 radio-bright stars to have WDS companions at separations $\le2.5$\arcsec, which corresponds to a close binary fraction of 38\% and is consistent with the close binary fraction based on WDS companions in our sample. 

The statistical lack of exact astrometric matches between VLA detections and star positions and the excess of \mbox{2-3\,$\sigma$} positional discrepancies revealed in Sect.\,\ref{ssec:res:astrom} (Fig.\,\ref{fig:sepdist3}) is unlikely to be caused by companion-related proper motion uncertainties since the GDR2 positions are derived within only about 1.5\,yrs from the VLA observations and the mean proper motion of our targets is only about 100\,mas/yr. This statistical anomaly could therefore hint at the presence of additional still unknown close companions and thus an even higher close companion fraction.
%
This suspicion is supported by the findings of \citet{forbrich2021} and \citet{dzib2021}, who carried out a VLBA survey for nonthermal emission toward 556 compact radio sources previously identified in a deep VLA survey of the Orion Nebula Cluster \citep[ONC;][]{forbrich2016,forbrich2017}, of which they detected 123 sources with the VLBA. Of the 34 VLBA radio sources that are associated with GDR2-listed stars within $0\farcs2$, 23 sources (68\%) are well separated ($>$4\,mas) from the associated GDR2 position and are likely close companions of the Gaia-detected optical stars.


The WDS was also probed for the 139 radio-quiet stars in our sample, for which we find 34 stars (24\%$\pm$4\%) to have known companions within $2\farcs5$\ listed, i.e., the close binary fraction of radio-bright stars is at least twice as high as that of radio-quiet stars. 
This significant difference in the close binary fraction between radio-loud and radio-quiet stars further suggests that binarity and gyro-synchrotron emission are closely related. 

We can now use these close binary fractions to correct our star system radio detection rate of $(18\pm3)$\% (Sect.\,\ref{ssec:res:overview}) and derive a star detection rate. Assuming for simplicity close binary fractions of $(25\pm5)$\% for the 139 radio-quiet systems in our sample, and $(50\pm10)$\% for the 31 radio-detected systems, and assuming further that we have no triple systems and that only one star per system is emitting at radio wavelengths, we derive a star detection rate of 31 out of 221 stars, i.e., $(14\pm2)$\%.



\subsection{Correlation between radio emission and other activity} \label{sec:dis:act}

19 out of 31 stars (61\%) detected at 6\,GHz are listed as photometrically variable in {\it Simbad} and/or the {\it Hipparcos} and {\it Tycho} catalogues \citep{esa1997}. Ten of these are classified as rotationally variable due to star spots (mostly BY\,Dra-type variability). 
Two stars are RS-Canum-Venaticorum variables (V875\,Per and SAO\,150676), which indicates the presence of a very close (interacting) companion in addition to the wide visual companions listed in the WDS \citep{mason2020}.
Two stars are flagged by {\it Hipparcos} to show duplicity-induced variability (HIP\,12635 and HD\,82159). Both stars have reported close companions (see Tab.\,\ref{tbl-det}).
Four more stars are listed as irregular or not further classified variables. 
One star (HD\,286264) is listed as a classical cepheid, i.e., with variability caused by radial pulsations.
Without having done this same assessment for the 139 radio-non-detected stars for comparison, this large fraction of documented photometric variability caused by both magnetic activity as well as close companions is consistent with the notion that non-thermal radio emission scales with various other indicators of variability.


\section{Summary and Conclusions} \label{sec:sum}

We have carried out a VLA 6\,GHz continuum survey for coronal radio emission from young low-mass stars in the immediate solar neighbourhood at distances between 8 and 130\,pc, i.e., inside the Local Bubble and Gould's Belt. In total, we have observed fields with FoV 7\farcm5 around 170 stars with spectral types ranging from F4 to M2 and with ages ranging from $\lesssim$5\,Myr to 12\,Gyr (median 149\,Myr). To ensure that we also obtain a measure of the long-term flux variability, each field was observed twice, with the two observations separated by 7 to 39 days. One short extra session with three stars could be scheduled only once. Thus, we observed 167 stars twice, and three more stars could be observed only once. 
The mean synthesized FWHM beam width (angular resolution) of the observations was $\approx$2\arcsec\ (major axis). We achieve a mean 3\,$\sigma$\ point source detection limit of $\approx$45\,$\mu$Jy. Typical (median) positional uncertainties of the centroids of detected radio sources are 0\farcs14 in right ascension and 0\farcs19 in declination, derived from image plane fitting. The typical uncertainty of the estimated radio flux densities is 12\%

Our survey is thus complementary to the Gould's Belt VLA survey \citep{loinard2011,dzib2013,dzib2015,kounkel2014,ortiz-leon2015,pech2016}, the hitherto largest survey for radio continuum emission from young stars. Our targets are all located inside the Local Bubble and Gould's Belt and are thus significantly closer to the Sun. Second, our survey covers stars with ages ranging from $\lesssim$5\,Myr to 12\,Gyr, while the oldest stars associated with the Gould's Belt are only about 30\,Myr old \citep[e.g.,][]{SF1974}. 

In total, we have identified $\ge$626 radio sources with fluxes above the 3\,$\sigma$\ threshold and with peak fluxes between 50\,$\mu$Jy and 900\,mJy (median 500\,$\mu$Jy). Of these, 31 radio sources with fluxes between 68\,$\mu$Jy and 13\,mJy (median 266\,mJy) are clearly associated with 31 out of our 170 target stars. The positional uncertainties of these radio sources are smaller than those of the entire sample (because they are brighter than the majority of the background sources) and amount to 70-80\,mas (median). None of the remaining 139 target stars was found to exhibit detectable radio emission at the time of the observations. We also find that, apart from UCAC4\,832-014013, none of the remaining $\ge$595 VLA radio sources is associated with another known star and conclude these radio sources are likely related to background sources (radio galaxies, quasars).

With 31 out of 170 surveyed stars (not counting companions) exhibiting 6\,GHz radio emission above the mean 3\,$\sigma$\ detection limit of $\approx$45\,$\mu$Jy, the overall system detection rate amounts to $(18\pm3)$\%. Of these, 22 sources (71\%) were detected in both epochs, eight sources were detected in only one out of two epochs, and one detected object was observed only once. Fluxes are varying for nearly all sources between the two epochs, with flux ratios ranging from 1.05 up to $>$12.5, a median flux ratio of $\approx$2.6, and one outlier with a flux ratio of 50. There is no significant correlation between the variability amplitude and the time lag between the two epochs, nor with stellar age or spectral type. Our system detection rate thus compares well with the system detection rate in the Ophiuchus complex \citep[$\approx$16\%;][]{dzib2013}, but is significantly lower than in Taurus \citep[$\approx$35\%;][]{dzib2015}. Our binarity-corrected star detection rate is slightly lower and amounts to $(14\pm2)$\%, assuming that the radio emission in binary systems arises from only one of the two stars.

We find a significant decline of the detection rate with age by a factor of five to six from 56$\pm$20\% for stars with ages $\le10$\,Myr to 10$\pm$3\% for stars with ages 100\,--\,200\,Myr. No star older than 200\,Myr was detected, although 45 such stars were observed.
The latest and earliest spectral types with detected radio emission in our sample are M0 and F8. 
We also find that the radio detection rate significantly declines with $T_{\rm eff}$\ by a factor of 2.5$\pm$1.4 from 36$\pm$10\% for stars with $T_{\rm eff}<4000$\,K (spT later than K8) to 12.6$\pm$3\% for stars with $T_{\rm eff}>5000$\,K (spT earlier than K2). No star with $T_{\rm eff}>6000$\,K was detected, although 22 such stars were observed. 

We find that the fraction of known visual binarity among the radio-bright stars is at least twice as high as that of radio-quiet stars (50$\pm$10\% vs. 24$\pm$4\%). Both some indirect binarity indicators like certain types of variability or PMa, as well as a statistical lack of exact positional matches between radio sources and star positions together with an excess of 2-3\,$\sigma$ positional discrepancies (corresponding to projected separations of 10-20\,au) suggest that the actual binary fraction among radio-bright stars could be significantly higher ($>$60\%). We may thus have detected radio emission from still unknown companions in at least a few cases, although none of our radio-bright stars has a known planetary companion.
The significant difference in the close binary fraction between radio-loud and radio-quiet stars confirms that binarity and gyro-synchrotron emission are closely related, i.e., that gyro-synchrotron emission is triggered by the presence of a close companion.

All but three of the radio-detected stars have X-ray luminosities between $8\times10^{-6}$\ and $2\times10^{-3}$\,L$_{\odot}$. 
No star with $L_{\rm X}<8\times10^{-6}$\,L$_{\odot}$\ was detected at 6\,GHz, although more than 70 such stars were observed.
Owing to their youth and the likely existence of debris disks, our target stars have $4.6-2.2\,\mu$m IR excesses of 0\,--\,1\,mag. However, the radio-bright stars have relatively small $4.6-2.2\,\mu$m excesses of 0\,--\,0.4\,mag, and no star with IR excess $>0.4$\,mag are detected at 6\,GHz. We conclude that the non-detection of stars with large IR excess is primarily related to the mass of the disks, which could in turn be related to the larger fraction of close binarity as compared to radio-quiet stars.
Our findings suggest that the efficiency of future searches for stars with detectable gyro-synchrotron emission can be greatly enhanced when the target list is restricted to stars with spectral types later than G0 ($T_{\rm eff}<6000$\,K), \mbox{ages $<$\,200-300\,Myr}, X-ray luminosity \mbox{$L_{\rm X}>5\times10^{-6}\,{\rm L}_{\odot}$}, and only moderate $4.6-2.2\,\mu$m IR excesses of $\lesssim0.5$\,mag.

The radio-bright nearby young stars identified here provide an interesting sample for future astrometric studies using VLBI arrays aimed at searching for hitherto unknown tight binary components or even exoplanets. The promising potential of such an approach, for both identifying hitherto unknown companions and obtaining high-precision astrometry, has been demonstrated recently by \citet{forbrich2021} and \citet{dzib2021}. But, these studies also showed that, owing to the strongly variable nature of the coronal gyro-synchrotron emission, not all VLA-detected sources might be detectable later with the VLBA.



\acknowledgments

We thank Neil Zimmerman for helping with the initial target list compilation, Ingo Stilz for setting up and maintaining our target star data base, and Grant Kennedy for providing his stellar atmosphere and black-body fitting toolbox.
The National Radio Astronomy Observatory is a facility of the National Science Foundation operated under cooperative agreement by Associated Universities, Inc.
This research has made use of the SIMBAD database and the VizieR catalogue access tool, both operated at CDS, Strasbourg, France. 
The original description of the VizieR service was published in A\&AS 143, 23.
This publication makes use of data products from the Two Micron All Sky Survey, which is a joint project of the University of Massachusetts and the Infrared Processing and Analysis Center/California Institute of Technology, funded by the National Aeronautics and Space Administration and the National Science Foundation.
This work has made use of data from the European Space Agency (ESA) mission
{\it Gaia} (\url{https://www.cosmos.esa.int/gaia}), processed by the {\it Gaia}
Data Processing and Analysis Consortium (DPAC,
\url{https://www.cosmos.esa.int/web/gaia/dpac/consortium}). Funding for the DPAC
has been provided by national institutions, in particular the institutions
participating in the {\it Gaia} Multilateral Agreement.
This research has made use of the Washington Double Star Catalog maintained at the U.S. Naval Observatory. 
This research has made use of the NASA Exoplanet Archive, which is operated by the California Institute of Technology, under contract with the National Aeronautics and Space Administration under the Exoplanet Exploration Program.
L.L. acknowledges the support of CONACyT-AEM grant 275201, CONACyT-CF grant 263356, and UNAM DGAPA/PAPIIT grants IN112417 and IN112820.
T.H. acknowledges support from the European Research Council under the Horizon 2020 Framework Program via the ERC Advanced Grant Origins\,83\,24,28.
S.R.\ acknowledges support through the DFG Research Unit FOR2544
'Blue Planets around Red Stars', program RE 2694/8-1.
We thank the anonymous referee for comments and suggestions that have helped to improve the clarity of the paper.



\facility{VLA, Exoplanet Archive}

\appendix

\section{List of observed stars} \label{sec:app:targetlist}

\startlongtable
\begin{deluxetable*}{lllllllllll}
\tablecaption{List of observed stars \label{tbl-sourcelist}}
\tablewidth{0pt}
\tablehead{
\colhead{No.}       		&
\colhead{Star ID\tablenotemark{a}} & 
\colhead{RA(J2000)}         & 
\colhead{DEC(J2000)}        & 
\colhead{Dist.\tablenotemark{b}} &
\colhead{SpT}       		&
\colhead{Age}		      	& 
\colhead{Assoc.\tablenotemark{c}} 	    	&
\colhead{S\tablenotemark{d}}&
\colhead{P\tablenotemark{e}}&
\colhead{Age} 	            \\
\colhead{} 	      		    &
\colhead{} 			        & 
\colhead{}			        & 
\colhead{}			        & 
\colhead{[pc]}		      	&
\colhead{}		       		&
\colhead{[Myr]}		      	& 
\colhead{}		 		    &
\colhead{} 			        &
\colhead{} 			        &
\colhead{ref.} 			    
}
\startdata
1   & HD\,166       	 	& 00:06:36.785 & $+$29:01:17.40 	& 13.8	& K0V   	& 250 	& field 	& 1 & 2  & 1  \\  
2   & HD\,377              	& 00:08:25.746 & $+$06:37:00.49 	& 38.5	& G2V   	& 220	& field     & 1 & 15 & 1   \\ 
3   & HD\,691     	 		& 00:11:22.438 & $+$30:26:58.47 	& 34.3	& K0V   	& 339 	& field     & 1 & 1  & 2   \\ 
4   & HD\,984     	 		& 00:14:10.254 & $-$07:11:56.84  	& 45.9	& F5V   	& 444 	& field  	& 1 & 16 & 2  \\ 
5   & HD\,1405	 	 		& 00:18:20.890 & $+$30:57:22.23 	& 28.3	& K2V   	& 149 	& AB\,Dor  	& 1 & 3  & 3   \\ 
6   & HD\,4277         	 	& 00:45:50.889 & $+$54:58:40.20 	& 57.6	& F8V   	& 149 	& AB\,Dor  	& 7 & 23 & 3  \\ 
7   & HD\,5996    	 		& 01:02:57.223 & $+$69:13:37.42 	& 26.4	& G5V   	& 914 	& field     & 7 & 24 & 2    \\ 
8   & HD\,6569         	 	& 01:06:26.153 & $-$14:17:47.11  	& 45.4	& K1V   	& 149 	& AB\,Dor  	& 1 & 17 & 3    \\ 
9   & HIP\,6276        	 	& 01:20:32.268 & $-$11:28:03.74  	& 35.3	& G9V   	& 149 	& AB\,Dor  	& 1 & 18 & 3   \\ 
10   & HD\,10008     	 	& 01:37:35.466 & $-$06:45:37.53  	& 24.0	& G5V	    &  24 	& beta\,Pic & 1 & 19 & 3    \\ 
11   & {\bf BD+17\,232} 	& 01:37:39.410 & $+$18:35:33.26 	& 52.1	& K3VE  	& 10 	& field  	& 1 & 10 & 4   \\ 
12   & HD\,10195   	 		& 01:42:06.406 & $+$69:05:09.60 	& 45.4	& F5    	& 682 	& field  	& 7 & 25 & 1    \\ 
13   & HD\,10780      	 	& 01:47:44.835 & $+$63:51:09.00 	& 10.0	& K0V     	& 200 	& CARN  	& 7 & 26 & 5   \\ 
14   & BD--16\,351 			& 02:01:35.610 & $-$16:10:00.68 	& 88.2	& K1V(e)	&  42 	& field (COL) & 1 & 20  & 6, 3   \\ 
15   & HD\,13357   	 		& 02:10:52.079 & $+$13:40:59.79 	& 42.9	& G5IV    	& 2940  & field  	& 1 & 11  & 2    \\ 
16   & HD\,13482        	& 02 12:15.410 & $+$23:57:29.53 	& 32.1	& K1V     	& 149   & AB\,Dor  	& 1 & 12  & 3  \\ 
17   & HD\,14082\,B    		& 02:17:24.734 & $+$28:44:30.33 	& 39.7	& G2V       &  24	& beta\,Pic	& 1 & 13b & 3  \\ 
18   & HD\,14082\,A    		& 02:17:25.287 & $+$28:44:42.16 	& 39.7	& F5V   	&  24 	& beta\,Pic & 1 & 13a & 3    \\ 
19   & HD\,14062        	& 02:18:24.478 & $+$54:16:45.33 	& 406	& K0 	    &  24 	& beta\,Pic & 2 & 1   & 2    \\ 
20   & HD\,15013   	 		& 02:26:09.591 & $+$34:28:10.03 	& 44.4	& G5V   	& 112 	& field (PLE) & 1 &  5  & 7, 21, 22    \\ 
21   & HD\,15115        	& 02:26:16.245 & $+$06:17:33.19 	& 49.0 	& F4IV 	    &  45 	& THA  	    & 1 & 24 & 3    \\ 
22   & BD+30\,397\,B 		& 02:27:28.048 & $+$30:58:40.53   	& 41.1\tablenotemark{1}	& M2Ve 	& 24 & beta\,Pic & 1 & 4b & 3  \\ 
23   & HIP\,11437           & 02:27:29.254 & $+$30:58:24.61 	& 41.1	& K7V   	&  24  	& beta\,Pic & 1 & 4a & 3    \\ 
24   & HD\,15526        	& 02:29:35.032 & $-$12:24:08.63 	& 116	& G5/6V 	&  60  	& field  	& 1 & 25 & 8    \\ 
25   & HD\,16765      	 	& 02:41:13.997 & $-$00:41:44.38 	& 22.2	& F7IV	    & 611 	& field  	& 1 & 21 & 2    \\ 
26   & {\bf HIP\,12545}		& 02:41:25.888 & $+$05:59:18.42 	& 44.4	& K6Ve  	&  10 	& field 	& 1 & 22 & 8    \\ 
27   & {\bf HIP\,12635}   	& 02:42:20.949 & $+$38:37:21.16  	& 49.3	& K2V 	    & 149	& AB\,Dor	& 1 & 6b & 3  \\ 
28   & HD\,16760        	& 02:42:21.311 & $+$38:37:07.23 	& 69.5	& G2V 	    & 6420	& field  	& 1 & 6a & 2    \\ 
29   & HD\,17250   	 		& 02:46:14.609 & $+$05:35:33.33 	& 57.1  & F8    	&  45 	& THA  	    & 1 & 23 & 3    \\ 
30   & HD\,17190     	 	& 02:46:15.208 & $+$25:38:59.65 	& 25.3	& K2IV  	& 6710 	& field     & 1 & 9  & 2    \\ 
31   & HD\,17332\,A    		& 02:47:27.227 & $+$19:22:20.82 	& 33.1	& G6V   	& 149 	& AB\,Dor  	& 2 & 11a & 3    \\ 
32   & HD\,17332\,B    		& 02:47:27.421 & $+$19:22:18.56 	& 33.2	& G1V   	& 149 	& AB\,Dor  	& 2 & 11b & 3   \\ 
33   & {\bf V875\,Per} 		& 02:52:17.597 & $+$36:16:48.19 	& 250 	& K2IV  	&  63 	& field  	& 1 & 7  & 9   \\ 
34   & HD\,17925     	 	& 02:52:32.128 & $-$12:46:10.97 	& 10.4	& K1V	    & 200 	& field  	& 1 & 26 & 1   \\ 
35a   & {\bf TYC\,3301-2585-1} 	& 02:55:43.821 & $+$47:46:46.46 & 50.0	& K5Ve 	    &  42 	& COL\tablenotemark{4}  		& 2 & 2  & 3    \\ 
35b  & {\bf WDS\,02557+4746\,B} & 02:55:43.621	& $+$47:46:46.47\tablenotemark{2}	&  50.1 & ... & 42\tablenotemark{3} & COL & 2 & 2 & 3	\\
36  & TYC\,1794-419-1  		& 02:58:28.763 & $+$29:47:53.78 	& 187	& K0IV  	&  100 	& field  	& 1 & 8  & 9   \\ 
37   & HD\,18803      	 	& 03:02:26.026 & $+$26:36:33.26 	& 21.2	& G8V   	& 6540 	& field  	& 2 & 12 & 2    \\ 
38   & BD+29\,525     		& 03:07:59.210 & $+$30:20:26.07 	& 90.0  & G5IV  	& 160 	& field  	& 2 & 5  & 9    \\ 
39   & HD\,19668	 		& 03:09:42.288 & $-$09:34:46.58 	& 38.7	& G0V   	&  149 	& AB\,Dor  	& 1 & 27 & 3    \\ 
40   & TYC\,654-1274-1\tablenotemark{5}  & 03:10:12.545 & $+$14:36:03.00  & 110	& G6V & 63 	& field     & 2 & 14 & 9   \\ 
41   & HIP\,14809       	& 03:11:13.841 & $+$22:24:57.11 	& 50.7	& G5V 	    &  149  & AB\,Dor  	& 2 & 13 & 3   \\ 
42   & HD\,19994	 		& 03:12:46.437 & $-$01:11:45.96 	& 22.5	& F8V   	&  45 	& ARG  	    & 3 & 1  & 10   \\ 
43   & HD\,20385        	& 03:16:40.671 & $-$03:31:48.92 	& 48.8	& F6V   	&  45 	& THA  	    & 3 & 2  & 3    \\ 
44   & HD\,20367   	 		& 03:17:40.045 & $+$31:07:37.36 	& 26.1	& G0    	&  613 	& field  	& 2 & 6  & 2    \\ 
45   & BD-19\,660  			& 03:20:50.711 & $-$19:16:08.76 	& 44.1	& K7V   	&  42 	& COL  		& 3 & 6  & 3   \\ 
46   & HD\,21845        	& 03:33:13.491 & $+$46:15:26.53 	& 36.4	& G5V 	    &  149 	& AB\,Dor  	& 2 & 3a & 3    \\ 
47   & HD\,21845\,B     	& 03:33:14.04  	& $+$46:15:19.4  	& 36.4	& M0Ve  	&  149 	& AB\,Dor  	& 2 & 3b & 3   \\ 
48   & {\bf HD\,22213}      & 03:34:16.357 & $-$12:04:07.27 	& 51.5	& G7V 	    &  45 	& THA  	    & 3 & 7  & 3    \\ 
49   & HD\,22179        	& 03:35:29.904 & $+$31:13:37.44 	& 70.4	& G5IV  	&  30 	& field  	& 2 & 7  & 2    \\ 
50   & BD+27\,555     		& 03:44:24.243 & $+$28:12:23.19 	& 65.3	& G7V   	&  63 	& field  	& 2 & 8  & 9    \\ 
51a   & HD\,23524   	 	& 03:48:23.113 & $+$52:02:15.01 	& 51.7\tablenotemark{2}	& G8 & 42 & COL & 2 & 4  & 3    \\ 
51b  & {\bf HD\,23524\,B}   & 03:48:23.113 & $+$52:02:14.79\tablenotemark{6} & 51.7\tablenotemark{9} & K1V & 42\tablenotemark{3}	 & COL  & 2	& 4  & 3 \\
52   & {\bf HD\,24681}      & 03:55:20.406 & $-$01:43:45.21 	& 56.1	& G8V   	&   149  & AB\,Dor  & 3 & 3  & 3    \\  
53   & {\bf HD\,285281}     & 04:00:31.069 & $+$19:35:20.85 	& 135	& K1    	&   1.5  &  TAU 	& 2 & 15 & 9    \\ 
54   & BD-15\,705     		& 04:02:16.487 & $-$15:21:29.82 	& 54.3	& K3/4  	&   45 	& THA    	& 3 & 8  & 3    \\ 
55   & HD\,25457			& 04:02:36.745 & $-$00:16:08.12 	& 18.8	& F6V   	&   149 & AB\,Dor  	& 3 & 4  & 3    \\ 
56   & HD\,25680      	 	& 04:05:20.258 & $+$22:00:32.05 	& 16.9	& G5V	    &   45 	& ARG  	    & 2 & 17 & 10    \\ 
57   & {\bf HD\,284135}     & 04:05:40.579 & $+$22:48:12.03 	& 80\tablenotemark{7} & G3V: & 1.5 & TAU & 2 & 16 & 11    \\
58   & HD\,25953        	& 04:06:41.534 & $+$01:41:02.08 	& 57.0	& F5    	&   149 & AB\,Dor  	& 3 & 5  & 3   \\ 
59   & HD\,281691       	& 04:09:09.740 & $+$29:01:30.34 	& 110	& G8III 	&  16	& field  	& 2 & 19 & 2    \\ 
60   & HD\,25665     	 	& 04:09:35.039 & $+$69:32:29.01 	& 18.8  & G5V       &  149 	& AB\,Dor\tablenotemark{4}  	& 2 & 18 & 3    \\ 
61   & HD\,26182        	& 04:10:04.692 & $+$36:39:12.25 	& 107	& G0V  	    &  63 	& field  	& 2 & 9  & 9    \\ 
62   & HD\,284266       	& 04:15:22.917 & $+$20:44:16.90 	& 120	& K0V:  	&  1.5 	& TAU  		& 2 & 18 & 11    \\ 
63   & HD\,285751       	& 04:23:41.321 & $+$15:37:54.87 	& 104	& K2V:  	&  1.5 	& TAU  		& 2 & 20 & 11    \\ 
64   & HD\,279788       	& 04:26:37.399 & $+$38:45:02.28 	& 170	& G5V   	&   40 	& field  	& 2 & 10 & 9    \\ 
65   & HD\,28344   	 		& 04:28:48.297 & $+$17:17:07.67 	& 46.7	& G2V   	&  750 	& HYA  		& 2 & 21 & 12    \\ 
66   & HD\,285840       	& 04:32:42.433 & $+$18:55:10.23 	& 90.5	& K1V:  	&  1.5 	& TAU\tablenotemark{4}  		& 2 & 22 & 11   \\ 
67   & BD-12\,943     		& 04:36:47.102 & $-$12:09:20.67 	& 68.9	& K0V 	    &   45 	& THA  	    & 3 & 9  & 3    \\ 
68   & TYC\,91-82-1	 		& 04:37:51.493 & $+$05:03:08.62 	& 93.9	& K2V 	    &  150 	& field (AB\,Dor) & 3 & 26 & 6, 3  \\ 
69   & HD\,29623   	 		& 04:39:23.768 & $-$12:31:47.91 	& 68.8	& G0V   	&   90 	& field 	& 3 & 11 & 8    \\ 
70   & HD\,29697     	 	& 04:41:18.856 & $+$20:54:05.45 	& 13.2	& K3V   	&   257 & field  	& 2 & 23 & 2    \\ 
71   & HD\,29883    	 	& 04:43:35.436 & $+$27:41:14.64 	& 21.9	& K5V   	&   112 & field (PLE) & 2 & 24 & 7, 21, 22  \\ 
72   & HD\,30495      	 	& 04:47:36.291 & $-$16:56:04.04 	& 13.2	& G1.5V	    &    45 & ARG   	& 3 & 11 & 10    \\ 
73   & HD\,30652      	 	& 04:49:50.411 & $+$06:57:40.59 	& 8.0 	& F6V   	&  1570 & field  	& 3 & 27 & 2    \\ 
74   & {\bf HD\,31281}      & 04:55:09.622 & $+$18:26:31.13 	& 122 	& G1V:  	&   1.5 & TAU  		& 2 & 25 & 11   \\ 
75   & HD\,286179       	& 04:57:00.645 & $+$15:17:53.14 	& 124 	& G3V:  	&   1.5 & TAU  		& 2 & 26 & 11    \\ 
76   & HD\,31652        	& 04:57:22.323 & $-$09:07:59.63 	& 91.1	& G8V 	    &   150 & field (AB\,Dor) & 3 & 21 & 6, 3   \\ 
77   & {\bf BD-08\,995}     & 04:58:48.534 & $-$08:43:39.79 	& 87.3	& K0V 	    &   42  & field (COL) & 3 & 20 & 6, 3    \\ 
78   & BD-19\,1062    		& 04:59:32.027 & $-$19:17:41.66 	& 66.9	& K3V(e)    &    45 & field (THA) & 3 & 12 & 6, 3    \\ 
79   & {\bf HD\,286264}     & 05:00:49.287 & $+$15:27:00.71 	& 53.4	& K2IV      &    24 & beta\,Pic & 2 & 27 & 3    \\ 
80   & HD\,32981        	& 05:06:27.688 & $-$15:49:30.38 	& 85.1	& F8V       &   149 & AB\,Dor\tablenotemark{4}  	& 3 & 13 & 3    \\ 
81   & {\bf HD\,293857}     & 05:11:09.676 & $-$04:10:54.37 	& 78\tablenotemark{8}& G8V & 24 & beta\,Pic & 3 & 22 & 3    \\ 
82   & BD-09\,1108    		& 05:15:36.518 & $-$09:30:51.55 	& 82.5	& G5V 	    &    45 & field (THA) & 3 & 23 & 6, 3    \\ 
83   & HD\,35112   	 		& 05:22:37.491 & $+$02:36:11.49 	& 19.8	& G5V       &    45 & field (IC\,2391)& 3 & 28 & 13    \\ 
84   & BD-08\,1115    		& 05:24:37.249 & $-$08:42:01.76 	& 158	& G7V(e)    &    42 & field (COL) & 3 & 24 & 6, 3   \\ 
85   & HD\,35850	 		& 05:27:04.763 & $-$11:54:03.48 	& 26.9	& F8V(n)k:  &    24 & beta\,Pic & 3 & 14 & 3   \\ 
86   & BD-19\,1194    		& 05:30:19.075 & $-$19:16:31.85 	& 113	& G5V       &    42 & field (COL) & 3 & 16 & 6, 3    \\ 
87   & HD\,36869     	 	& 05:34:09.162 & $-$15:17:03.18 	& 57.9	& G3V       &    42 & COL  	    & 3 & 15 & 3    \\ 
88   & {\bf TYC\,713-661-1} & 05:36:50.055 & $+$13:37:56.11 	& 57.0	& K0V       &   149 & AB\,Dor  	& 3 & 30 & 3    \\ 
89   & TYC\,119-1242-1		& 05:37:45.335 & $+$02:30:57.52 	& 68.2	& K4V	    &    42 & COL\tablenotemark{4}   	    & 3 & 29a & 3   \\ 
90   & TYC\,119-497-1		& 05:37:46.501 & $+$02:31:26.44 	& 68.4	& K5V       &    42 & field (COL) & 3 & 29b & 6, 3   \\ 
91   & TYC\,4779-394-1  	& 05:38:56.636 & $-$06:24:40.97 	& 124	& G8V 	    &   150 & field (AB\,Dor) 	& 3 & 25 & 6, 3    \\ 
92   & {\bf TYC\,5925-1547-1}& 05:39:23.169 & $-$19:33:29.45 	& 77.9	& K1V       &   150 & field (AB\,Dor) 	& 3 & 17 & 6, 3    \\ 
93   & {\bf SAO\,150676}    & 05:40:20.732 & $-$19:40:10.99 	& 73.0	& G2V       &    42 & COL  	    & 3 & 18 & 3   \\ 
94   & BD-13\,1328    		& 06:02:21.897 & $-$13:55:32.59 	& 45.2	& K4V(e)    &   149 & AB\,Dor  	& 3 & 19 & 3    \\ 
95   & HD\,48737	 		& 06:45:17.364 & $+$12:53:44.13 	& 18.2	& F5IV	    &  1700 & field  	& 4 & 1  & 14     \\ 
96   & TYC\,1355-214-1  	& 07:23:43.592 & $+$20:24:58.66 	& 27.8	& K5Ve 	    &   149 & AB\,Dor  	& 4 & 2  & 3   \\ 
97   & GJ\,281     	 		& 07:39:23.039 & $+$02:11:01.19 	& 15.1	& K7        &   300 & field  	& 4 & 3  & 15   \\ 
98   & {\bf HD\,62237}      & 07:42:26.574 & $-$16:17:00.37 	& 124	& G5V 	    &    42	& field (COL) & 4 & 4  & 6, 3    \\ 
99   & GJ\,9251\,B   	 	& 08:07:08.777 & $+$07:22:58.39 	& 43.9	& K5        &   112 & field (PLE) 	& 4 & 8b & 7, 21, 22    \\ 
100  & GJ\,9251\,A 	 		& 08:07:09.095 & $+$07:23:00.13 	& 41.5	& K8   	    &   112 & field (PLE)	& 4 & 8a & 7, 21,22    \\ 
101a  & {\bf SAO\,135659}   & 08:13:50.993 & $-$07:38:24.61 	& 53.9	& K0	    &    42 & COL  	    & 4 & 5  & 3    \\ 
101b  & {\bf WDS\,08138-0738\,B} & 08:13:50.998 & $-$07:38:24.52\tablenotemark{6} & 53.9 & ...	& 42\tablenotemark{3} &	COL	& 4	& 5 & 3 \\
102   & HIP\,40774  	 	& 08 19 19.051 & $+$01 20 19.91 	& 22.4	& G5V	   &   200	& CARN  	& 4 & 6  & 1, 5   \\ 
103   & HD\,70573   	 	& 08 22 49.951 & $+$01 51 33.55 	& 59.3	& G6V   &   66 	& field 		& 4 & 7  & 2   \\ 
104   & HD\,70516        	& 08 24 15.656 & $+$44 56 58.96 	& 36.9	& G0V   &  280 	& field  		& 8 & 1  & 2    \\ 
105   & HIP\,42253     		& 08 36 55.782 & $+$23 14 47.95 	& 39.2	& K5V   &    45 & field (IC\,2391)    	& 4 & 17 & 7    \\ 
106a   & {\bf HD\,77407}   	& 09 03 26.973 & $+$37 50 24.98 	& 30.2	& G0V:  &   42 	& COL   		& 8 & 3  & 3    \\ 
106b  & {\bf WDS\,09035+3750\,B}& 09 03 26.977	& $+$37 50 26.70\tablenotemark{2}& 33.9 & M3-6V & 42\tablenotemark{3} & COL & 8 & 3 & 3 \\
107   & HD\,78141   	 	& 09 07 18.077 & $+$22 52 21.57 	& 25.3	& K0    &   232 & field  		& 4 & 18 & 2    \\ 
108   & {\bf HD\,82159}	    & 09 30 35.834 & $+$10 36 06.31 	& 35.2	& G9V   &  150  & field   	    & 4 & 14 & 16    \\ 
109   & {\bf HD\,82558}     & 09 32 25.568 & $-$11 11 04.70 	& 18.3	& K0V	&  43 	& field  	    & 4 & 9  & 2    \\ 
110   & HD\,82443     		& 09 32 43.759 & $+$26 59 18.70 	& 18.1	& K0V	&  204 	& field  	    & 4 & 19 & 2   \\ 
111   & HD\,82939   	 	& 09 36 04.278 & $+$37 33 10.36 	& 38.8	& G5V   &  112 	& field (PLE)	& 8 & 3  & 7, 21, 22    \\ 
112a  & {\bf GJ\,2079}     	& 10 14 19.177 & $+$21 04 29.55 	& 23.4	& M0.7V	&  24 	& field (beta\,Pic) & 4 & 15 & 24, 3    \\ 
112b  & {\bf WDS\,10143+2104\,B}& 10 14 19.180 & $+$21 04 29.48\tablenotemark{6}	& 23.4 & ... & 24\tablenotemark{3} & field (beta\,Pic) & 4 & 15 & 17	\\
113   & HD\,89449      		& 10 19 44.167 & $+$19 28 15.29 	& 20.9	& F6IV	&  3100 & field  	    & 4 & 16 & 18    \\ 
114   & HD\,90905        	& 10 29 42.229 & $+$01 29 28.04 	& 31.0	& G1V   &  350 	& field  		& 4 & 11 & 1   \\ 
115   & HD\,91901   	 	& 10 36 30.792 & $-$13 50 35.82 	& 31.9	& K2V   &  966 	& field  	    & 4 & 10 & 2    \\ 
116   & HD\,94765     		& 10 56 30.798 & $+$07 23 18.51 	& 17.3	& K0V	&  200	& field (Castor)& 4 & 12 & 7     \\ 
117   & HD\,96064     		& 11 04 41.474 & $-$04 13 15.92 	& 26.2	& G8V	&  233 	& field  	    & 4 & 13 & 2    \\ 
118   & HD\,98736   	 	& 11 21 49.343 & $+$18 11 24.02 	& 32.5	& G5    & 8920 	& field 		& 5 & 1  & 2    \\ 
119   & HD\,99303   	 	& 11 25 39.948 & $+$20 00 07.68 	& 31.5	& G5V   &  112 	& field	(PLE)	& 5 & 2  & 7, 21, 22  \\ 
120   & HD\,102195  		& 11 45 42.292 & $+$02 49 17.33 	& 29.4	& K0V   & 1650 	& field 		& 5 & 4  & 2    \\ 
121   & HD\,102392     		& 11 47 03.836 & $-$11 49 26.58 	& 25.9	& K4.5V & 5940	& field	        & 5 & 12 & 19   \\ 
122   & TYC\,870-798-1   	& 11 47 45.730 & $+$12 54 03.40 	& 61.4	& K5Ve  &  126 	& field  		& 5 & 3  & 9     \\ 
123   & HD\,104576       	& 12 02 39.454 & $-$10 42 49.10 	& 54.4	& G3V   &  451 	& field 	 	& 5 & 13 & 2    \\ 
124   & HD\,104860       	& 12 04 33.731 & $+$66 20 11.72 	& 45.2	& F8    &  249 	& field  		& 5 & 21 & 2    \\ 
125   & HD\,105631    		& 12 09 37.257 & $+$40 15 07.40 	& 25.1	& K0V   &  200 	& CARN  	    & 5 & 18 & 5    \\ 
126   & HD\,105963    		& 12 11 27.754 & $+$53 25 17.45 	& 39.9	& K0V	&  609	& field  	    & 5 & 19 & 2    \\ 
127   & HD\,107146       	& 12 19 06.502 & $+$16 32 53.86 	& 27.5	& G2V   &  150 	& field  		& 5 & 5  & 1    \\ 
128   & HD\,108574  		& 12 28 04.447 & $+$44 47 39.53 	& 45.6	& G5V   &   200 & CARN  	    & 5 & 20a & 5    \\ 
129   & HD\,108575  		& 12 28 04.800 & $+$44 47 30.48 	& 45.6	& K V   &   200 & CARN  	    & 5 & 20b & 5    \\ 
130   & HD\,108767  		& 12 29 51.855 & $-$16 30 55.55 	& 26.0	& K0V   &   128 & field  		& 5 & 14  & 2   \\ 
131   & HD\,108944       	& 12 31 00.736 & $+$31 25 25.80 	& 45.0	& F9V   &   160 & field  		& 5 & 6  & 9    \\ 
132   & HD\,109157  		& 12 32 27.436 & $+$28 05 04.64 	& 43.8	& G7V   &   112 & field (PLE)  	& 5 & 7  & 7, 21, 22    \\ 
133   & BD+60\,1417 		& 12 43 33.272 & $+$60 00 52.66 	& 45.0  & K0    &   203	& field  		& 5 & 22 & 2    \\ 
134   & HD\,111395     		& 12 48 47.048 & $+$24 50 24.82 	& 17.1	& G5V	&  1430 & field  	    & 5 & 8  & 2    \\ 
135   & HIP\,62686  	 	& 12 50 41.858 & $+$20 32 05.07 	& 38.3\tablenotemark{9}	& K5 &  45 & field (IC\,2391) & 5 & 9 & 7     \\ 
136   & HD\,111813     		& 12 51 38.409 & $+$25 30 31.81 	& 38.7	& K1V   &   45 	& field (IC\,2391) & 5 & 10 & 7     \\ 
137   & HD\,112196       	& 12 54 40.016 & $+$22 06 28.56 	& 35.2	& F8V   &  80 	& field 	 	& 5 & 11 & 9    \\ 
138   & HD\,113449     		& 13 03 49.655 & $-$05 09 42.52 	& 20.5	& K1V 	&  187 	& field  	    & 4 & 20 & 2    \\ 
139   & HD\,116956     		& 13 25 45.533 & $+$56 58 13.78 	& 21.7	& G9V	&  597 	& field  	    & 5 & 23 & 2    \\ 
140   & HD\,118100     		& 13 34 43.207 & $-$08 20 31.34 	& 20.5	& K5Ve  &  353 	& field  	    & 5 & 15 & 2   \\ 
141   & HIP\,67092  	 	& 13 45 05.340 & $-$04 37 13.23 	& 29.4	& K5    & 5200 	& field  		& 5 & 16 & 19   \\ 
142   & HD\,120352  		& 13 48 58.192 & $-$01 35 34.64 	& 33.6	& K0    &   45 	& field (IC\,2391) & 5 & 17 & 7     \\ 
143   & HD\,121979      	& 13 56 17.761 & $+$66 56 41.02 	& 46.0	& K0V   &   45 	& field (IC\,2391) & 5 & 24 & 7     \\ 
144   & {\bf HD\,135363}  	& 15 07 56.262 & $+$76 12 02.68 	& 29.6	& G5V   &   45 	& field (IC\,2391) & 5 & 25a & 7     \\ 
145   & HD\,139777			& 15 29 11.186 & $+$80 26 54.97	    & 21.8	& G1.5V	&  12000& field			& 5 & 26  & 20   \\ 
146	 & HD\,139813			& 15 29 23.594 & $+$80 27 00.97	    & 21.8	& G9V	&  235	& field			& 5 & 27  & 2  \\ 
147   & TYC\,486-4943-1 	& 19 33 03.758 & $+$03 45 39.67 	& 70.2	& K3V 	&  150 	& field (AB\,Dor) & 7 & 1  & 6, 3    \\ 
148   & HD\,189285       	& 19 59 24.103 & $-$04 32 06.20 	& 71.6	& G7V 	&  150 	& field (AB\,Dor) & 7 & 2  & 6, 3    \\ 
149   & BD-03\,4778    		& 20 04 49.361 & $-$02 39 20.31 	& 66.9	& K1V 	&  150 	& field (AB\,Dor) & 7 & 3  & 6, 3    \\ 
150   & HIP\,101262 	 	& 20 31 32.072 & $+$33 46 33.14 	& 26.9	& K5    &   45 	& field (IC\,2391)& 7 & 17 & 7    \\ 
151   & HD\,199058       	& 20 54 21.083 & $+$09 02 23.83 	& 75.6	& G5V 	&  244 	& field  	    & 7 & 5  & 2    \\ 
152   & TYC\,1090-543-1 	& 20 54 28.006 & $+$09 06 06.66 	& 74.9	& K4Ve  &  150 	& field (AB\,Dor) & 7 & 6 & 23, 3    \\ 
153a  & HD\,199143          & 20 55 47.674 & $-$17 06 51.04 	& 46.0	& F7V   &   24 	& beta\,Pic  	& 7 & 7 & 3    \\ 
153b &  {\bf HD\,199143\,B}	& 20 55 47.637 & $-$17 06 51.58\tablenotemark{6} &	46.0	& M2 & 24\tablenotemark{3} & beta\,Pic & 7 & 7 & 3	\\
154   & {\bf HD\,358623}	& 20 56 02.738 & $-$17 10 53.73 	& 45.9	& K6Ve  &   24 	& beta\,Pic  	& 7 & 8 & 3    \\ 
155   & {\bf SAO\,50350} 	& 21 00 47.108 & $+$45 30 10.91 	& 51.8	& F8 	&  200 	& field  	    & 7 & 18 & 2    \\ 
156   & HD\,201919       	& 21 13 05.271 & $-$17 29 12.61 	& 38.2	& K6Ve  &  149 	& AB\,Dor   	& 7 & 9  & 3    \\ 
157   & HD\,202575     		& 21 16 32.468 & $+$09 23 37.77 	& 16.2	& K3V   &  457 	& field 		& 7 & 4  & 2    \\ 
158   & HD\,203030  		& 21 18 58.219 & $+$26 13 49.95 	& 39.3	& G8V   &  445 	& field  	    & 7 & 13 & 2    \\ 
159   & {\bf GJ\,4199}     	& 21 31 01.713 & $+$23 20 07.37 	& 24.2	& K5Ve 	&  149 	& AB\,Dor  	    & 7 & 14 & 3   \\ 
160   & HD\,206860     		& 21 44 31.329 & $+$14 46 18.98 	& 18.1	& G0V	&  400	& field  	    & 7 & 15 & 2    \\ 
161   & HD\,209458  		& 22 03 10.772 & $+$18 53 03.54 	& 48.4	& G0V   & 6850 	& field  		& 7 & 16 & 2    \\ 
162   & HD\,209779     		& 22 06 05.336 & $-$05 21 28.91 	& 36.4	& G0V   &  838 	& field  	    & 7 & 10 & 2    \\ 
163   & HD\,211472    		& 22 15 54.139 & $+$54 40 22.40 	& 22.0	& K1V   &   45 	& ARG  	        & 7 & 20 & 10    \\ 
164   & {\bf SAO\,51891} 	& 22 20 07.026 & $+$49 30 11.76 	& 34.4	& K1V   &   37	& field		    & 7 & 21 & 2    \\ 
165   & {\bf SAO\,108142} 	& 22 44 41.545 & $+$17 54 18.30 	& 49.7	& K0    &  149 	& AB\,Dor  		& 7 & 11 & 3   \\ 
166   & HD\,217014       	& 22 57 27.980 & $+$20 46 07.79 	& 15.5	& G2.5IVa& 8810 & field  		& 7 & 12 & 2    \\ 
167   & HD\,218866  		& 23 10 24.624 & $+$64 31 47.58 	& 36.1	& F8    &  5000 & field 		& 7 & 22 & 20    \\ 
168   & HIP\,115162      	& 23 19 39.561 & $+$42 15 09.83 	& 51.3	& G4 	&   149	& AB\,Dor  	    & 7 & 19 & 3    \\ 
169   & GJ\,900     	 	& 23 35 00.276 & $+$01 36 19.44 	& 20.8	& K7V   &   200	& CARN  	    & 1 & 14 & 5   \\ 
170   & {\bf UCAC4\,832-014013}\tablenotemark{10}  & 15 07 57.226 & $+$76 13 59.15 & 29.7 & M4.5V   &  ... & field  & 5 & 25b & ... \\ 
\enddata
\tablenotetext{a}{Stars with VLA detections are marked boldface and listed in Tab.\,\ref{tbl-det} with additional information.}
\tablenotetext{b}{Distances are derived from GDR2 parallaxes unless noted otherwise. Values $\ge100$\,pc are rounded to integer pc.}
\tablenotetext{c}{Association if membership probability is $>$80\%\ according to Banyan\,$\Sigma$\ \citep{gagne2018}. Individual cases, in which membership probabilities are only 55\%\,--\,75\%, are marked. Association memberships adopted from other authors are listed in brackets.}
\tablenotetext{d}{Session No. (see Tab.\,\ref{tbl-obssessions}).}
\tablenotetext{e}{Pointing No. in Session. If multiple targets in one pointing, star at phase center is marked "a" and off-center star is marked "b".}
\tablenotetext{1}{GDR2 distance from HIP\,11437 adopted.} 
\tablenotetext{2}{Independent GDR2 coordinates and parallax.}
\tablenotetext{3}{Age estimate of primary adopted.}
\tablenotetext{4}{Membership probablility 55\%\,--\,80\%\ acccording to Banyan\,$\Sigma$, but association and MG age widely used in the literature.}
\tablenotetext{5}{The originally selected pointing center was star 2E\,0307.4$+$1424 at 03:10:14.20, $+$14:35:47.0. However, this star is not visible on any VIS or NIR images and the properties that are assigned to it resemble suspiciously clearly those of the nearby visible star TYC\,654-1274-1. No radio continuum emission was detected at either of the two positions.}
\tablenotetext{6}{No independent GDR2 coordinates, WDS relative coordinates used.}
\tablenotetext{7}{No GDR2 nor {\it Hipparcos} parallax, distance from \citet{carpenter2009}.}   
\tablenotetext{8}{No GDR2 nor {\it Hipparcos} parallax, distance from \citet{dasilva2009}.} 
\tablenotetext{9}{No GDR2 parallax, distance from \it{Hipparcos} \citep{leeuwen2010}.} 
\tablenotetext{10}{UCAC4\,832-014013 was not in our original target list, but turned out to be the only non-targeted star from which we detected 6\,GHz radio emission (see Sect.\,\ref{ssec:res:overview}).}   
\tablerefs{
(1)\,\cite{pearce2022};
(2)\,\cite{baffles2020};
(3)\,\cite{bell2015}; 
(4)\,\cite{galicher2016};
(5)\,\cite{zuckerman2006};
(6)\,\cite{dasilva2009};
(7)\,\cite{montes2001};
(8)\,\cite{weise2010};
(9)\,\cite{carpenter2009};
(10)\,\cite{zuckerman2018};
(11)\,\cite{KH1995};
(12)\,\cite{BH2015};
(13)\,\cite{maldonado2010}; 
(14)\,\cite{DH2015};
(15)\,\cite{nielsen2019};
(16)\,\cite{desidera2015};
(17)\,\cite{nakajima2012};
(18)\,\cite{gaspar2016};
(19)\,\cite{brandt2014b};
(20)\,\cite{holmberg2009};
(21)\,\cite{gagne2018};
(22)\,\cite{dahm2015};
(23)\,\cite{desidera2021};
(24)\,\cite{binks2016}.
}
\end{deluxetable*}


\section{Analysis of proper motion anomalies} \label{sec:app:pma}

Here we analyze the the PMa of those five target stars with 6\,GHz detections for which \citet{kervella2019} found a significant ($>3\,\sigma$) discrepancy between the long-term ({\it Hipparcos}\,--\,{\it Gaia}) proper motion vector and the GDR2 measurements, which could be indicative of the presence of a perturbing secondary object. In this analysis, we follow the formalism laid out by  \citet{kervella2019}. Masses and their uncertainties of the secondary components are estimated from the spectral type of the secondary (where listed in the WDS) and from the {\it V} magnitude difference together with BT-Settl evolutionary models \mbox{\citep{allard2014,baraffe2015}}. The current angular separations are adopted from the individual measurements obtained from the WDS \citep{mason2020}. 

\begin{figure}[ht!]
\epsscale{.6}
\plotone{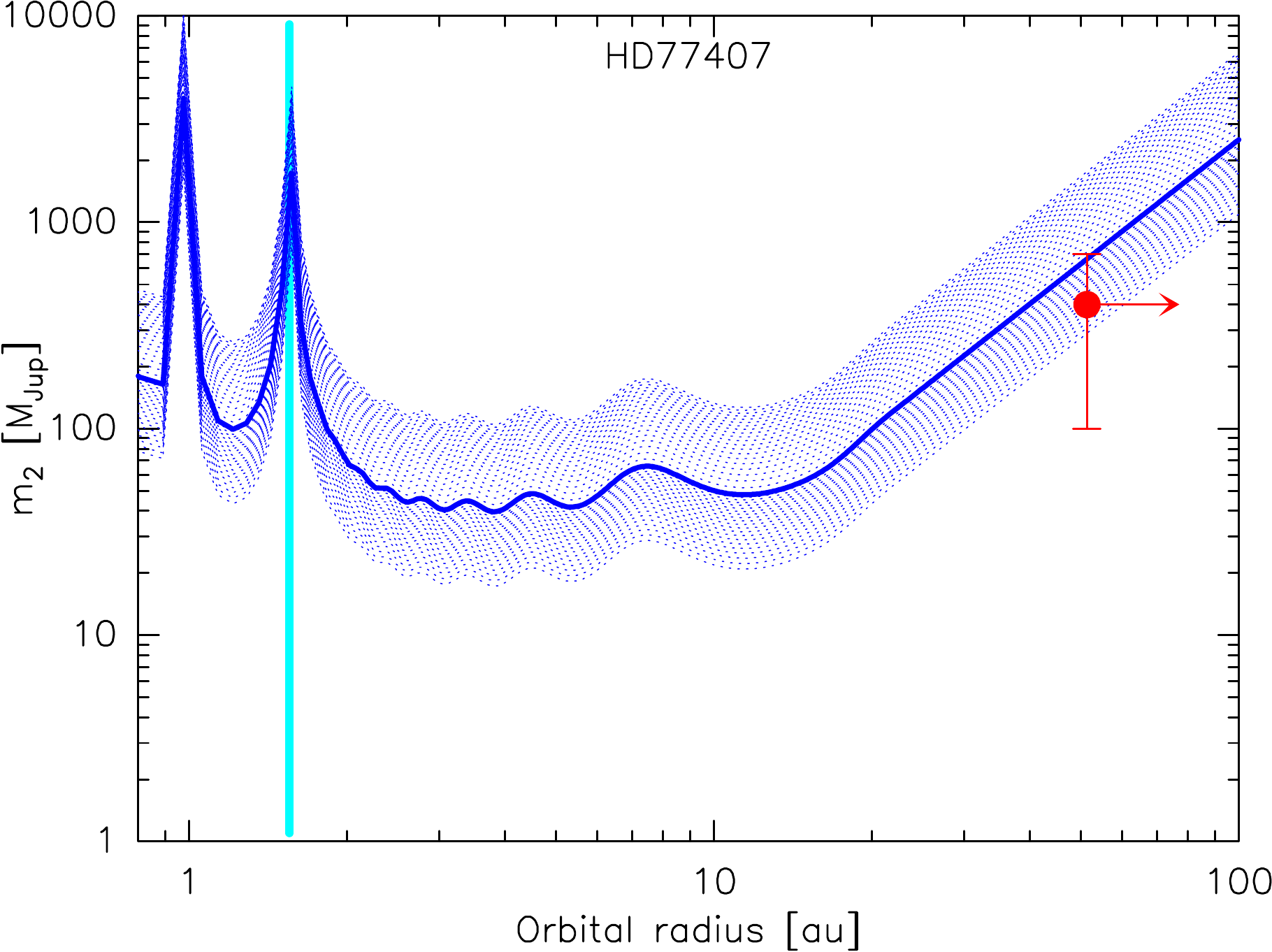}
\caption{\label{fig:pma_hd77407} 
Analaysis of PMa after \citet{kervella2019} for HD\,77407. The blue line and shaded area (1\,$\sigma$ uncertainty domain) show the relation between a companion’s mass (in Jupiter masses) and its orbital radius (in au) that can explain the observed PMa, taking into account the GDR2 time window (668\,d). The cyan vertical line marks the orbital radius the period of which corresponds to the GDR2 time window. The red dot marks the parameters of the known visual companion WDS\,09035+3750\,B. The arrow indicates that the current projected separation is a lower limit to the orbital radius (assuming zero or low eccentricity).}
\end{figure}

\begin{figure}[ht!]
\epsscale{.6}
\plotone{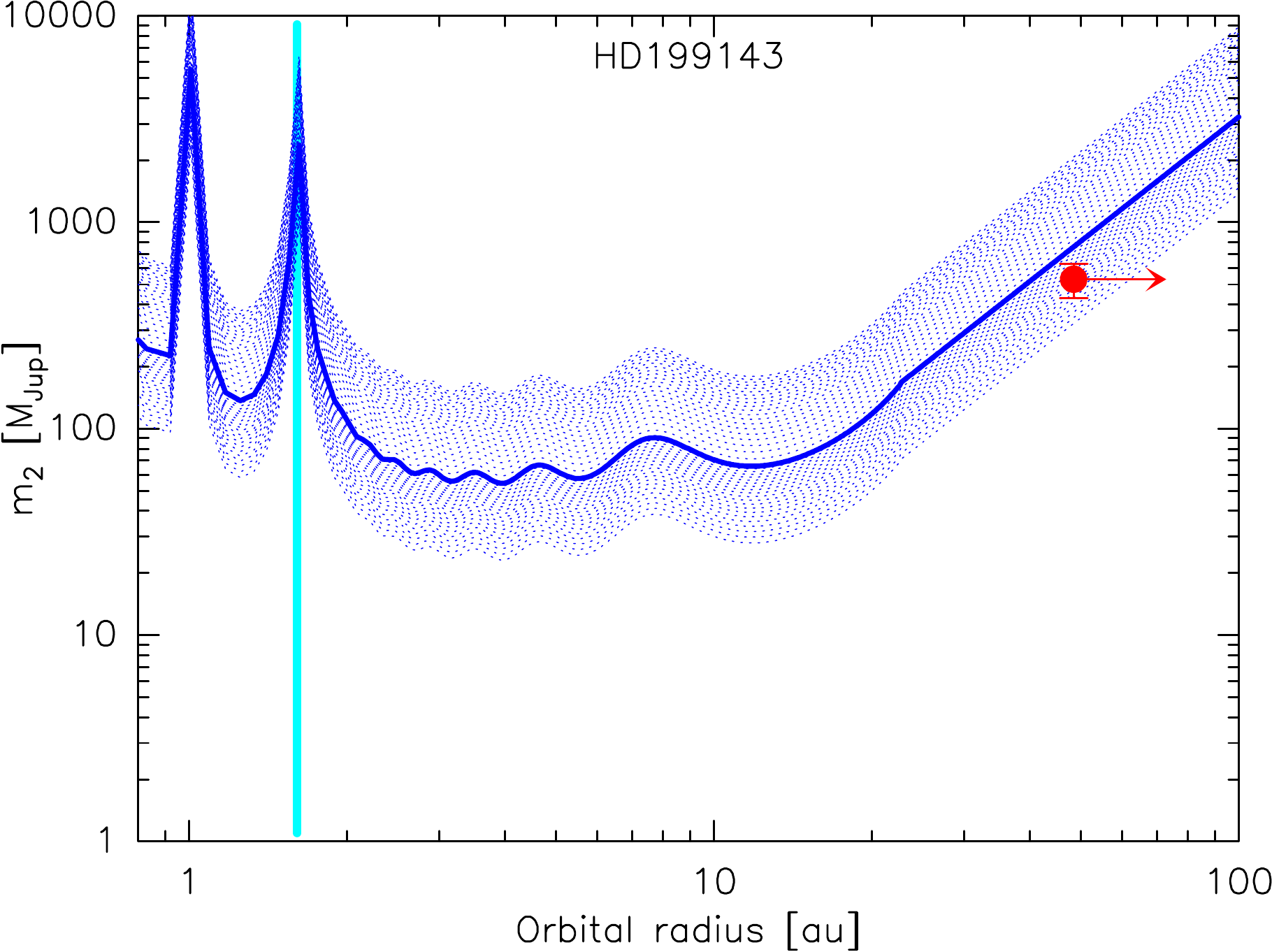}
\caption{\label{fig:pma_hd199143} 
Same as Fig.\,\ref{fig:pma_hd77407} for HD\,199143. The red dot marks the parameters of the known visual companion WDS\,09035+3750\,B. 
}
\end{figure}

\begin{figure}[ht!]
\epsscale{.6}
\plotone{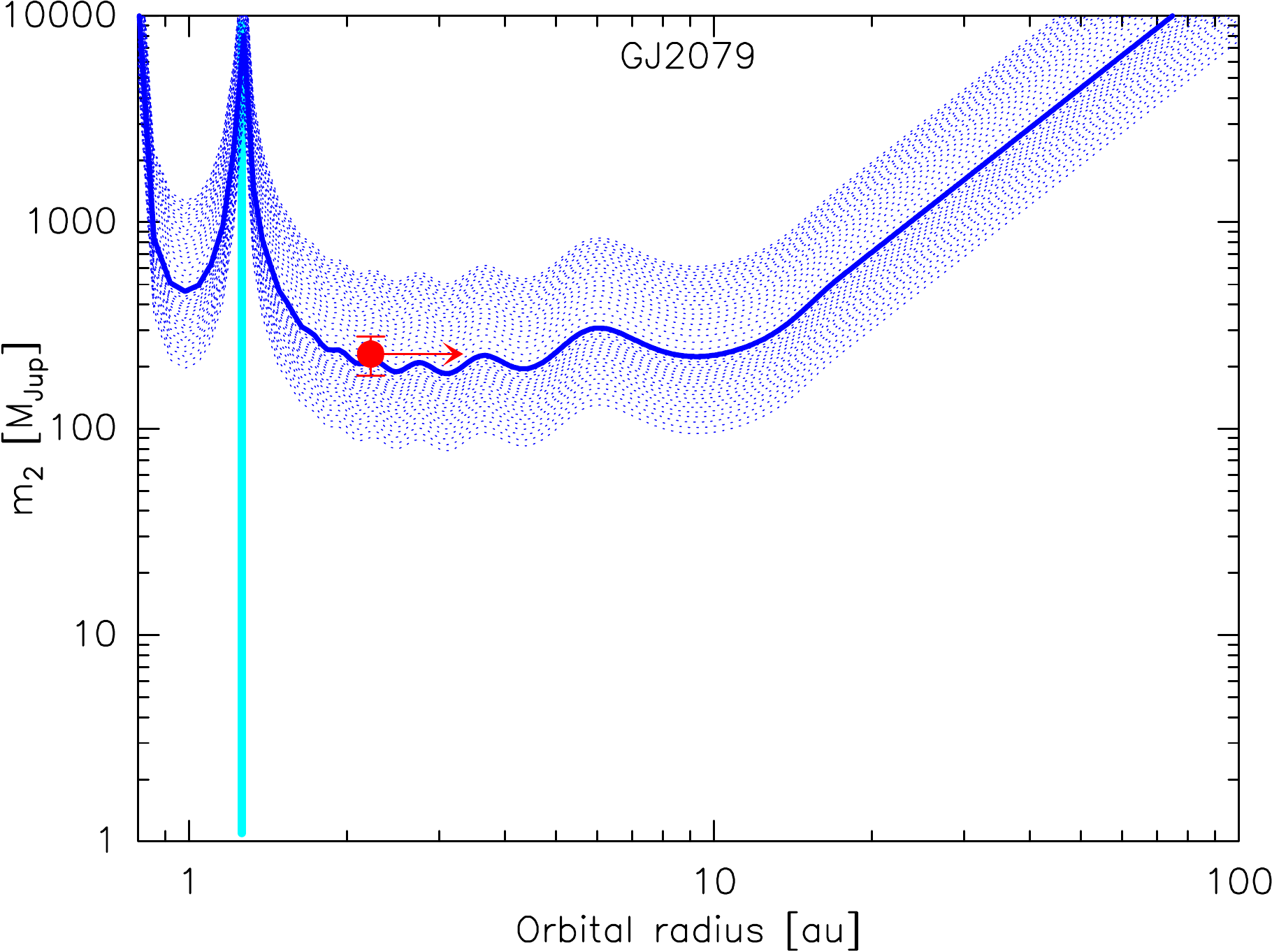}
\caption{\label{fig:pma_gj2079} 
Same as Fig.\,\ref{fig:pma_hd77407} for GJ\,2079. The red dot marks the parameters of the known visual companion WDS\,09035+3750\,B. 
}
\end{figure}

\begin{figure}[ht!]
\epsscale{.6}
\plotone{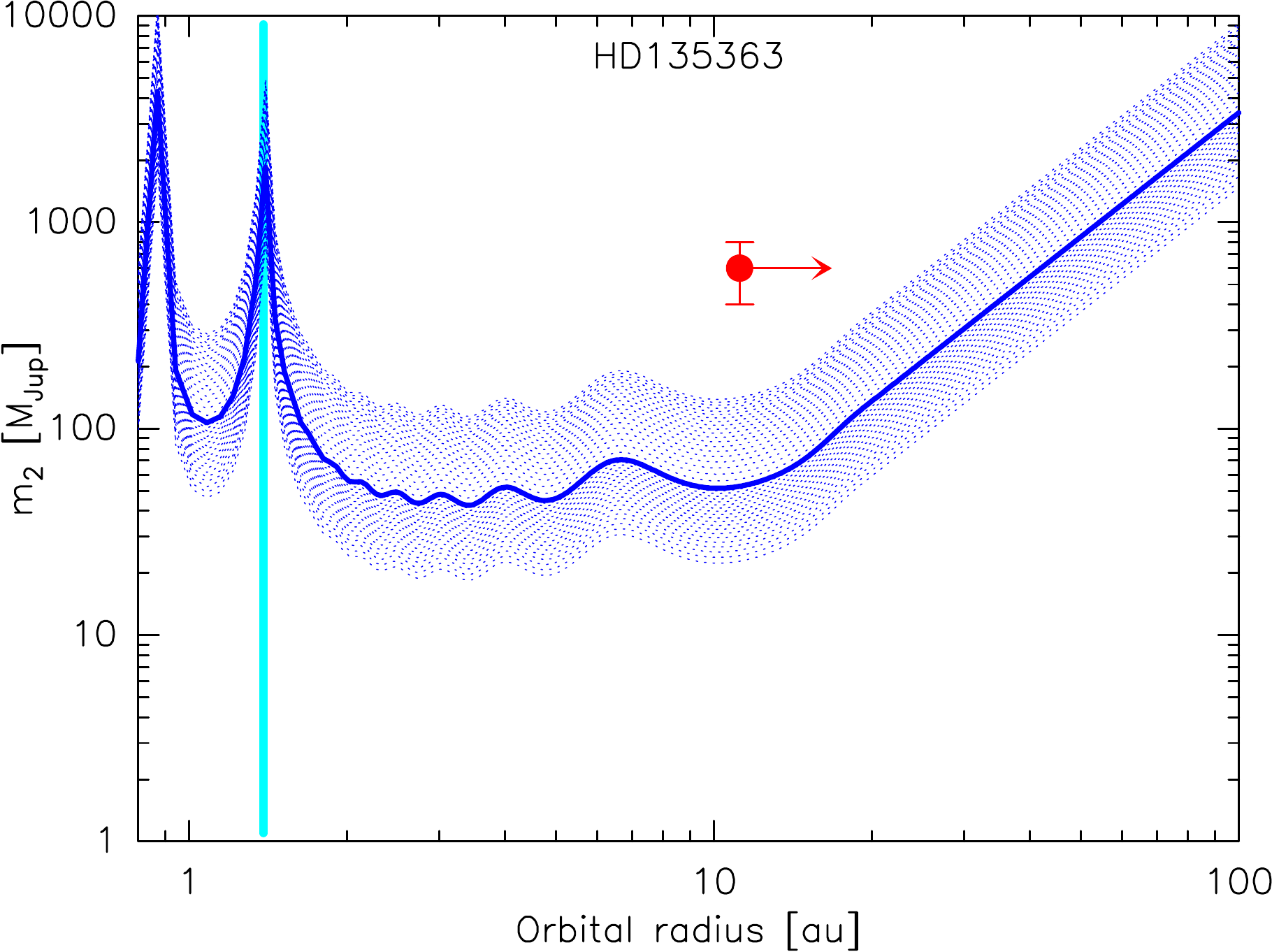}
\caption{\label{fig:pma_hd135363} 
Same as Fig.\,\ref{fig:pma_hd77407} for HD\,135363. The red dot marks the parameters of the known visual companion WDS\,09035+3750\,B. 
}
\end{figure}

\begin{figure}[ht!]
\epsscale{.6}
\plotone{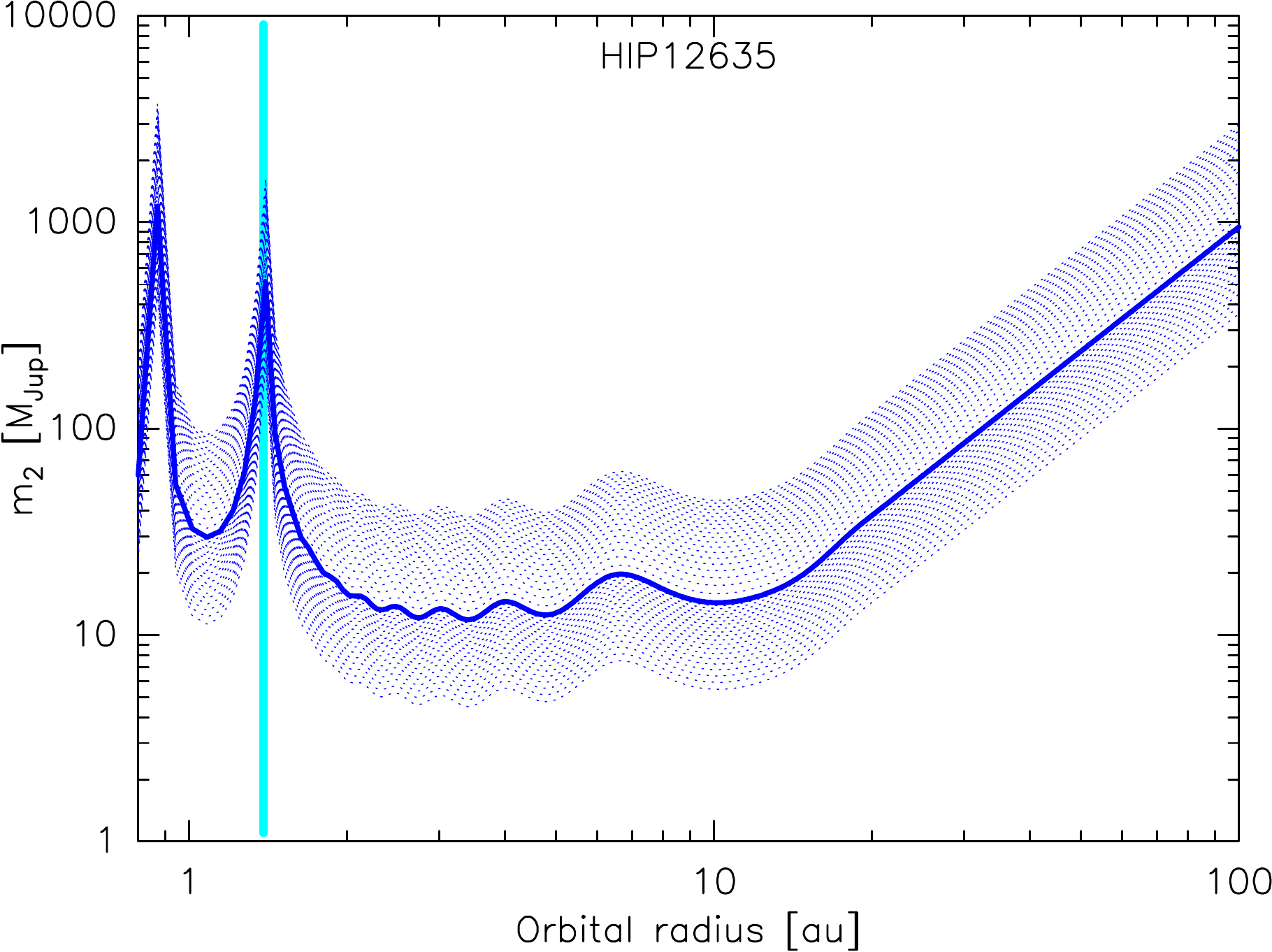}
\caption{\label{fig:pma_hip12635} 
Same as Fig.\,\ref{fig:pma_hd77407} for HIP\,12635. There is no known companion that could explain the observed PMa.}
\end{figure}


\bibliography{rlbib}
\bibliographystyle{aasjournal}




\end{document}